\documentclass[aps,prl,nobibnotes,twocolumn,superscriptaddress,bibliography]{revtex4-2}

\usepackage{amsfonts}
\usepackage{mathrsfs}
\usepackage{amsmath}% needed for subequations
\usepackage{color}
\usepackage{graphicx}
\usepackage{bm}% bold maths
\usepackage{amssymb}
\usepackage{xspace}
\usepackage{epstopdf}
\usepackage{dcolumn}% Align table columns on decimal point
\usepackage{longtable}
\usepackage{multirow}
\usepackage{float}
\usepackage{comment}
\usepackage{tabularx}
\usepackage{dsfont}
\usepackage{fontenc}
\usepackage{verbatim}
\usepackage{color}
\usepackage{natbib}
\usepackage{bm}% bold maths
\usepackage{amssymb}
\usepackage{xspace}
\usepackage{epstopdf}
\usepackage{dcolumn}
\usepackage{multirow}
\usepackage{tabularx}
\usepackage{bbding}
\usepackage{booktabs}
\usepackage{placeins}
\usepackage[colorlinks=true, letterpaper=true, pdfstartview=FitV, linkcolor=blue, citecolor=blue, urlcolor=blue]{hyperref}

\usepackage[colorlinks=true, letterpaper=true, pdfstartview=FitV, linkcolor=blue, citecolor=blue, urlcolor=blue]{hyperref}

\makeatother

\begin{document}

\title{Ferroelectrically Controlled Chirality Switching of Weyl Quasiparticles}

\author{Zeling Li}
%\thanks{These authors contributed equally to this work.}
\address{Research Center for Quantum Physics and Technologies, School of Physical Science and Technology, Inner Mongolia University, Hohhot 010021, China}
\address{Key Laboratory of Semiconductor Photovoltaic Technology and Energy Materials at Universities of Inner Mongolia Autonomous Region, Inner Mongolia University, Hohhot 010021, China}

\author{Yu Liu}
%\thanks{These authors contributed equally to this work.}
\address{Research Center for Quantum Physics and Technologies, School of Physical Science and Technology, Inner Mongolia University, Hohhot 010021, China}
\address{Key Laboratory of Semiconductor Photovoltaic Technology and Energy Materials at Universities of Inner Mongolia Autonomous Region, Inner Mongolia University, Hohhot 010021, China}

\author{Le Du}
\address{Research Center for Quantum Physics and Technologies, School of Physical Science and Technology, Inner Mongolia University, Hohhot 010021, China}
\address{Key Laboratory of Semiconductor Photovoltaic Technology and Energy Materials at Universities of Inner Mongolia Autonomous Region, Inner Mongolia University, Hohhot 010021, China}

\author{Fengyu Li}
\address{Research Center for Quantum Physics and Technologies, School of Physical Science and Technology, Inner Mongolia University, Hohhot 010021, China}
\address{Key Laboratory of Semiconductor Photovoltaic Technology and Energy Materials at Universities of Inner Mongolia Autonomous Region, Inner Mongolia University, Hohhot 010021, China}

\author{Zhifeng Liu}
\address{Research Center for Quantum Physics and Technologies, School of Physical Science and Technology, Inner Mongolia University, Hohhot 010021, China}
\address{Inner Mongolia Key Laboratory of Microscale Physics and Atom Innovation, Inner Mongolia University, Hohhot 010021, China}

\author{Lei Li}
\address{Research Center for Quantum Physics and Technologies, School of Physical Science and Technology, Inner Mongolia University, Hohhot 010021, China}

\author{Lei Wang}
\email{lwang@imu.edu.cn}
\address{Research Center for Quantum Physics and Technologies, School of Physical Science and Technology, Inner Mongolia University, Hohhot 010021, China}
\address{Inner Mongolia Key Laboratory of Microscale Physics and Atom Innovation, Inner Mongolia University, Hohhot 010021, China}

\author{Botao Fu}
\email{fubotao2008@gmail.com}
\address{College of Physics and Electronic Engineering, Center for Computational Sciences, Sichuan Normal University, Chengdu 610068, China}

\author{Xiao-Ping Li}
\email{xpli@imu.edu.cn}
\address{Research Center for Quantum Physics and Technologies, School of Physical Science and Technology, Inner Mongolia University, Hohhot 010021, China}
\address{Key Laboratory of Semiconductor Photovoltaic Technology and Energy Materials at Universities of Inner Mongolia Autonomous Region, Inner Mongolia University, Hohhot 010021, China}

\begin{abstract}
Weyl quasiparticles, as gapless low-energy excitations with nontrivial chirality, have garnered extensive interest in recent years. However, archieving effective and reversible control over their chirality (topological charge) remains a major challenge due to topological protection. 
In this Letter, we propose a ferroelectric mechanism to switch the chirality of Weyl phonons, where the reversal of ferroelectric polarization is intrinsically coupled to a simultaneous reversal of the chirality of Weyl points. This enables electric-field-driven control over the topological properties of phonon excitations.
Through a comprehensive symmetry analysis of polar space groups, we identify 27 groups capable of hosting symmetry-protected Weyl phonons with chiral charges $C = 1$, $2$, and $3$, whose chirality can be reversed via polarization switching. The first-principles calculations are performed to screen feasible material candidates for each type of chirality, yielding a set of prototypical ferroelectric compounds that realize the proposed mechanism.
As a representative example, K$_2$ZnBr$_4$ hosts the minimal configuration of two pairs of Weyl phonons. Upon polarization reversal, the chirality of all Weyl points is inverted, accompanied by a reversal of associated topological features such as Berry curvature and surface arcs. These findings provide a viable pathway for dynamic, electrical control of topological band crossings and open different avenues for chirality-based phononic applications.
\end{abstract}

\maketitle

\textit{\textcolor{blue}{Introduction}}\textit{.}
The past decade has witnessed the rapid development of topological semimetals and emergent quasiparticles, among which the most important type is Weyl quasiparticles~\cite{RevModPhys.90.015001, herring1937accidental, murakami2007phase, PhysRevB.83.205101}. In Weyl semimetals, the low-energy bands cross to form a singular twofold degenerate point in three-dimensional (3D) momentum space, known as Weyl points. The existence of Weyl points requires the breaking of either time-reversal symmetry $\mathcal{T}$ or inversion symmetry $\mathcal{I}$, so as to ensure their two fold degeneracy. In addition, the existence of Weyl points does not require any other symmetry. Therefore, Weyl points can exist not only in electronic systems~\cite{PhysRevLett.107.186806, PhysRevLett.107.127205, PhysRevB.84.235126, PhysRevLett.108.266802, PhysRevB.85.035103, PhysRevB.90.155316, PhysRevB.90.155316, PhysRevX.5.011029, yu2022encyclopedia, soluyanov2015type}, but also in bosonic systems such as phonons~\cite{PhysRevB.96.064106, ji2017topological, PhysRevLett.120.016401, PhysRevB.97.054305, PhysRevLett.123.065501, PhysRevB.100.081204}. A large number of materials hosting Weyl electrons~\cite{PhysRevB.100.081204, tang2019comprehensive, vergniory2019complete, huang2015weyl, PhysRevB.92.161107, PhysRevLett.116.226801, PhysRevLett.117.066402, PhysRevLett.117.236401, PhysRevB.97.060406, PhysRevB.97.235416, xu2017discovery, PhysRevB.93.201101, chang2018topological, PhysRevB.99.245147, huang2016new, PhysRevB.103.L081402, chang2016room, yan2017topological} and Weyl phonons~\cite{xu2024catalog, liu2020symmetry, li2021computation, yang2024topological, qin2024diverse, ding2024topological, PhysRevLett.124.105303, PhysRevB.102.125148, PhysRevB.103.L161303, PhysRevB.103.094306, PhysRevB.106.214308, PhysRevB.106.214309, PhysRevB.106.195129, PhysRevB.108.104110, PhysRevB.109.235415, PhysRevB.109.045203, lange2024negative} have been predicted, and some of them have been experimentally verified~\cite{RevModPhys.93.025002, lv2015observation, PhysRevX.5.031013, xu2015discovery, yang2015weyl, deng2016experimental, PhysRevB.94.121113, PhysRevLett.121.035302, PhysRevLett.131.116602, PhysRevLett.123.245302, PhysRevB.106.224304}. 

As the most prominent topological feature, the Weyl point can be regarded as a point source of the Berry curvature field~\cite{RevModPhys.90.015001}. The Berry flux integral of Weyl point enclosed by a Gaussian sphere is a quantized topological charge, or chirality. 
The nontrivial chirality can give rise to fascinating phenomena, such as Fermi arc surface states~\cite{PhysRevB.83.205101, PhysRevX.5.031013, xu2015discovery, yang2015weyl}, the chiral anomaly~\cite{PhysRevB.86.115133, PhysRevB.88.104412, PhysRevLett.111.027201, PhysRevX.4.031035, PhysRevB.87.235306, PhysRevX.5.031023}, chiral Landau levels~\cite{PhysRevLett.111.246603, PhysRevLett.126.046401, xiong2015evidence} etc. This has motivated many efforts to modulate the chirality of Weyl points, reverse their sign, and thereby control the associated physical properties. 
For example, mechanical strain has been proposed to control the annihilation of Weyl points to change their chirality~\cite{PhysRevResearch.3.L042017}. Other methods such as magnetic field and optical approaches have also been reported to modulate chirality~\cite{thareja, yoshikawa2022non}. 
Unfortunately, these methods all rely on external perturbations. 
An intrinsic mechanism and means of control are still highly desired.

Ferroelectricity, an intrinsic material property characterized by a switchable spontaneous polarization, is often coupled with diverse material functionalities and has been widely employed to modulate topological states. However, most of these studies have focused on ferroelectric control of topological insulators~\cite{liu2016strain, PhysRevB.92.220101, PhysRevLett.117.076401, PhysRevLett.119.036802, PhysRevB.107.045125, jsm, zhao2018reversible}, while investigations into the ferroelectric modulation of Weyl semimetals remain scarce~\cite{sharma2019room, li2016weyl}. This scarcity is primarily due to the strong Coulomb screening effect in metallic or semimetallic systems, which suppresses the formation of long-range ferroelectric order, thereby hindering the coexistence of ferroelectricity and topological semimetal phases in real materials.
Crucially, this limitation applies mainly to electronic Weyl states. In contrast, the coupling between ferroelectricity and Weyl phonons is not constrained by electronic screening, thus opening up a promising avenue for the realization and control of Weyl phonons via ferroelectric polarization.

In this Letter, we propose a general mechanism for controlling the chirality of Weyl phonons via ferroelectric switching. Using symmetry analysis, we show that both ferroelectric polarization and Weyl point chirality transform identically under improper symmetry operations. This symmetry correspondence enables polarization reversal to induce a chirality flip, offering a universal route for electrical control of topological phonon states. Guided by symmetry analysis, we systematically screen 27 polar space groups that can host symmetry-protected Weyl phonons with switchable chiralities of $C = \pm1, \pm2, \pm3$.
A set of prototypical ferroelectric (FE) materials exhibiting tunable phonon chirality is identified through high-throughput first-principles calculations. As a representative example, K$_2$ZnBr$_4$ hosts two pairs of Weyl phonons with chirality $C = \pm1$ along a high-symmetry path. Upon polarization reversal, their chirality and associated topological features—including Berry curvature and surface states—undergo a complete sign change. Our work thus establishes an alternative principle for manipulating topological phonons and provides a realistic material platform for exploring electrically tunable chirality in crystalline solids.
%%%%%%%%%%%

%
\begin{figure}[t]
\includegraphics[width=8.4cm]{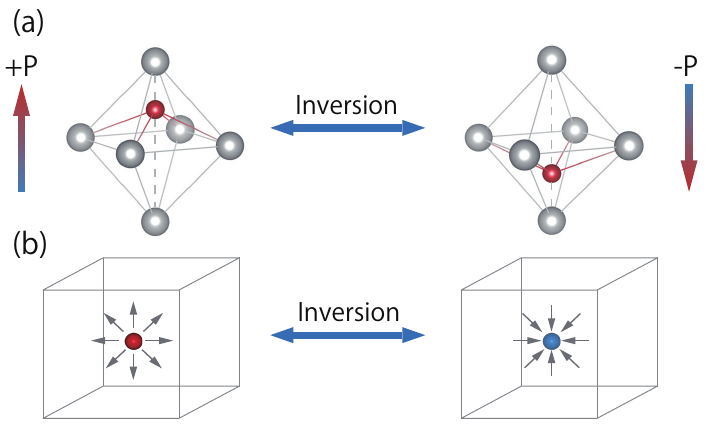}
\caption{Schematic of ferroelectric switchable chirality of Weyl points. (a) denotes two ferroelectric structures connected by inversion symmetry, with opposite electric polarizations $\mathbf{P}$. (b) represents the Weyl points in these structures, respectively, with opposite chiralities.
\label{fig1}}
\end{figure}
\begin{figure}[t]
\includegraphics[width=8.4cm]{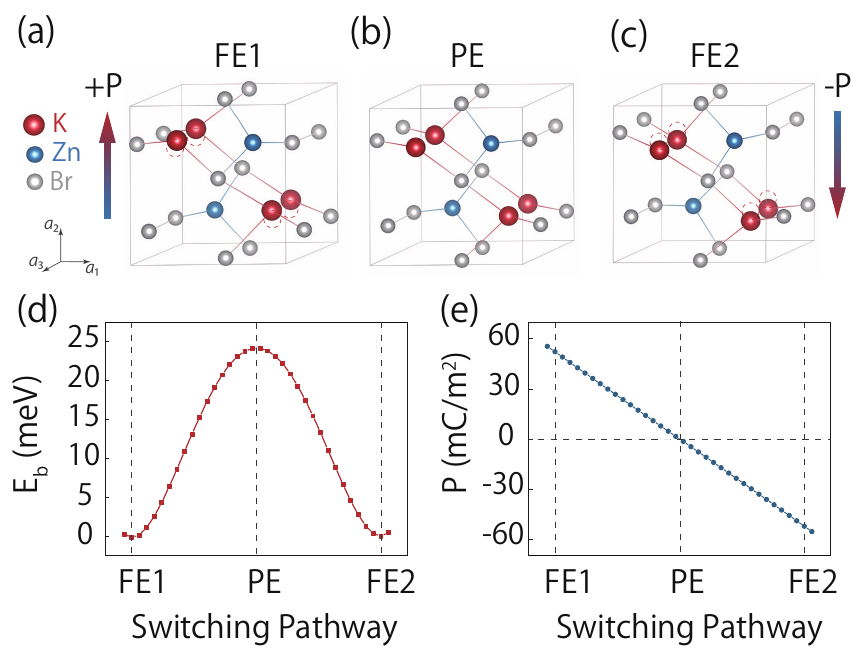}
\caption{(a) Structures of K$_{2}$ZnBr$_{4}$ in the FE1 state, (b) in the PE state, and (c) in the FE2 state. (d) and (e) show the calculated ferroelectric switching pathway and corresponding polarization in K$_{2}$ZnBr$_{4}$.  
\label{fig2}}
\end{figure}

\begin{table*}[tb]
		\centering
		\renewcommand\arraystretch{1.3}
		\caption{The candidate polar space groups that can host symmetry-protected Weyl points in spinless systems. SGs denotes the space group, and position denotes the high-symmetry point or high-symmetry line, where the Weyl points are located.}
		\begin{tabular}{llr} % 
			\hline\hline % 
			Type & \multicolumn{1}{c}{SGs (position)} & \multicolumn{1}{c}{Species}   \\ \hline
			& 3,4($\Gamma$-$Z$,$B$-$D$,$Y$-$C$,$A$-$E$); 5($\Gamma$-$Z$,$A$-$M$); 35-37($S$-$R$); 44-46($T$-$W$); 75-78($\Gamma$-$Z$,$M$-$A$,$X$-$R$); 79-80($\Gamma$-$Z$,$Z$-$V$,$X$-$P$); & \multirow{2}{*}{C-1 WP}\\
			&143-145($\Gamma$-$A$,$K$-$H$); 146($\Gamma$-$Z$,$Z$-$P$); 156,158($K$-$H$); 168-173($\Gamma$-$A$,$M$-$L$,$K$-$H$). &  \\
			& 75-78($\Gamma$-$Z$,$M$-$A$); 79,80($\Gamma$-$Z$,$Z$-$V$); 168-173($\Gamma$-$A$). & \multicolumn{1}{r}{C-2 WP}  \\
			\multirow[t]{-4}{*}{\raggedright HSL}  & 168-173($\Gamma$-$A$). & \multicolumn{1}{r}{C-3 WP}   \\
			 \hline
			& 80($P$); 168,171,172($K$,$H$); 169,170,173($K$). & \multicolumn{1}{r}{C-1 WP}   \\
			& 75,77($\Gamma$,$M$,$Z$,$A$); 76,78($\Gamma$,$M$); 79($\Gamma$,$Z$,$P$); 80($\Gamma$,$Z$); 143-145($\Gamma$,$A$); & \multirow{2}{*}{C-2 WP}  \\
			 \multirow[t]{-3}{*}{\raggedright HSP} & 146($\Gamma$,$A$); 168,171,172($\Gamma$,$A$); 169,170,173($\Gamma$). &    \\
			\hline\hline % 
		\end{tabular}
		\label{table1}
\end{table*}

\textit{\textcolor{blue}{Rationale and candidate space group.}} 
We analyze the principle of switching the chirality of Weyl points via ferroelectricity. Generally, two FE phases with opposite ferroelectric polarization strengths are connected by inversion symmetry $\mathcal{I}$ (or mirror operation $\mathcal{M}$). Accordingly, it is natural to consider that the relevant physical quantities in these two FE phases would also be correlated through $\mathcal{I}$, which provides a foundation for tuning their properties. It should be noted that polar tensors and axial tensors of physical tensors exhibit distinct properties under symmetric transformations. For instance, as a type of polar vector, the transformation of the electric polarization vector is given by $\mathbf{P}\stackrel{\mathcal{I}}{\rightarrow}-\mathbf{P}$ under $\mathcal{I}$ [as shown in Fig.~\ref{fig1}(a)]. %In contrast, the Berry curvature, being an axial vector in three dimension, remains unchanged in direction under $\mathcal{I}$, i.e., $\mathbf{\Omega}(\mathbf{k})\rightarrow\mathbf{\Omega}(-\mathbf{k})$.
In contrast, the Berry curvature, as an axial vector, transforms between the two ferroelectric phases under $\mathcal{I}$ as $\mathbf{\Omega}_{1}(\mathbf{k})\stackrel{\mathcal{I}}{\rightarrow}\mathbf{\Omega}_{2}(-\mathbf{k})$ (where 1 and 2 denote the two FE phases), with its sign unchanged. Since our target is the chirality switching of Weyl points, we need to focus on the transformation properties of the integral of the Berry curvature over a closed surface. Chirality can be expressed as $\mathcal{C}=\frac{1}{2\pi}\oint_{s}\mathbf{\Omega}(\mathbf{k})\cdot d\mathbf{S}(\mathbf{k})$. Under the operation $\mathcal{I}$, although the direction of the Berry curvature does not change, the differential surface element vector undergoes a sign change, $d\mathbf{S}_{1}(\mathbf{k})\stackrel{\mathcal{I}}{\rightarrow}-d\mathbf{S}_{2}(-\mathbf{k})$. As a result, the integral value $\mathcal{C}$ undergos a sign reversal under $\mathcal{I}$ [see Fig.~\ref{fig1}(b)]. The chirality reversal can also be explained by a simple $k\cdot p$ model. In the FE phase, the Weyl point is described by $H_{1}=\eta\boldsymbol{k}\cdot\boldsymbol{\sigma}$ with chirality $C=\textrm{sgn}(\eta)$. Under inversion $\mathcal{I}$, the polarization reverses, and the equation becomes $H_{2}=-\eta\boldsymbol{k}\cdot\boldsymbol{\sigma}$, causing the chirality to flip to $C=\textrm{sgn}(-\eta)$.  The scheme for controlling the chirality of Weyl points through ferroelectric polarization reversal has now been established‌.

We turn to investigate the candidate space groups for the ferroelectric control scheme. Fundamentally, the coexistence of the FE phase and the Weyl phase is crucial for achieving ferroelectric control. FE materials typically reside in specific polar space groups, all of which break inversion symmetry, thereby providing the possibility for the simultaneous existence of both the FE phase and the Weyl phase. Moreover, the existence of 3D Weyl points does not rely on additional symmetry protection. This implies that the effect of ferroelectric switching of phonon Weyl chirality can be realized in all polar space groups.

On the other hand, crystalline symmetries constrain Weyl points to specific high-symmetry points and high-symmetry paths within the Brillouin zone, giving rise to symmetry-protected Weyl points. This constraint not only facilitates theoretical analysis but also greatly benefits high-throughput material screening and subsequent experimental exploration. Building on this idea, we systematically investigate the presence of symmetry-protected Weyl phonons in polar space groups and identify potential candidate groups. Specifically, for each polar space group, we examine the two-dimensional irreducible representations (IRRs) of the little groups associated with all high-symmetry points in the Brillouin zone, as well as the relevant one-dimensional IRRs along high-symmetry paths~\cite{Bradley2009Mathematical-Oxford}. These IRRs serve as key indicators for the emergence of symmetry-protected Weyl points~\cite{yu2022encyclopedia}. Based on the IRR characteristics, we construct corresponding $k \cdot p$ models and identify three types of Weyl points within the candidate groups: charge-1 (C-1), charge-2 (C-2), and charge-3 (C-3) Weyl points. The main results are summarized in Table~\ref{table1}, with a more detailed version provided in the Supplemental Material (SM)~\cite{sm}, offering a guiding framework for identifying candidate materials.

\begin{figure*}[t]
\includegraphics[width=16.2cm]{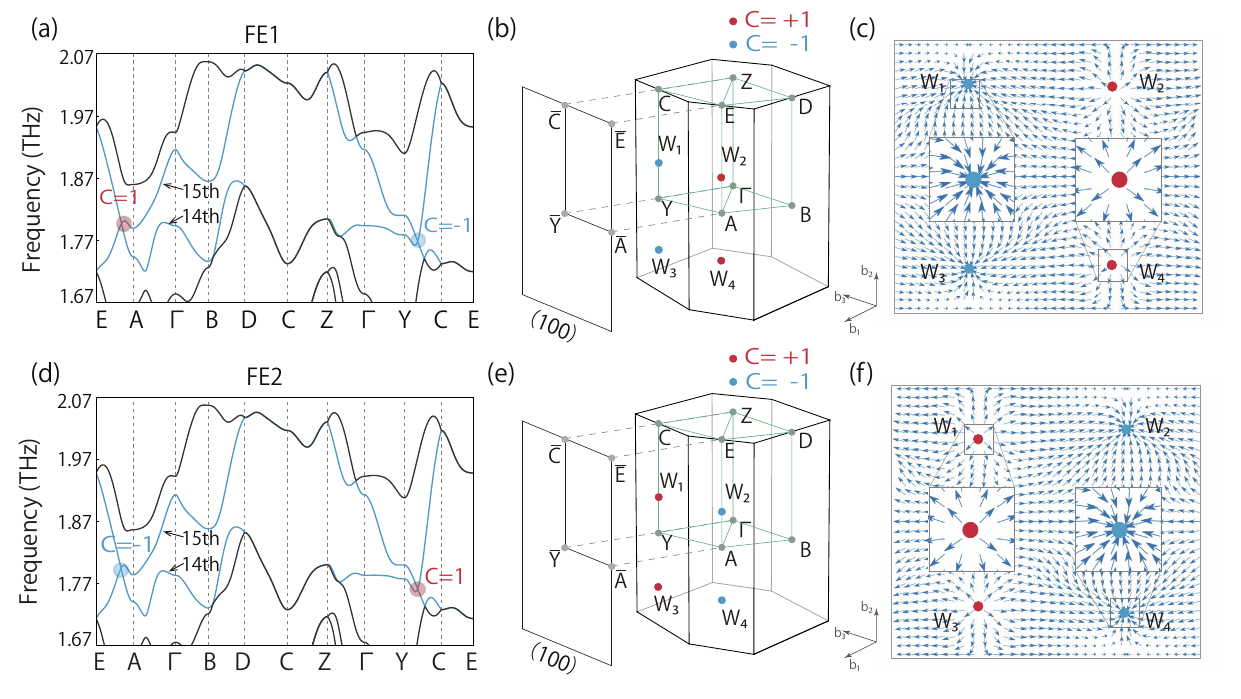}
\caption{(a)-(c) respectively show the calculated phonon spectrum, the bulk and surface BZs, and the distribution of Berry curvature in the $k_{x}=\pi$ for K$_{2}$ZnBr$_{4}$ in FE1 state. Similarly, (d)-(f) correspond to the same properties of K$_{2}$ZnBr$_{4}$ in the FE2 state. The phonon branches forming the Weyl points are highlighted by blue curves.
\label{fig3}}
\end{figure*}

\textit{\textcolor{blue}{High-quality material realization}}\textit{.}
Building on the symmetry-guided classification above, we further search for realizations of these Weyl phonons in experimentally synthesized ferroelectric materials with diverse crystal symmetries. Through high-throughput first-principles calculations, we identify high-quality candidate compounds for each type of Weyl phonon, and demonstrate their ferroelectrically switchable topological properties. Without loss of generality, we select K$_{2}$ZnBr$_{4}$, which hosts C-1 Weyl phonons, as a representative example to demonstrate the electric-field-driven chirality switching of Weyl points in the main text, with other candidates of ferroelectric-switchable Weyl phonon materials with chirality $C = 1, 2,$ and 3 presented in the SM~\cite{sm}.

K$_{2}$ZnBr$_{4}$ belongs to the ferroelectric family of $A$$_{2}$$BX$$_{4}$-type compounds ($A$ = Rb, K; $B$ = Zn, Co; $X$ = Cl, Br, I), which were successfully synthesized via aqueous solution methods and confirmed to exhibit ferroelectricity in 1990~\cite{shimizu1990new}. Its ferroelectric phase adopts a monoclinic structure with space group $P$2$_1$ (No. 4), which is among the candidate groups listed in Table~\ref{table1}. We now focus on the reversible chirality switching behavior in K$_{2}$ZnBr$_{4}$ as a concrete demonstration of the proposed mechanism.

To examine the feasibility of polarization switching of K$_{2}$ZnBr$_{4}$, we calculated the FE switching pathway and identified a centrosymmetric paraelectric (PE) phase (see Fig.~\ref{fig2}).
The crystal structure of PE phase is shown in Fig.~\ref{fig2}(b), where the K atoms are surrounded by ZnBr$_{4}$ units and possess the symmetry of space group $P$2$_{1}$/$m$ (No. 11). Upon displacement of the K atoms along the $b$ axis, the system presents two energetically degenerate FE ground states (FE1 and FE2) with opposite polarizations [see Figs.~\ref{fig2}(a) and~\ref{fig2}(c)]. Moreover, the two FE states are related by the inversion operation $\mathcal{I}$. Once Weyl points emerge in the phonon spectra of the FE phase, the effect of switching the chirality of Weyl points through ferroelectric switching can be realized.
%Furthermore, the intermediate states and activation barrier between FE1 and FE2 are presented in Fig~\ref{fig2}(b).

%
\begin{figure}[t]
\includegraphics[width=8.4cm]{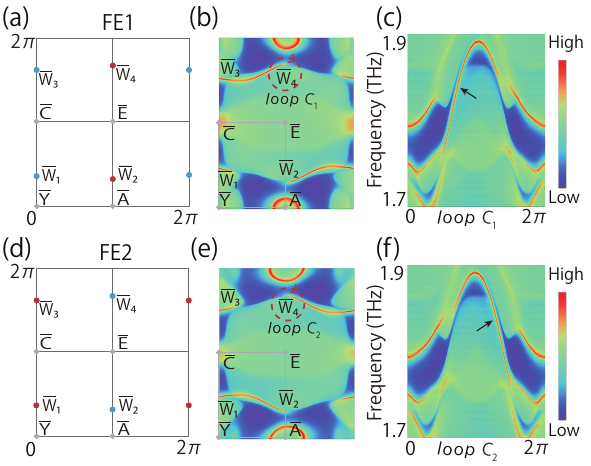}
\caption{(a) and (d) show the (100) surface BZ of K$_{2}$ZnBr$_{4}$ for FE1 and FE2 states, with $\overline{W}_{i=1,2,3,4}$ indicating projected Weyl points. (b) and (e) show the projected spectra on the (001) surface for the FE1 and FE2 states, with red dashed circles indicating the closed loops. (c) and (f) present the corresponding surface energy dispersion along these loops.
\label{fig4}}
\end{figure}

We then calculate the phonon  band structure of the FE1 state along high-symmetry paths in Brillouin zone (BZ). The FE1 state of K$_{2}$ZnBr$_{4}$ belongs to the SG 4, which, according to Table~\ref{table1}, may host Weyl points on the $E$-$A$, $Y$-$C$, $B$-$D$, or $\Gamma$-$Z$ paths owing to accidental band crossings.  Here, we plot the phonon bands within the frequency range of 1.6-2.1 THz, as illustrated in Fig.~\ref{fig3}(a) (the whole spectrum, including temperature and lattice anharmonicity effects, can be found in SM~\cite{sm}). One observes that there are indeed two WPs located on the $E$-$A$ and $Y$-$C$  paths, formed by the crossing of two phonon branches No. 14 and No. 15, %which are represented by red and blue balls, respectively. 
which are protected by a twofold screw rotation $\widetilde{C_{2y}}=\left\{ C_{2y}|0\frac{1}{2}0\right\} $~\cite{sm}. In addition, time-reversal symmetry results in the existence of two additional Weyl points on the opposite paths.
A careful scan reveals that apart from these four WPs, the 14th and 15th
bands do not form any other band degeneracies. Thus, the system contains only  two pairs of Weyl points, which is the minimal number allowed in nonmagnetic materials. 

The four Weyl points are labeled as $W_{i=1,2,3,4}$, and their distribution in the BZ is shown in Fig.~\ref{fig3}(b), where the red and blue balls represent chiralities of $+1$ and $-1$, respectively~\cite{sm}. Moreover, the $k\cdot p$ effective Hamiltonian of the four Weyl points $W_{i=1,2,3,4}$ share an identical form, given by~\cite{sm}
\begin{equation}\label{kp1}
\mathcal{H}(\boldsymbol{k})	=	\sigma_{0}(c_{1}+c_{2}k_{y})+c_{3}\sigma_{3}k_{y}+\sum_{j=1}^{2}\sigma_{j}(c_{j,1}k_{x}+c_{j,2}k_{z}),
\end{equation}
with $\sigma_{i}(i=1,2,3)$ the Pauli matrix and $\sigma_{0}$ the identity
matrix, where $c_{i=1,2,3}$, $c_{j,1,}$ and $c_{j,2}$ ($j$ = 1, 2) are real parameters.  According to Eq. (\ref{kp1}), the band splitting around the WPs is linear, implying that these Weyl points have a chirality of $\pm1$, which agrees with our calculations.

Furthermore, we turn to investigating the change of chirality of Weyl points in K$_{2}$ZnBr$_{4}$ under ferroelectric switching. The phonon spectrum of the FE2 state  is presented in Fig.~\ref{fig3}(d). It is important to note that the two polarization states give identical dispersions, as it should be. Naturally, the distribution of Weyl points in the two polarization states is also completely identical, but the chirality of the Weyl points is different. The calculated chirality of the Weyl points in the FE2 state is shown in Fig.~\ref{fig3}(e), where one can observe that their value has flipped signs compared to those in FE1, indicating the realization of the sign reversal of Weyl phonon chirality under ferroelectric switching.

\textit{\textcolor{blue}{Switching of topological properties.}} 
In addition to chirality, other topological quantities associated with Weyl points may also exhibit a flip during the ferroelectric switching. Among them, the Berry curvature field, as a significant quantity that characterizes the singularity of Weyl points with a ``source'' or a ``sink,'' exhibits a divergent geometric feature in the vicinity of the Weyl points~\cite{RevModPhys.90.015001}. Thus, it will inevitably undergo redistribution with the change of Weyl point chirality. We then plot the Berry curvature fields for two FE states on the $E$-$C$-$Y$-$A$ plane in the BZ. The results are shown in Figs.~\ref{fig3}(c) and~\ref{fig3}(f), respectively.

One can see that the two polarization phases exhibit completely opposite distribution characteristics. Specifically, near the Weyl point $W_{1}$, the Berry curvature in the FE1 state emanates from the Weyl point [see Fig.~\ref{fig3}(c)], while in the FE2 state, it is absorbed and converges at the same Weyl point [see Fig.~\ref{fig3}(f)]. The Berry curvatures of the other three Weyl points also exhibit similar opposite distributions in the two FE states and are consistent with their respective chiralities.

Another important quantity associated with the chirality of Weyl points is the topological surface state. It has been recognized that surface states of Weyl points exhibit a helicoid band structure surrounding the surface projections of the points, which can be described by a non compact Riemann surface~\cite{fang2016topological}. Notably, the helicoid direction of surface states is determined by the chirality of the Weyl point, and thus it may be reversed by changing chirality. %Moreover, the equal-energy contours of such helicoid surface states are Fermi arcs connecting the surface projections of nodes with opposite chirality.  
To verify this idea, we calculated the surface states of the (100) surface for K$_{2}$ZnBr$_{4}$ in two FE states (see Fig.~\ref{fig4}). In Figs.~\ref{fig4}(a) and~\ref{fig4}(d), the surface BZ with projected Weyl points is presented, while Figs.~\ref{fig4}(b) and~\ref{fig4}(e) show the corresponding Fermi arc states. One can observe that in both FE1 and FE2 states, a Fermi arc connects the projection of each Weyl point to another with opposite chirality, indicating a $\pm1$ chirality of bulk Weyl points. However, the surface states exhibit opposite helicoid structures in FE1 and FE2, owing to the reversed chirality of the projected Weyl points. For instance, consider a closed loop $C_{1}$ surrounding $\overline{W}_{4}$ in the FE1 phase, whose surface energy dispersion exhibits a right-moving chiral edge mode emerging inside the band gap [see Fig~\ref{fig4}(c)]. By contrast, the identical loop $C_{2}$ applied to the FE2 phase reveals a left-moving chiral edge mode in the surface dispersion [see Fig~\ref{fig4}(f)]. This switching with opposite helicoid is also applicable to other projected Weyl points, enabling the ferroelectric switching of the helicoid direction of topological surface states.

\begin{figure}[t]
\includegraphics[width=8.4cm]{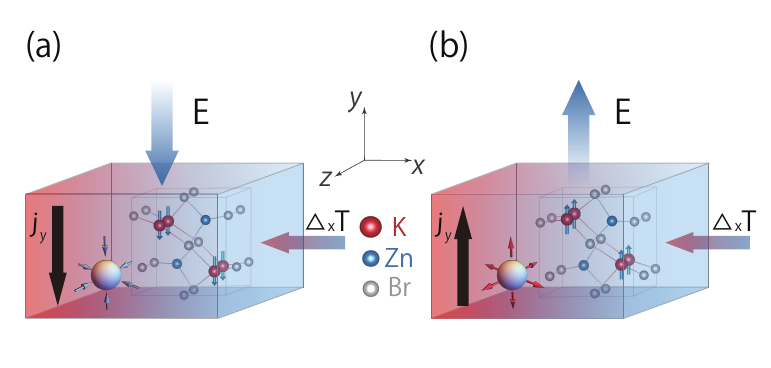}
\caption{Schematic diagram of the NPHE under the temperature gradient, where $E$ represents the applied external electric field. The black arrows indicate the deflection of the nonlinear thermal current.
\label{fig5}}
\end{figure}

%%%
It is particularly noteworthy that the topological charge of Weyl phonons plays a crucial role when they interact with other quasiparticles, such as electrons and photons. This makes the electric control of phonon chirality not only a fundamental advance in topological phononics, but also a powerful tool for manipulating diverse physical processes in ferroelectric systems~\cite{PhysRevB.107.L241107, chen2024electrically}. As illustrated in Figs.~\ref{fig5}(a) and~\ref{fig5}(b), we designed a device for the electrically controlled nonlinear phonon Hall effect (NPHE) based on K$_{2}$ZnBr$_{4}$ material~\cite{sm}. Applying a transverse temperature gradient ($\Delta_x \mathbf{T}$) to K$_{2}$ZnBr$_{4}$ induces a longitudinal nonlinear thermal current ($\mathbf{j_y}$), whose direction is directly linked to the Berry curvature of the Weyl phonons. Remarkably, reversing the ferroelectric polarization via an external electric field flips the phonon chirality and thus reverses the nonlinear thermal current $\mathbf{j_y}$. This electrically driven reversal provides a clear and experimentally accessible signature of chirality switching, highlighting the potential of ferroelectric Weyl systems for functional topological control and phonon-based information processing.

\textit{\textcolor{blue}{Conclusion.}} 
In summary, we present a general symmetry-based framework for electrically controlling the topological properties of Weyl phonons---including chirality, Berry curvature, and helicoid surface states---through displacement-type ferroelectricity. By systematically analyzing polar space groups, we identify 27 candidate symmetries that can host symmetry-protected Weyl phonons with tunable topological charges, providing a concrete guideline for the targeted search of real material realizations. Our high-throughput first-principles screening further reveals that experimentally synthesized ferroelectrics, such as K$_2$ZnBr$_4$ and its structural family, as well as Y(OH)$_{3}$ and LiIO$_{3}$~\cite{sm}, are ideal platforms for realizing electric-field-driven Weyl chirality switching in phononic systems. In contrast to conventional tuning methods (e.g., magnetic fields, strain, or doping), our approach utilizes the intrinsic, reversible polarization of ferroelectrics for robust and scalable topological control. The concept of chirality switching can be readily extended across a spectrum of systems, including electronic Weyl semiconductors~\cite{PhysRevLett.114.206401, zhang2020magnetotransport} and Weyl chiral magnons~\cite{li2016weylmagnons, PhysRevLett.117.157204}. Our work offers a promising strategy for developing phononic and information devices, with potential applications in phononic logic, spin-heat conversion, and nonreciprocal thermal transport.

\textit{\textcolor{blue}{Acknowledgment.}} 
This work is supported by the National Natural Science Foundation of China (Grants No. 12304086, No. 12304165, and No. 12204330). 

\textit{\textcolor{blue}{Data availability.}}
The data that support the findings of this article are not publicly available. The data are available from the authors upon reasonable request.

\bibliography{ref}

\begin{thebibliography}{104}%
\makeatletter
\providecommand \@ifxundefined [1]{%
 \@ifx{#1\undefined}
}%
\providecommand \@ifnum [1]{%
 \ifnum #1\expandafter \@firstoftwo
 \else \expandafter \@secondoftwo
 \fi
}%
\providecommand \@ifx [1]{%
 \ifx #1\expandafter \@firstoftwo
 \else \expandafter \@secondoftwo
 \fi
}%
\providecommand \natexlab [1]{#1}%
\providecommand \enquote  [1]{``#1''}%
\providecommand \bibnamefont  [1]{#1}%
\providecommand \bibfnamefont [1]{#1}%
\providecommand \citenamefont [1]{#1}%
\providecommand \href@noop [0]{\@secondoftwo}%
\providecommand \href [0]{\begingroup \@sanitize@url \@href}%
\providecommand \@href[1]{\@@startlink{#1}\@@href}%
\providecommand \@@href[1]{\endgroup#1\@@endlink}%
\providecommand \@sanitize@url [0]{\catcode `\\12\catcode `\$12\catcode
  `\&12\catcode `\#12\catcode `\^12\catcode `\_12\catcode `\%12\relax}%
\providecommand \@@startlink[1]{}%
\providecommand \@@endlink[0]{}%
\providecommand \url  [0]{\begingroup\@sanitize@url \@url }%
\providecommand \@url [1]{\endgroup\@href {#1}{\urlprefix }}%
\providecommand \urlprefix  [0]{URL }%
\providecommand \Eprint [0]{\href }%
\providecommand \doibase [0]{https://doi.org/}%
\providecommand \selectlanguage [0]{\@gobble}%
\providecommand \bibinfo  [0]{\@secondoftwo}%
\providecommand \bibfield  [0]{\@secondoftwo}%
\providecommand \translation [1]{[#1]}%
\providecommand \BibitemOpen [0]{}%
\providecommand \bibitemStop [0]{}%
\providecommand \bibitemNoStop [0]{.\EOS\space}%
\providecommand \EOS [0]{\spacefactor3000\relax}%
\providecommand \BibitemShut  [1]{\csname bibitem#1\endcsname}%
\let\auto@bib@innerbib\@empty
%</preamble>
\bibitem [{\citenamefont {Armitage}\ \emph {et~al.}(2018)\citenamefont
  {Armitage}, \citenamefont {Mele},\ and\ \citenamefont
  {Vishwanath}}]{RevModPhys.90.015001}%
  \BibitemOpen
  \bibfield  {author} {\bibinfo {author} {\bibfnamefont {N.~P.}\ \bibnamefont
  {Armitage}}, \bibinfo {author} {\bibfnamefont {E.~J.}\ \bibnamefont {Mele}},\
  and\ \bibinfo {author} {\bibfnamefont {A.}~\bibnamefont {Vishwanath}},\
  }\bibfield  {title} {\bibinfo {title} {{Weyl and Dirac semimetals in
  three-dimensional solids}},\ }\href@noop {} {\bibfield  {journal} {\bibinfo
  {journal} {Rev. Mod. Phys.}\ }\textbf {\bibinfo {volume} {90}},\ \bibinfo
  {pages} {015001} (\bibinfo {year} {2018})}\BibitemShut {NoStop}%
\bibitem [{\citenamefont {Herring}(1937)}]{herring1937accidental}%
  \BibitemOpen
  \bibfield  {author} {\bibinfo {author} {\bibfnamefont {C.}~\bibnamefont
  {Herring}},\ }\bibfield  {title} {\bibinfo {title} {{Accidental degeneracy in
  the energy bands of crystals}},\ }\href@noop {} {\bibfield  {journal}
  {\bibinfo  {journal} {Phys. Rev.}\ }\textbf {\bibinfo {volume} {52}},\
  \bibinfo {pages} {365} (\bibinfo {year} {1937})}\BibitemShut {NoStop}%
\bibitem [{\citenamefont {Murakami}(2007)}]{murakami2007phase}%
  \BibitemOpen
  \bibfield  {author} {\bibinfo {author} {\bibfnamefont {S.}~\bibnamefont
  {Murakami}},\ }\bibfield  {title} {\bibinfo {title} {{Phase transition
  between the quantum spin Hall and insulator phases in 3D: emergence of a
  topological gapless phase}},\ }\href@noop {} {\bibfield  {journal} {\bibinfo
  {journal} {New J. Phys.}\ }\textbf {\bibinfo {volume} {9}},\ \bibinfo {pages}
  {356} (\bibinfo {year} {2007})}\BibitemShut {NoStop}%
\bibitem [{\citenamefont {Wan}\ \emph {et~al.}(2011)\citenamefont {Wan},
  \citenamefont {Turner}, \citenamefont {Vishwanath},\ and\ \citenamefont
  {Savrasov}}]{PhysRevB.83.205101}%
  \BibitemOpen
  \bibfield  {author} {\bibinfo {author} {\bibfnamefont {X.}~\bibnamefont
  {Wan}}, \bibinfo {author} {\bibfnamefont {A.~M.}\ \bibnamefont {Turner}},
  \bibinfo {author} {\bibfnamefont {A.}~\bibnamefont {Vishwanath}},\ and\
  \bibinfo {author} {\bibfnamefont {S.~Y.}\ \bibnamefont {Savrasov}},\
  }\bibfield  {title} {\bibinfo {title} {{Topological semimetal and Fermi-arc
  surface states in the electronic structure of pyrochlore iridates}},\
  }\href@noop {} {\bibfield  {journal} {\bibinfo  {journal} {Phys. Rev. B}\
  }\textbf {\bibinfo {volume} {83}},\ \bibinfo {pages} {205101} (\bibinfo
  {year} {2011})}\BibitemShut {NoStop}%
\bibitem [{\citenamefont {Xu}\ \emph {et~al.}(2011)\citenamefont {Xu},
  \citenamefont {Weng}, \citenamefont {Wang}, \citenamefont {Dai},\ and\
  \citenamefont {Fang}}]{PhysRevLett.107.186806}%
  \BibitemOpen
  \bibfield  {author} {\bibinfo {author} {\bibfnamefont {G.}~\bibnamefont
  {Xu}}, \bibinfo {author} {\bibfnamefont {H.}~\bibnamefont {Weng}}, \bibinfo
  {author} {\bibfnamefont {Z.}~\bibnamefont {Wang}}, \bibinfo {author}
  {\bibfnamefont {X.}~\bibnamefont {Dai}},\ and\ \bibinfo {author}
  {\bibfnamefont {Z.}~\bibnamefont {Fang}},\ }\bibfield  {title} {\bibinfo
  {title} {{Chern semimetal and the quantized anomalous Hall effect in
  ${\mathrm{HgCr}}_{2}{\mathrm{Se}}_{4}$}},\ }\href@noop {} {\bibfield
  {journal} {\bibinfo  {journal} {Phys. Rev. Lett.}\ }\textbf {\bibinfo
  {volume} {107}},\ \bibinfo {pages} {186806} (\bibinfo {year}
  {2011})}\BibitemShut {NoStop}%
\bibitem [{\citenamefont {Burkov}\ and\ \citenamefont
  {Balents}(2011)}]{PhysRevLett.107.127205}%
  \BibitemOpen
  \bibfield  {author} {\bibinfo {author} {\bibfnamefont {A.~A.}\ \bibnamefont
  {Burkov}}\ and\ \bibinfo {author} {\bibfnamefont {L.}~\bibnamefont
  {Balents}},\ }\bibfield  {title} {\bibinfo {title} {{Weyl Semimetal in a
  Topological Insulator Multilayer}},\ }\href@noop {} {\bibfield  {journal}
  {\bibinfo  {journal} {Phys. Rev. Lett.}\ }\textbf {\bibinfo {volume} {107}},\
  \bibinfo {pages} {127205} (\bibinfo {year} {2011})}\BibitemShut {NoStop}%
\bibitem [{\citenamefont {Burkov}\ \emph {et~al.}(2011)\citenamefont {Burkov},
  \citenamefont {Hook},\ and\ \citenamefont {Balents}}]{PhysRevB.84.235126}%
  \BibitemOpen
  \bibfield  {author} {\bibinfo {author} {\bibfnamefont {A.~A.}\ \bibnamefont
  {Burkov}}, \bibinfo {author} {\bibfnamefont {M.~D.}\ \bibnamefont {Hook}},\
  and\ \bibinfo {author} {\bibfnamefont {L.}~\bibnamefont {Balents}},\
  }\bibfield  {title} {\bibinfo {title} {{Topological nodal semimetals}},\
  }\href@noop {} {\bibfield  {journal} {\bibinfo  {journal} {Phys. Rev. B}\
  }\textbf {\bibinfo {volume} {84}},\ \bibinfo {pages} {235126} (\bibinfo
  {year} {2011})}\BibitemShut {NoStop}%
\bibitem [{\citenamefont {Fang}\ \emph {et~al.}(2012)\citenamefont {Fang},
  \citenamefont {Gilbert}, \citenamefont {Dai},\ and\ \citenamefont
  {Bernevig}}]{PhysRevLett.108.266802}%
  \BibitemOpen
  \bibfield  {author} {\bibinfo {author} {\bibfnamefont {C.}~\bibnamefont
  {Fang}}, \bibinfo {author} {\bibfnamefont {M.~J.}\ \bibnamefont {Gilbert}},
  \bibinfo {author} {\bibfnamefont {X.}~\bibnamefont {Dai}},\ and\ \bibinfo
  {author} {\bibfnamefont {B.~A.}\ \bibnamefont {Bernevig}},\ }\bibfield
  {title} {\bibinfo {title} {{Multi-Weyl Topological Semimetals Stabilized by
  Point Group Symmetry}},\ }\href@noop {} {\bibfield  {journal} {\bibinfo
  {journal} {Phys. Rev. Lett.}\ }\textbf {\bibinfo {volume} {108}},\ \bibinfo
  {pages} {266802} (\bibinfo {year} {2012})}\BibitemShut {NoStop}%
\bibitem [{\citenamefont {Hal\'asz}\ and\ \citenamefont
  {Balents}(2012)}]{PhysRevB.85.035103}%
  \BibitemOpen
  \bibfield  {author} {\bibinfo {author} {\bibfnamefont {G.~B.}\ \bibnamefont
  {Hal\'asz}}\ and\ \bibinfo {author} {\bibfnamefont {L.}~\bibnamefont
  {Balents}},\ }\bibfield  {title} {\bibinfo {title} {{Time-reversal invariant
  realization of the Weyl semimetal phase}},\ }\href@noop {} {\bibfield
  {journal} {\bibinfo  {journal} {Phys. Rev. B}\ }\textbf {\bibinfo {volume}
  {85}},\ \bibinfo {pages} {035103} (\bibinfo {year} {2012})}\BibitemShut
  {NoStop}%
\bibitem [{\citenamefont {Liu}\ and\ \citenamefont
  {Vanderbilt}(2014)}]{PhysRevB.90.155316}%
  \BibitemOpen
  \bibfield  {author} {\bibinfo {author} {\bibfnamefont {J.}~\bibnamefont
  {Liu}}\ and\ \bibinfo {author} {\bibfnamefont {D.}~\bibnamefont
  {Vanderbilt}},\ }\bibfield  {title} {\bibinfo {title} {{Weyl semimetals from
  noncentrosymmetric topological insulators}},\ }\href@noop {} {\bibfield
  {journal} {\bibinfo  {journal} {Phys. Rev. B}\ }\textbf {\bibinfo {volume}
  {90}},\ \bibinfo {pages} {155316} (\bibinfo {year} {2014})}\BibitemShut
  {NoStop}%
\bibitem [{\citenamefont {Weng}\ \emph {et~al.}(2015)\citenamefont {Weng},
  \citenamefont {Fang}, \citenamefont {Fang}, \citenamefont {Bernevig},\ and\
  \citenamefont {Dai}}]{PhysRevX.5.011029}%
  \BibitemOpen
  \bibfield  {author} {\bibinfo {author} {\bibfnamefont {H.}~\bibnamefont
  {Weng}}, \bibinfo {author} {\bibfnamefont {C.}~\bibnamefont {Fang}}, \bibinfo
  {author} {\bibfnamefont {Z.}~\bibnamefont {Fang}}, \bibinfo {author}
  {\bibfnamefont {B.~A.}\ \bibnamefont {Bernevig}},\ and\ \bibinfo {author}
  {\bibfnamefont {X.}~\bibnamefont {Dai}},\ }\bibfield  {title} {\bibinfo
  {title} {{Weyl Semimetal Phase in Noncentrosymmetric Transition-Metal
  Monophosphides}},\ }\href@noop {} {\bibfield  {journal} {\bibinfo  {journal}
  {Phys. Rev. X}\ }\textbf {\bibinfo {volume} {5}},\ \bibinfo {pages} {011029}
  (\bibinfo {year} {2015})}\BibitemShut {NoStop}%
\bibitem [{\citenamefont {Yu}\ \emph {et~al.}(2022)\citenamefont {Yu},
  \citenamefont {Zhang}, \citenamefont {Liu}, \citenamefont {Wu}, \citenamefont
  {Li}, \citenamefont {Zhang}, \citenamefont {Yang},\ and\ \citenamefont
  {Yao}}]{yu2022encyclopedia}%
  \BibitemOpen
  \bibfield  {author} {\bibinfo {author} {\bibfnamefont {Z.-M.}\ \bibnamefont
  {Yu}}, \bibinfo {author} {\bibfnamefont {Z.}~\bibnamefont {Zhang}}, \bibinfo
  {author} {\bibfnamefont {G.-B.}\ \bibnamefont {Liu}}, \bibinfo {author}
  {\bibfnamefont {W.}~\bibnamefont {Wu}}, \bibinfo {author} {\bibfnamefont
  {X.-P.}\ \bibnamefont {Li}}, \bibinfo {author} {\bibfnamefont {R.-W.}\
  \bibnamefont {Zhang}}, \bibinfo {author} {\bibfnamefont {S.~A.}\ \bibnamefont
  {Yang}},\ and\ \bibinfo {author} {\bibfnamefont {Y.}~\bibnamefont {Yao}},\
  }\bibfield  {title} {\bibinfo {title} {{Encyclopedia of emergent particles in
  three-dimensional crystals}},\ }\href@noop {} {\bibfield  {journal} {\bibinfo
   {journal} {Sci. Bull.}\ }\textbf {\bibinfo {volume} {67}},\ \bibinfo {pages}
  {375} (\bibinfo {year} {2022})}\BibitemShut {NoStop}%
\bibitem [{\citenamefont {Soluyanov}\ \emph {et~al.}(2015)\citenamefont
  {Soluyanov}, \citenamefont {Gresch}, \citenamefont {Wang}, \citenamefont
  {Wu}, \citenamefont {Troyer}, \citenamefont {Dai},\ and\ \citenamefont
  {Bernevig}}]{soluyanov2015type}%
  \BibitemOpen
  \bibfield  {author} {\bibinfo {author} {\bibfnamefont {A.~A.}\ \bibnamefont
  {Soluyanov}}, \bibinfo {author} {\bibfnamefont {D.}~\bibnamefont {Gresch}},
  \bibinfo {author} {\bibfnamefont {Z.}~\bibnamefont {Wang}}, \bibinfo {author}
  {\bibfnamefont {Q.}~\bibnamefont {Wu}}, \bibinfo {author} {\bibfnamefont
  {M.}~\bibnamefont {Troyer}}, \bibinfo {author} {\bibfnamefont
  {X.}~\bibnamefont {Dai}},\ and\ \bibinfo {author} {\bibfnamefont {B.~A.}\
  \bibnamefont {Bernevig}},\ }\bibfield  {title} {\bibinfo {title} {{Type-II
  Weyl semimetals}},\ }\href@noop {} {\bibfield  {journal} {\bibinfo  {journal}
  {Nature}\ }\textbf {\bibinfo {volume} {527}},\ \bibinfo {pages} {495}
  (\bibinfo {year} {2015})}\BibitemShut {NoStop}%
\bibitem [{\citenamefont {Liu}\ \emph {et~al.}(2017)\citenamefont {Liu},
  \citenamefont {Xu}, \citenamefont {Zhang},\ and\ \citenamefont
  {Duan}}]{PhysRevB.96.064106}%
  \BibitemOpen
  \bibfield  {author} {\bibinfo {author} {\bibfnamefont {Y.}~\bibnamefont
  {Liu}}, \bibinfo {author} {\bibfnamefont {Y.}~\bibnamefont {Xu}}, \bibinfo
  {author} {\bibfnamefont {S.-C.}\ \bibnamefont {Zhang}},\ and\ \bibinfo
  {author} {\bibfnamefont {W.}~\bibnamefont {Duan}},\ }\bibfield  {title}
  {\bibinfo {title} {{Model for topological phononics and phonon diode}},\
  }\href@noop {} {\bibfield  {journal} {\bibinfo  {journal} {Phys. Rev. B}\
  }\textbf {\bibinfo {volume} {96}},\ \bibinfo {pages} {064106} (\bibinfo
  {year} {2017})}\BibitemShut {NoStop}%
\bibitem [{\citenamefont {Ji}\ and\ \citenamefont
  {Shi}(2017)}]{ji2017topological}%
  \BibitemOpen
  \bibfield  {author} {\bibinfo {author} {\bibfnamefont {W.-C.}\ \bibnamefont
  {Ji}}\ and\ \bibinfo {author} {\bibfnamefont {J.-R.}\ \bibnamefont {Shi}},\
  }\bibfield  {title} {\bibinfo {title} {{Topological phonon modes in a
  two-dimensional Wigner crystal}},\ }\href@noop {} {\bibfield  {journal}
  {\bibinfo  {journal} {Chin. Phys. Lett.}\ }\textbf {\bibinfo {volume} {34}},\
  \bibinfo {pages} {036301} (\bibinfo {year} {2017})}\BibitemShut {NoStop}%
\bibitem [{\citenamefont {Zhang}\ \emph {et~al.}(2018)\citenamefont {Zhang},
  \citenamefont {Song}, \citenamefont {Alexandradinata}, \citenamefont {Weng},
  \citenamefont {Fang}, \citenamefont {Lu},\ and\ \citenamefont
  {Fang}}]{PhysRevLett.120.016401}%
  \BibitemOpen
  \bibfield  {author} {\bibinfo {author} {\bibfnamefont {T.}~\bibnamefont
  {Zhang}}, \bibinfo {author} {\bibfnamefont {Z.}~\bibnamefont {Song}},
  \bibinfo {author} {\bibfnamefont {A.}~\bibnamefont {Alexandradinata}},
  \bibinfo {author} {\bibfnamefont {H.}~\bibnamefont {Weng}}, \bibinfo {author}
  {\bibfnamefont {C.}~\bibnamefont {Fang}}, \bibinfo {author} {\bibfnamefont
  {L.}~\bibnamefont {Lu}},\ and\ \bibinfo {author} {\bibfnamefont
  {Z.}~\bibnamefont {Fang}},\ }\bibfield  {title} {\bibinfo {title}
  {{Double-Weyl Phonons in Transition-Metal Monosilicides}},\ }\href@noop {}
  {\bibfield  {journal} {\bibinfo  {journal} {Phys. Rev. Lett.}\ }\textbf
  {\bibinfo {volume} {120}},\ \bibinfo {pages} {016401} (\bibinfo {year}
  {2018})}\BibitemShut {NoStop}%
\bibitem [{\citenamefont {Li}\ \emph {et~al.}(2018)\citenamefont {Li},
  \citenamefont {Xie}, \citenamefont {Ullah}, \citenamefont {Li}, \citenamefont
  {Ma}, \citenamefont {Li}, \citenamefont {Li},\ and\ \citenamefont
  {Chen}}]{PhysRevB.97.054305}%
  \BibitemOpen
  \bibfield  {author} {\bibinfo {author} {\bibfnamefont {J.}~\bibnamefont
  {Li}}, \bibinfo {author} {\bibfnamefont {Q.}~\bibnamefont {Xie}}, \bibinfo
  {author} {\bibfnamefont {S.}~\bibnamefont {Ullah}}, \bibinfo {author}
  {\bibfnamefont {R.}~\bibnamefont {Li}}, \bibinfo {author} {\bibfnamefont
  {H.}~\bibnamefont {Ma}}, \bibinfo {author} {\bibfnamefont {D.}~\bibnamefont
  {Li}}, \bibinfo {author} {\bibfnamefont {Y.}~\bibnamefont {Li}},\ and\
  \bibinfo {author} {\bibfnamefont {X.-Q.}\ \bibnamefont {Chen}},\ }\bibfield
  {title} {\bibinfo {title} {{Coexistent three-component and two-component Weyl
  phonons in TiS, ZrSe, and HfTe}},\ }\href@noop {} {\bibfield  {journal}
  {\bibinfo  {journal} {Phys. Rev. B}\ }\textbf {\bibinfo {volume} {97}},\
  \bibinfo {pages} {054305} (\bibinfo {year} {2018})}\BibitemShut {NoStop}%
\bibitem [{\citenamefont {Xia}\ \emph {et~al.}(2019)\citenamefont {Xia},
  \citenamefont {Wang}, \citenamefont {Chen}, \citenamefont {Zhao},\ and\
  \citenamefont {Xu}}]{PhysRevLett.123.065501}%
  \BibitemOpen
  \bibfield  {author} {\bibinfo {author} {\bibfnamefont {B.~W.}\ \bibnamefont
  {Xia}}, \bibinfo {author} {\bibfnamefont {R.}~\bibnamefont {Wang}}, \bibinfo
  {author} {\bibfnamefont {Z.~J.}\ \bibnamefont {Chen}}, \bibinfo {author}
  {\bibfnamefont {Y.~J.}\ \bibnamefont {Zhao}},\ and\ \bibinfo {author}
  {\bibfnamefont {H.}~\bibnamefont {Xu}},\ }\bibfield  {title} {\bibinfo
  {title} {{Symmetry-protected ideal type-II Weyl Phonons in CdTe}},\
  }\href@noop {} {\bibfield  {journal} {\bibinfo  {journal} {Phys. Rev. Lett.}\
  }\textbf {\bibinfo {volume} {123}},\ \bibinfo {pages} {065501} (\bibinfo
  {year} {2019})}\BibitemShut {NoStop}%
\bibitem [{\citenamefont {Liu}\ \emph {et~al.}(2019)\citenamefont {Liu},
  \citenamefont {Hou}, \citenamefont {Wang}, \citenamefont {Zhang},
  \citenamefont {Sun},\ and\ \citenamefont {Meng}}]{PhysRevB.100.081204}%
  \BibitemOpen
  \bibfield  {author} {\bibinfo {author} {\bibfnamefont {J.}~\bibnamefont
  {Liu}}, \bibinfo {author} {\bibfnamefont {W.}~\bibnamefont {Hou}}, \bibinfo
  {author} {\bibfnamefont {E.}~\bibnamefont {Wang}}, \bibinfo {author}
  {\bibfnamefont {S.}~\bibnamefont {Zhang}}, \bibinfo {author} {\bibfnamefont
  {J.-T.}\ \bibnamefont {Sun}},\ and\ \bibinfo {author} {\bibfnamefont
  {S.}~\bibnamefont {Meng}},\ }\bibfield  {title} {\bibinfo {title} {{Ideal
  type-II Weyl phonons in wurtzite CuI}},\ }\href@noop {} {\bibfield  {journal}
  {\bibinfo  {journal} {Phys. Rev. B}\ }\textbf {\bibinfo {volume} {100}},\
  \bibinfo {pages} {081204(R)} (\bibinfo {year} {2019})}\BibitemShut {NoStop}%
\bibitem [{\citenamefont {Tang}\ \emph {et~al.}(2019)\citenamefont {Tang},
  \citenamefont {Po}, \citenamefont {Vishwanath},\ and\ \citenamefont
  {Wan}}]{tang2019comprehensive}%
  \BibitemOpen
  \bibfield  {author} {\bibinfo {author} {\bibfnamefont {F.}~\bibnamefont
  {Tang}}, \bibinfo {author} {\bibfnamefont {H.~C.}\ \bibnamefont {Po}},
  \bibinfo {author} {\bibfnamefont {A.}~\bibnamefont {Vishwanath}},\ and\
  \bibinfo {author} {\bibfnamefont {X.}~\bibnamefont {Wan}},\ }\bibfield
  {title} {\bibinfo {title} {{Comprehensive search for topological materials
  using symmetry indicators}},\ }\href@noop {} {\bibfield  {journal} {\bibinfo
  {journal} {Nature}\ }\textbf {\bibinfo {volume} {566}},\ \bibinfo {pages}
  {486} (\bibinfo {year} {2019})}\BibitemShut {NoStop}%
\bibitem [{\citenamefont {Vergniory}\ \emph {et~al.}(2019)\citenamefont
  {Vergniory}, \citenamefont {Elcoro}, \citenamefont {Felser}, \citenamefont
  {Regnault}, \citenamefont {Bernevig},\ and\ \citenamefont
  {Wang}}]{vergniory2019complete}%
  \BibitemOpen
  \bibfield  {author} {\bibinfo {author} {\bibfnamefont {M.}~\bibnamefont
  {Vergniory}}, \bibinfo {author} {\bibfnamefont {L.}~\bibnamefont {Elcoro}},
  \bibinfo {author} {\bibfnamefont {C.}~\bibnamefont {Felser}}, \bibinfo
  {author} {\bibfnamefont {N.}~\bibnamefont {Regnault}}, \bibinfo {author}
  {\bibfnamefont {B.~A.}\ \bibnamefont {Bernevig}},\ and\ \bibinfo {author}
  {\bibfnamefont {Z.}~\bibnamefont {Wang}},\ }\bibfield  {title} {\bibinfo
  {title} {{A complete catalogue of high-quality topological materials}},\
  }\href@noop {} {\bibfield  {journal} {\bibinfo  {journal} {Nature}\ }\textbf
  {\bibinfo {volume} {566}},\ \bibinfo {pages} {480} (\bibinfo {year}
  {2019})}\BibitemShut {NoStop}%
\bibitem [{\citenamefont {Huang}\ \emph
  {et~al.}(2015{\natexlab{a}})\citenamefont {Huang}, \citenamefont {Xu},
  \citenamefont {Belopolski}, \citenamefont {Lee}, \citenamefont {Chang},
  \citenamefont {Wang}, \citenamefont {Alidoust}, \citenamefont {Bian},
  \citenamefont {Neupane}, \citenamefont {Zhang} \emph
  {et~al.}}]{huang2015weyl}%
  \BibitemOpen
  \bibfield  {author} {\bibinfo {author} {\bibfnamefont {S.-M.}\ \bibnamefont
  {Huang}}, \bibinfo {author} {\bibfnamefont {S.-Y.}\ \bibnamefont {Xu}},
  \bibinfo {author} {\bibfnamefont {I.}~\bibnamefont {Belopolski}}, \bibinfo
  {author} {\bibfnamefont {C.-C.}\ \bibnamefont {Lee}}, \bibinfo {author}
  {\bibfnamefont {G.}~\bibnamefont {Chang}}, \bibinfo {author} {\bibfnamefont
  {B.}~\bibnamefont {Wang}}, \bibinfo {author} {\bibfnamefont {N.}~\bibnamefont
  {Alidoust}}, \bibinfo {author} {\bibfnamefont {G.}~\bibnamefont {Bian}},
  \bibinfo {author} {\bibfnamefont {M.}~\bibnamefont {Neupane}}, \bibinfo
  {author} {\bibfnamefont {C.}~\bibnamefont {Zhang}}, \emph {et~al.},\
  }\bibfield  {title} {\bibinfo {title} {{A Weyl Fermion semimetal with surface
  Fermi arcs in the transition metal monopnictide TaAs class}},\ }\href@noop {}
  {\bibfield  {journal} {\bibinfo  {journal} {Nat. Commun.}\ }\textbf {\bibinfo
  {volume} {6}},\ \bibinfo {pages} {7373} (\bibinfo {year}
  {2015}{\natexlab{a}})}\BibitemShut {NoStop}%
\bibitem [{\citenamefont {Sun}\ \emph {et~al.}(2015)\citenamefont {Sun},
  \citenamefont {Wu}, \citenamefont {Ali}, \citenamefont {Felser},\ and\
  \citenamefont {Yan}}]{PhysRevB.92.161107}%
  \BibitemOpen
  \bibfield  {author} {\bibinfo {author} {\bibfnamefont {Y.}~\bibnamefont
  {Sun}}, \bibinfo {author} {\bibfnamefont {S.-C.}\ \bibnamefont {Wu}},
  \bibinfo {author} {\bibfnamefont {M.~N.}\ \bibnamefont {Ali}}, \bibinfo
  {author} {\bibfnamefont {C.}~\bibnamefont {Felser}},\ and\ \bibinfo {author}
  {\bibfnamefont {B.}~\bibnamefont {Yan}},\ }\bibfield  {title} {\bibinfo
  {title} {{Prediction of Weyl semimetal in orthorhombic
  ${\mathrm{MoTe}}_{2}$}},\ }\href@noop {} {\bibfield  {journal} {\bibinfo
  {journal} {Phys. Rev. B}\ }\textbf {\bibinfo {volume} {92}},\ \bibinfo
  {pages} {161107(R)} (\bibinfo {year} {2015})}\BibitemShut {NoStop}%
\bibitem [{\citenamefont {Ruan}\ \emph {et~al.}(2016)\citenamefont {Ruan},
  \citenamefont {Jian}, \citenamefont {Zhang}, \citenamefont {Yao},
  \citenamefont {Zhang}, \citenamefont {Zhang},\ and\ \citenamefont
  {Xing}}]{PhysRevLett.116.226801}%
  \BibitemOpen
  \bibfield  {author} {\bibinfo {author} {\bibfnamefont {J.}~\bibnamefont
  {Ruan}}, \bibinfo {author} {\bibfnamefont {S.-K.}\ \bibnamefont {Jian}},
  \bibinfo {author} {\bibfnamefont {D.}~\bibnamefont {Zhang}}, \bibinfo
  {author} {\bibfnamefont {H.}~\bibnamefont {Yao}}, \bibinfo {author}
  {\bibfnamefont {H.}~\bibnamefont {Zhang}}, \bibinfo {author} {\bibfnamefont
  {S.-C.}\ \bibnamefont {Zhang}},\ and\ \bibinfo {author} {\bibfnamefont
  {D.}~\bibnamefont {Xing}},\ }\bibfield  {title} {\bibinfo {title} {{Ideal
  Weyl semimetals in the chalcopyrites ${\mathrm{CuTlSe}}_{2}$,
  ${\mathrm{AgTlTe}}_{2}$, ${\mathrm{AuTlTe}}_{2}$, and
  ${\mathrm{ZnPbAs}}_{2}$}},\ }\href@noop {} {\bibfield  {journal} {\bibinfo
  {journal} {Phys. Rev. Lett.}\ }\textbf {\bibinfo {volume} {116}},\ \bibinfo
  {pages} {226801} (\bibinfo {year} {2016})}\BibitemShut {NoStop}%
\bibitem [{\citenamefont {Aut\`es}\ \emph {et~al.}(2016)\citenamefont
  {Aut\`es}, \citenamefont {Gresch}, \citenamefont {Troyer}, \citenamefont
  {Soluyanov},\ and\ \citenamefont {Yazyev}}]{PhysRevLett.117.066402}%
  \BibitemOpen
  \bibfield  {author} {\bibinfo {author} {\bibfnamefont {G.}~\bibnamefont
  {Aut\`es}}, \bibinfo {author} {\bibfnamefont {D.}~\bibnamefont {Gresch}},
  \bibinfo {author} {\bibfnamefont {M.}~\bibnamefont {Troyer}}, \bibinfo
  {author} {\bibfnamefont {A.~A.}\ \bibnamefont {Soluyanov}},\ and\ \bibinfo
  {author} {\bibfnamefont {O.~V.}\ \bibnamefont {Yazyev}},\ }\bibfield  {title}
  {\bibinfo {title} {{Robust type-II Weyl semimetal phase in transition metal
  diphosphides $X{\mathrm{P}}_{2}$ ($X=\mathrm{Mo}$, W)}},\ }\href@noop {}
  {\bibfield  {journal} {\bibinfo  {journal} {Phys. Rev. Lett.}\ }\textbf
  {\bibinfo {volume} {117}},\ \bibinfo {pages} {066402} (\bibinfo {year}
  {2016})}\BibitemShut {NoStop}%
\bibitem [{\citenamefont {Wang}\ \emph {et~al.}(2016)\citenamefont {Wang},
  \citenamefont {Vergniory}, \citenamefont {Kushwaha}, \citenamefont
  {Hirschberger}, \citenamefont {Chulkov}, \citenamefont {Ernst}, \citenamefont
  {Ong}, \citenamefont {Cava},\ and\ \citenamefont
  {Bernevig}}]{PhysRevLett.117.236401}%
  \BibitemOpen
  \bibfield  {author} {\bibinfo {author} {\bibfnamefont {Z.}~\bibnamefont
  {Wang}}, \bibinfo {author} {\bibfnamefont {M.~G.}\ \bibnamefont {Vergniory}},
  \bibinfo {author} {\bibfnamefont {S.}~\bibnamefont {Kushwaha}}, \bibinfo
  {author} {\bibfnamefont {M.}~\bibnamefont {Hirschberger}}, \bibinfo {author}
  {\bibfnamefont {E.~V.}\ \bibnamefont {Chulkov}}, \bibinfo {author}
  {\bibfnamefont {A.}~\bibnamefont {Ernst}}, \bibinfo {author} {\bibfnamefont
  {N.~P.}\ \bibnamefont {Ong}}, \bibinfo {author} {\bibfnamefont {R.~J.}\
  \bibnamefont {Cava}},\ and\ \bibinfo {author} {\bibfnamefont {B.~A.}\
  \bibnamefont {Bernevig}},\ }\bibfield  {title} {\bibinfo {title}
  {{Time-reversal-breaking Weyl Fermions in Magnetic Heusler alloys}},\
  }\href@noop {} {\bibfield  {journal} {\bibinfo  {journal} {Phys. Rev. Lett.}\
  }\textbf {\bibinfo {volume} {117}},\ \bibinfo {pages} {236401} (\bibinfo
  {year} {2016})}\BibitemShut {NoStop}%
\bibitem [{\citenamefont {Shi}\ \emph {et~al.}(2018)\citenamefont {Shi},
  \citenamefont {Muechler}, \citenamefont {Manna}, \citenamefont {Zhang},
  \citenamefont {Koepernik}, \citenamefont {Car}, \citenamefont {van~den
  Brink}, \citenamefont {Felser},\ and\ \citenamefont
  {Sun}}]{PhysRevB.97.060406}%
  \BibitemOpen
  \bibfield  {author} {\bibinfo {author} {\bibfnamefont {W.}~\bibnamefont
  {Shi}}, \bibinfo {author} {\bibfnamefont {L.}~\bibnamefont {Muechler}},
  \bibinfo {author} {\bibfnamefont {K.}~\bibnamefont {Manna}}, \bibinfo
  {author} {\bibfnamefont {Y.}~\bibnamefont {Zhang}}, \bibinfo {author}
  {\bibfnamefont {K.}~\bibnamefont {Koepernik}}, \bibinfo {author}
  {\bibfnamefont {R.}~\bibnamefont {Car}}, \bibinfo {author} {\bibfnamefont
  {J.}~\bibnamefont {van~den Brink}}, \bibinfo {author} {\bibfnamefont
  {C.}~\bibnamefont {Felser}},\ and\ \bibinfo {author} {\bibfnamefont
  {Y.}~\bibnamefont {Sun}},\ }\bibfield  {title} {\bibinfo {title} {{Prediction
  of a magnetic Weyl semimetal without spin-orbit coupling and strong anomalous
  Hall effect in the Heusler compensated ferrimagnet
  ${\mathrm{Ti}}_{2}\mathrm{MnAl}$}},\ }\href@noop {} {\bibfield  {journal}
  {\bibinfo  {journal} {Phys. Rev. B}\ }\textbf {\bibinfo {volume} {97}},\
  \bibinfo {pages} {060406(R)} (\bibinfo {year} {2018})}\BibitemShut {NoStop}%
\bibitem [{\citenamefont {Xu}\ \emph {et~al.}(2018)\citenamefont {Xu},
  \citenamefont {Liu}, \citenamefont {Shi}, \citenamefont {Muechler},
  \citenamefont {Gayles}, \citenamefont {Felser},\ and\ \citenamefont
  {Sun}}]{PhysRevB.97.235416}%
  \BibitemOpen
  \bibfield  {author} {\bibinfo {author} {\bibfnamefont {Q.}~\bibnamefont
  {Xu}}, \bibinfo {author} {\bibfnamefont {E.}~\bibnamefont {Liu}}, \bibinfo
  {author} {\bibfnamefont {W.}~\bibnamefont {Shi}}, \bibinfo {author}
  {\bibfnamefont {L.}~\bibnamefont {Muechler}}, \bibinfo {author}
  {\bibfnamefont {J.}~\bibnamefont {Gayles}}, \bibinfo {author} {\bibfnamefont
  {C.}~\bibnamefont {Felser}},\ and\ \bibinfo {author} {\bibfnamefont
  {Y.}~\bibnamefont {Sun}},\ }\bibfield  {title} {\bibinfo {title}
  {{Topological surface Fermi arcs in the magnetic Weyl semimetal
  ${\mathrm{Co}}_{3}{\mathrm{Sn}}_{2}{\mathrm{S}}_{2}$}},\ }\href@noop {}
  {\bibfield  {journal} {\bibinfo  {journal} {Phys. Rev. B}\ }\textbf {\bibinfo
  {volume} {97}},\ \bibinfo {pages} {235416} (\bibinfo {year}
  {2018})}\BibitemShut {NoStop}%
\bibitem [{\citenamefont {Xu}\ \emph {et~al.}(2017)\citenamefont {Xu},
  \citenamefont {Alidoust}, \citenamefont {Chang}, \citenamefont {Lu},
  \citenamefont {Singh}, \citenamefont {Belopolski}, \citenamefont {Sanchez},
  \citenamefont {Zhang}, \citenamefont {Bian}, \citenamefont {Zheng} \emph
  {et~al.}}]{xu2017discovery}%
  \BibitemOpen
  \bibfield  {author} {\bibinfo {author} {\bibfnamefont {S.-Y.}\ \bibnamefont
  {Xu}}, \bibinfo {author} {\bibfnamefont {N.}~\bibnamefont {Alidoust}},
  \bibinfo {author} {\bibfnamefont {G.}~\bibnamefont {Chang}}, \bibinfo
  {author} {\bibfnamefont {H.}~\bibnamefont {Lu}}, \bibinfo {author}
  {\bibfnamefont {B.}~\bibnamefont {Singh}}, \bibinfo {author} {\bibfnamefont
  {I.}~\bibnamefont {Belopolski}}, \bibinfo {author} {\bibfnamefont {D.~S.}\
  \bibnamefont {Sanchez}}, \bibinfo {author} {\bibfnamefont {X.}~\bibnamefont
  {Zhang}}, \bibinfo {author} {\bibfnamefont {G.}~\bibnamefont {Bian}},
  \bibinfo {author} {\bibfnamefont {H.}~\bibnamefont {Zheng}}, \emph {et~al.},\
  }\bibfield  {title} {\bibinfo {title} {{Discovery of Lorentz-violating type
  II Weyl fermions in LaAlGe}},\ }\href@noop {} {\bibfield  {journal} {\bibinfo
   {journal} {Sci. Adv.}\ }\textbf {\bibinfo {volume} {3}},\ \bibinfo {pages}
  {e1603266} (\bibinfo {year} {2017})}\BibitemShut {NoStop}%
\bibitem [{\citenamefont {Koepernik}\ \emph {et~al.}(2016)\citenamefont
  {Koepernik}, \citenamefont {Kasinathan}, \citenamefont {Efremov},
  \citenamefont {Khim}, \citenamefont {Borisenko}, \citenamefont {B\"uchner},\
  and\ \citenamefont {van~den Brink}}]{PhysRevB.93.201101}%
  \BibitemOpen
  \bibfield  {author} {\bibinfo {author} {\bibfnamefont {K.}~\bibnamefont
  {Koepernik}}, \bibinfo {author} {\bibfnamefont {D.}~\bibnamefont
  {Kasinathan}}, \bibinfo {author} {\bibfnamefont {D.~V.}\ \bibnamefont
  {Efremov}}, \bibinfo {author} {\bibfnamefont {S.}~\bibnamefont {Khim}},
  \bibinfo {author} {\bibfnamefont {S.}~\bibnamefont {Borisenko}}, \bibinfo
  {author} {\bibfnamefont {B.}~\bibnamefont {B\"uchner}},\ and\ \bibinfo
  {author} {\bibfnamefont {J.}~\bibnamefont {van~den Brink}},\ }\bibfield
  {title} {\bibinfo {title} {{${\mathrm{TaIrTe}}_{4}$: A ternary type-II Weyl
  semimetal}},\ }\href@noop {} {\bibfield  {journal} {\bibinfo  {journal}
  {Phys. Rev. B}\ }\textbf {\bibinfo {volume} {93}},\ \bibinfo {pages}
  {201101(R)} (\bibinfo {year} {2016})}\BibitemShut {NoStop}%
\bibitem [{\citenamefont {Chang}\ \emph {et~al.}(2018)\citenamefont {Chang},
  \citenamefont {Wieder}, \citenamefont {Schindler}, \citenamefont {Sanchez},
  \citenamefont {Belopolski}, \citenamefont {Huang}, \citenamefont {Singh},
  \citenamefont {Wu}, \citenamefont {Chang}, \citenamefont {Neupert} \emph
  {et~al.}}]{chang2018topological}%
  \BibitemOpen
  \bibfield  {author} {\bibinfo {author} {\bibfnamefont {G.}~\bibnamefont
  {Chang}}, \bibinfo {author} {\bibfnamefont {B.~J.}\ \bibnamefont {Wieder}},
  \bibinfo {author} {\bibfnamefont {F.}~\bibnamefont {Schindler}}, \bibinfo
  {author} {\bibfnamefont {D.~S.}\ \bibnamefont {Sanchez}}, \bibinfo {author}
  {\bibfnamefont {I.}~\bibnamefont {Belopolski}}, \bibinfo {author}
  {\bibfnamefont {S.-M.}\ \bibnamefont {Huang}}, \bibinfo {author}
  {\bibfnamefont {B.}~\bibnamefont {Singh}}, \bibinfo {author} {\bibfnamefont
  {D.}~\bibnamefont {Wu}}, \bibinfo {author} {\bibfnamefont {T.-R.}\
  \bibnamefont {Chang}}, \bibinfo {author} {\bibfnamefont {T.}~\bibnamefont
  {Neupert}}, \emph {et~al.},\ }\bibfield  {title} {\bibinfo {title}
  {{Topological quantum properties of chiral crystals}},\ }\href@noop {}
  {\bibfield  {journal} {\bibinfo  {journal} {Nat. Mater.}\ }\textbf {\bibinfo
  {volume} {17}},\ \bibinfo {pages} {978} (\bibinfo {year} {2018})}\BibitemShut
  {NoStop}%
\bibitem [{\citenamefont {Wang}\ \emph {et~al.}(2019)\citenamefont {Wang},
  \citenamefont {Jo}, \citenamefont {Kuthanazhi}, \citenamefont {Wu},
  \citenamefont {McQueeney}, \citenamefont {Kaminski},\ and\ \citenamefont
  {Canfield}}]{PhysRevB.99.245147}%
  \BibitemOpen
  \bibfield  {author} {\bibinfo {author} {\bibfnamefont {L.-L.}\ \bibnamefont
  {Wang}}, \bibinfo {author} {\bibfnamefont {N.~H.}\ \bibnamefont {Jo}},
  \bibinfo {author} {\bibfnamefont {B.}~\bibnamefont {Kuthanazhi}}, \bibinfo
  {author} {\bibfnamefont {Y.}~\bibnamefont {Wu}}, \bibinfo {author}
  {\bibfnamefont {R.~J.}\ \bibnamefont {McQueeney}}, \bibinfo {author}
  {\bibfnamefont {A.}~\bibnamefont {Kaminski}},\ and\ \bibinfo {author}
  {\bibfnamefont {P.~C.}\ \bibnamefont {Canfield}},\ }\bibfield  {title}
  {\bibinfo {title} {{Single pair of Weyl fermions in the half-metallic
  semimetal $\mathrm{EuC}{\mathrm{d}}_{2}\mathrm{A}{\mathrm{s}}_{2}$}},\
  }\href@noop {} {\bibfield  {journal} {\bibinfo  {journal} {Phys. Rev. B}\
  }\textbf {\bibinfo {volume} {99}},\ \bibinfo {pages} {245147} (\bibinfo
  {year} {2019})}\BibitemShut {NoStop}%
\bibitem [{\citenamefont {Huang}\ \emph {et~al.}(2016)\citenamefont {Huang},
  \citenamefont {Xu}, \citenamefont {Belopolski}, \citenamefont {Lee},
  \citenamefont {Chang}, \citenamefont {Chang}, \citenamefont {Wang},
  \citenamefont {Alidoust}, \citenamefont {Bian}, \citenamefont {Neupane} \emph
  {et~al.}}]{huang2016new}%
  \BibitemOpen
  \bibfield  {author} {\bibinfo {author} {\bibfnamefont {S.-M.}\ \bibnamefont
  {Huang}}, \bibinfo {author} {\bibfnamefont {S.-Y.}\ \bibnamefont {Xu}},
  \bibinfo {author} {\bibfnamefont {I.}~\bibnamefont {Belopolski}}, \bibinfo
  {author} {\bibfnamefont {C.-C.}\ \bibnamefont {Lee}}, \bibinfo {author}
  {\bibfnamefont {G.}~\bibnamefont {Chang}}, \bibinfo {author} {\bibfnamefont
  {T.-R.}\ \bibnamefont {Chang}}, \bibinfo {author} {\bibfnamefont
  {B.}~\bibnamefont {Wang}}, \bibinfo {author} {\bibfnamefont {N.}~\bibnamefont
  {Alidoust}}, \bibinfo {author} {\bibfnamefont {G.}~\bibnamefont {Bian}},
  \bibinfo {author} {\bibfnamefont {M.}~\bibnamefont {Neupane}}, \emph
  {et~al.},\ }\bibfield  {title} {\bibinfo {title} {{New type of Weyl semimetal
  with quadratic double Weyl fermions}},\ }\href@noop {} {\bibfield  {journal}
  {\bibinfo  {journal} {Proc. Natl. Acad. Sci. U.S.A.}\ }\textbf {\bibinfo
  {volume} {113}},\ \bibinfo {pages} {1180} (\bibinfo {year}
  {2016})}\BibitemShut {NoStop}%
\bibitem [{\citenamefont {Li}\ \emph {et~al.}(2021{\natexlab{a}})\citenamefont
  {Li}, \citenamefont {Deng}, \citenamefont {Fu}, \citenamefont {Li},
  \citenamefont {Ma}, \citenamefont {Han}, \citenamefont {Zhou}, \citenamefont
  {Zhou},\ and\ \citenamefont {Yao}}]{PhysRevB.103.L081402}%
  \BibitemOpen
  \bibfield  {author} {\bibinfo {author} {\bibfnamefont {X.-P.}\ \bibnamefont
  {Li}}, \bibinfo {author} {\bibfnamefont {K.}~\bibnamefont {Deng}}, \bibinfo
  {author} {\bibfnamefont {B.}~\bibnamefont {Fu}}, \bibinfo {author}
  {\bibfnamefont {Y.~K.}\ \bibnamefont {Li}}, \bibinfo {author} {\bibfnamefont
  {D.-S.}\ \bibnamefont {Ma}}, \bibinfo {author} {\bibfnamefont {J.~F.}\
  \bibnamefont {Han}}, \bibinfo {author} {\bibfnamefont {J.}~\bibnamefont
  {Zhou}}, \bibinfo {author} {\bibfnamefont {S.}~\bibnamefont {Zhou}},\ and\
  \bibinfo {author} {\bibfnamefont {Y.}~\bibnamefont {Yao}},\ }\bibfield
  {title} {\bibinfo {title} {{Type-III Weyl semimetals:
  ${({\mathrm{TaSe}}_{4})}_{2}\mathrm{I}$}},\ }\href@noop {} {\bibfield
  {journal} {\bibinfo  {journal} {Phys. Rev. B}\ }\textbf {\bibinfo {volume}
  {103}},\ \bibinfo {pages} {L081402} (\bibinfo {year}
  {2021}{\natexlab{a}})}\BibitemShut {NoStop}%
\bibitem [{\citenamefont {Chang}\ \emph {et~al.}(2016)\citenamefont {Chang},
  \citenamefont {Xu}, \citenamefont {Zheng}, \citenamefont {Singh},
  \citenamefont {Hsu}, \citenamefont {Bian}, \citenamefont {Alidoust},
  \citenamefont {Belopolski}, \citenamefont {Sanchez}, \citenamefont {Zhang}
  \emph {et~al.}}]{chang2016room}%
  \BibitemOpen
  \bibfield  {author} {\bibinfo {author} {\bibfnamefont {G.}~\bibnamefont
  {Chang}}, \bibinfo {author} {\bibfnamefont {S.-Y.}\ \bibnamefont {Xu}},
  \bibinfo {author} {\bibfnamefont {H.}~\bibnamefont {Zheng}}, \bibinfo
  {author} {\bibfnamefont {B.}~\bibnamefont {Singh}}, \bibinfo {author}
  {\bibfnamefont {C.-H.}\ \bibnamefont {Hsu}}, \bibinfo {author} {\bibfnamefont
  {G.}~\bibnamefont {Bian}}, \bibinfo {author} {\bibfnamefont {N.}~\bibnamefont
  {Alidoust}}, \bibinfo {author} {\bibfnamefont {I.}~\bibnamefont
  {Belopolski}}, \bibinfo {author} {\bibfnamefont {D.~S.}\ \bibnamefont
  {Sanchez}}, \bibinfo {author} {\bibfnamefont {S.}~\bibnamefont {Zhang}},
  \emph {et~al.},\ }\bibfield  {title} {\bibinfo {title} {{Room-temperature
  magnetic topological Weyl fermion and nodal line semimetal states in
  half-metallic Heusler Co$_{2}$Ti$X$ ($X$= Si, Ge, or Sn)}},\ }\href@noop {}
  {\bibfield  {journal} {\bibinfo  {journal} {Sci. Rep.}\ }\textbf {\bibinfo
  {volume} {6}},\ \bibinfo {pages} {38839} (\bibinfo {year}
  {2016})}\BibitemShut {NoStop}%
\bibitem [{\citenamefont {Yan}\ and\ \citenamefont
  {Felser}(2017)}]{yan2017topological}%
  \BibitemOpen
  \bibfield  {author} {\bibinfo {author} {\bibfnamefont {B.}~\bibnamefont
  {Yan}}\ and\ \bibinfo {author} {\bibfnamefont {C.}~\bibnamefont {Felser}},\
  }\bibfield  {title} {\bibinfo {title} {{Topological materials: Weyl
  semimetals}},\ }\href@noop {} {\bibfield  {journal} {\bibinfo  {journal}
  {Annu. Rev. Condens. Matter Phys.}\ }\textbf {\bibinfo {volume} {8}},\
  \bibinfo {pages} {337} (\bibinfo {year} {2017})}\BibitemShut {NoStop}%
\bibitem [{\citenamefont {Xu}\ \emph {et~al.}(2024)\citenamefont {Xu},
  \citenamefont {Vergniory}, \citenamefont {Ma}, \citenamefont {Ma{\~n}es},
  \citenamefont {Song}, \citenamefont {Bernevig}, \citenamefont {Regnault},\
  and\ \citenamefont {Elcoro}}]{xu2024catalog}%
  \BibitemOpen
  \bibfield  {author} {\bibinfo {author} {\bibfnamefont {Y.}~\bibnamefont
  {Xu}}, \bibinfo {author} {\bibfnamefont {M.}~\bibnamefont {Vergniory}},
  \bibinfo {author} {\bibfnamefont {D.-S.}\ \bibnamefont {Ma}}, \bibinfo
  {author} {\bibfnamefont {J.~L.}\ \bibnamefont {Ma{\~n}es}}, \bibinfo {author}
  {\bibfnamefont {Z.-D.}\ \bibnamefont {Song}}, \bibinfo {author}
  {\bibfnamefont {B.~A.}\ \bibnamefont {Bernevig}}, \bibinfo {author}
  {\bibfnamefont {N.}~\bibnamefont {Regnault}},\ and\ \bibinfo {author}
  {\bibfnamefont {L.}~\bibnamefont {Elcoro}},\ }\bibfield  {title} {\bibinfo
  {title} {{Catalog of topological phonon materials}},\ }\href@noop {}
  {\bibfield  {journal} {\bibinfo  {journal} {Science}\ }\textbf {\bibinfo
  {volume} {384}},\ \bibinfo {pages} {eadf8458} (\bibinfo {year}
  {2024})}\BibitemShut {NoStop}%
\bibitem [{\citenamefont {Liu}\ \emph {et~al.}(2020)\citenamefont {Liu},
  \citenamefont {Qian}, \citenamefont {Fu},\ and\ \citenamefont
  {Wang}}]{liu2020symmetry}%
  \BibitemOpen
  \bibfield  {author} {\bibinfo {author} {\bibfnamefont {Q.-B.}\ \bibnamefont
  {Liu}}, \bibinfo {author} {\bibfnamefont {Y.}~\bibnamefont {Qian}}, \bibinfo
  {author} {\bibfnamefont {H.-H.}\ \bibnamefont {Fu}},\ and\ \bibinfo {author}
  {\bibfnamefont {Z.}~\bibnamefont {Wang}},\ }\bibfield  {title} {\bibinfo
  {title} {{Symmetry-enforced Weyl phonons}},\ }\href@noop {} {\bibfield
  {journal} {\bibinfo  {journal} {npj Comput. Mater.}\ }\textbf {\bibinfo
  {volume} {6}},\ \bibinfo {pages} {95} (\bibinfo {year} {2020})}\BibitemShut
  {NoStop}%
\bibitem [{\citenamefont {Li}\ \emph {et~al.}(2021{\natexlab{b}})\citenamefont
  {Li}, \citenamefont {Liu}, \citenamefont {Baronett}, \citenamefont {Liu},
  \citenamefont {Wang}, \citenamefont {Li}, \citenamefont {Chen}, \citenamefont
  {Li}, \citenamefont {Zhu},\ and\ \citenamefont {Chen}}]{li2021computation}%
  \BibitemOpen
  \bibfield  {author} {\bibinfo {author} {\bibfnamefont {J.}~\bibnamefont
  {Li}}, \bibinfo {author} {\bibfnamefont {J.}~\bibnamefont {Liu}}, \bibinfo
  {author} {\bibfnamefont {S.~A.}\ \bibnamefont {Baronett}}, \bibinfo {author}
  {\bibfnamefont {M.}~\bibnamefont {Liu}}, \bibinfo {author} {\bibfnamefont
  {L.}~\bibnamefont {Wang}}, \bibinfo {author} {\bibfnamefont {R.}~\bibnamefont
  {Li}}, \bibinfo {author} {\bibfnamefont {Y.}~\bibnamefont {Chen}}, \bibinfo
  {author} {\bibfnamefont {D.}~\bibnamefont {Li}}, \bibinfo {author}
  {\bibfnamefont {Q.}~\bibnamefont {Zhu}},\ and\ \bibinfo {author}
  {\bibfnamefont {X.-Q.}\ \bibnamefont {Chen}},\ }\bibfield  {title} {\bibinfo
  {title} {{Computation and data driven discovery of topological phononic
  materials}},\ }\href@noop {} {\bibfield  {journal} {\bibinfo  {journal} {Nat.
  Commun.}\ }\textbf {\bibinfo {volume} {12}},\ \bibinfo {pages} {1204}
  (\bibinfo {year} {2021}{\natexlab{b}})}\BibitemShut {NoStop}%
\bibitem [{\citenamefont {Yang}\ \emph {et~al.}(2024)\citenamefont {Yang},
  \citenamefont {Wang}, \citenamefont {Li}, \citenamefont {Wang}, \citenamefont
  {Cheng}, \citenamefont {Wang},\ and\ \citenamefont
  {Zhang}}]{yang2024topological}%
  \BibitemOpen
  \bibfield  {author} {\bibinfo {author} {\bibfnamefont {T.}~\bibnamefont
  {Yang}}, \bibinfo {author} {\bibfnamefont {J.}~\bibnamefont {Wang}}, \bibinfo
  {author} {\bibfnamefont {X.-P.}\ \bibnamefont {Li}}, \bibinfo {author}
  {\bibfnamefont {X.}~\bibnamefont {Wang}}, \bibinfo {author} {\bibfnamefont
  {Z.}~\bibnamefont {Cheng}}, \bibinfo {author} {\bibfnamefont
  {W.}~\bibnamefont {Wang}},\ and\ \bibinfo {author} {\bibfnamefont
  {G.}~\bibnamefont {Zhang}},\ }\bibfield  {title} {\bibinfo {title}
  {{Topological nodal-point phononic systems}},\ }\href@noop {} {\bibfield
  {journal} {\bibinfo  {journal} {Matter}\ }\textbf {\bibinfo {volume} {7}},\
  \bibinfo {pages} {320} (\bibinfo {year} {2024})}\BibitemShut {NoStop}%
\bibitem [{\citenamefont {Qin}\ \emph {et~al.}(2024)\citenamefont {Qin},
  \citenamefont {Liu}, \citenamefont {Wu},\ and\ \citenamefont
  {Xu}}]{qin2024diverse}%
  \BibitemOpen
  \bibfield  {author} {\bibinfo {author} {\bibfnamefont {P.}~\bibnamefont
  {Qin}}, \bibinfo {author} {\bibfnamefont {G.}~\bibnamefont {Liu}}, \bibinfo
  {author} {\bibfnamefont {P.}~\bibnamefont {Wu}},\ and\ \bibinfo {author}
  {\bibfnamefont {H.}~\bibnamefont {Xu}},\ }\bibfield  {title} {\bibinfo
  {title} {{Diverse degeneracy types in topological phonons: A perspective}},\
  }\href@noop {} {\bibfield  {journal} {\bibinfo  {journal} {Appl. Phys.
  Lett.}\ }\textbf {\bibinfo {volume} {124}} (\bibinfo {year}
  {2024})}\BibitemShut {NoStop}%
\bibitem [{\citenamefont {Ding}\ \emph {et~al.}(2024)\citenamefont {Ding},
  \citenamefont {Zeng}, \citenamefont {Liu}, \citenamefont {Tang},\ and\
  \citenamefont {Chen}}]{ding2024topological}%
  \BibitemOpen
  \bibfield  {author} {\bibinfo {author} {\bibfnamefont {Z.-K.}\ \bibnamefont
  {Ding}}, \bibinfo {author} {\bibfnamefont {Y.-J.}\ \bibnamefont {Zeng}},
  \bibinfo {author} {\bibfnamefont {W.}~\bibnamefont {Liu}}, \bibinfo {author}
  {\bibfnamefont {L.-M.}\ \bibnamefont {Tang}},\ and\ \bibinfo {author}
  {\bibfnamefont {K.-Q.}\ \bibnamefont {Chen}},\ }\bibfield  {title} {\bibinfo
  {title} {{Topological phonons and thermoelectric conversion in crystalline
  materials}},\ }\href@noop {} {\bibfield  {journal} {\bibinfo  {journal} {Adv.
  Funct. Mater.}\ }\textbf {\bibinfo {volume} {34}},\ \bibinfo {pages}
  {2401684} (\bibinfo {year} {2024})}\BibitemShut {NoStop}%
\bibitem [{\citenamefont {Wang}\ \emph {et~al.}(2020)\citenamefont {Wang},
  \citenamefont {Xia}, \citenamefont {Chen}, \citenamefont {Zheng},
  \citenamefont {Zhao},\ and\ \citenamefont {Xu}}]{PhysRevLett.124.105303}%
  \BibitemOpen
  \bibfield  {author} {\bibinfo {author} {\bibfnamefont {R.}~\bibnamefont
  {Wang}}, \bibinfo {author} {\bibfnamefont {B.~W.}\ \bibnamefont {Xia}},
  \bibinfo {author} {\bibfnamefont {Z.~J.}\ \bibnamefont {Chen}}, \bibinfo
  {author} {\bibfnamefont {B.~B.}\ \bibnamefont {Zheng}}, \bibinfo {author}
  {\bibfnamefont {Y.~J.}\ \bibnamefont {Zhao}},\ and\ \bibinfo {author}
  {\bibfnamefont {H.}~\bibnamefont {Xu}},\ }\bibfield  {title} {\bibinfo
  {title} {{Symmetry-protected Topological Triangular Weyl Complex}},\
  }\href@noop {} {\bibfield  {journal} {\bibinfo  {journal} {Phys. Rev. Lett.}\
  }\textbf {\bibinfo {volume} {124}},\ \bibinfo {pages} {105303} (\bibinfo
  {year} {2020})}\BibitemShut {NoStop}%
\bibitem [{\citenamefont {Zhang}\ \emph
  {et~al.}(2020{\natexlab{a}})\citenamefont {Zhang}, \citenamefont {Takahashi},
  \citenamefont {Fang},\ and\ \citenamefont {Murakami}}]{PhysRevB.102.125148}%
  \BibitemOpen
  \bibfield  {author} {\bibinfo {author} {\bibfnamefont {T.}~\bibnamefont
  {Zhang}}, \bibinfo {author} {\bibfnamefont {R.}~\bibnamefont {Takahashi}},
  \bibinfo {author} {\bibfnamefont {C.}~\bibnamefont {Fang}},\ and\ \bibinfo
  {author} {\bibfnamefont {S.}~\bibnamefont {Murakami}},\ }\bibfield  {title}
  {\bibinfo {title} {{Twofold quadruple Weyl nodes in chiral cubic crystals}},\
  }\href@noop {} {\bibfield  {journal} {\bibinfo  {journal} {Phys. Rev. B}\
  }\textbf {\bibinfo {volume} {102}},\ \bibinfo {pages} {125148} (\bibinfo
  {year} {2020}{\natexlab{a}})}\BibitemShut {NoStop}%
\bibitem [{\citenamefont {Liu}\ \emph {et~al.}(2021{\natexlab{a}})\citenamefont
  {Liu}, \citenamefont {Wang},\ and\ \citenamefont
  {Fu}}]{PhysRevB.103.L161303}%
  \BibitemOpen
  \bibfield  {author} {\bibinfo {author} {\bibfnamefont {Q.-B.}\ \bibnamefont
  {Liu}}, \bibinfo {author} {\bibfnamefont {Z.}~\bibnamefont {Wang}},\ and\
  \bibinfo {author} {\bibfnamefont {H.-H.}\ \bibnamefont {Fu}},\ }\bibfield
  {title} {\bibinfo {title} {{Charge-four Weyl phonons}},\ }\href@noop {}
  {\bibfield  {journal} {\bibinfo  {journal} {Phys. Rev. B}\ }\textbf {\bibinfo
  {volume} {103}},\ \bibinfo {pages} {L161303} (\bibinfo {year}
  {2021}{\natexlab{a}})}\BibitemShut {NoStop}%
\bibitem [{\citenamefont {Liu}\ \emph {et~al.}(2021{\natexlab{b}})\citenamefont
  {Liu}, \citenamefont {Li}, \citenamefont {Tu}, \citenamefont {Li},
  \citenamefont {Zhang}, \citenamefont {Zhang}, \citenamefont {Gao},\ and\
  \citenamefont {Wang}}]{PhysRevB.103.094306}%
  \BibitemOpen
  \bibfield  {author} {\bibinfo {author} {\bibfnamefont {P.-F.}\ \bibnamefont
  {Liu}}, \bibinfo {author} {\bibfnamefont {J.}~\bibnamefont {Li}}, \bibinfo
  {author} {\bibfnamefont {X.-H.}\ \bibnamefont {Tu}}, \bibinfo {author}
  {\bibfnamefont {H.}~\bibnamefont {Li}}, \bibinfo {author} {\bibfnamefont
  {J.}~\bibnamefont {Zhang}}, \bibinfo {author} {\bibfnamefont
  {P.}~\bibnamefont {Zhang}}, \bibinfo {author} {\bibfnamefont
  {Q.}~\bibnamefont {Gao}},\ and\ \bibinfo {author} {\bibfnamefont {B.-T.}\
  \bibnamefont {Wang}},\ }\bibfield  {title} {\bibinfo {title}
  {{First-principles prediction of ideal type-II Weyl phonons in wurtzite
  ZnSe}},\ }\href@noop {} {\bibfield  {journal} {\bibinfo  {journal} {Phys.
  Rev. B}\ }\textbf {\bibinfo {volume} {103}},\ \bibinfo {pages} {094306}
  (\bibinfo {year} {2021}{\natexlab{b}})}\BibitemShut {NoStop}%
\bibitem [{\citenamefont {Liu}\ \emph {et~al.}(2022)\citenamefont {Liu},
  \citenamefont {Chen}, \citenamefont {Wu},\ and\ \citenamefont
  {Xu}}]{PhysRevB.106.214308}%
  \BibitemOpen
  \bibfield  {author} {\bibinfo {author} {\bibfnamefont {G.}~\bibnamefont
  {Liu}}, \bibinfo {author} {\bibfnamefont {Z.}~\bibnamefont {Chen}}, \bibinfo
  {author} {\bibfnamefont {P.}~\bibnamefont {Wu}},\ and\ \bibinfo {author}
  {\bibfnamefont {H.}~\bibnamefont {Xu}},\ }\bibfield  {title} {\bibinfo
  {title} {{Triple hourglass Weyl phonons}},\ }\href@noop {} {\bibfield
  {journal} {\bibinfo  {journal} {Phys. Rev. B}\ }\textbf {\bibinfo {volume}
  {106}},\ \bibinfo {pages} {214308} (\bibinfo {year} {2022})}\BibitemShut
  {NoStop}%
\bibitem [{\citenamefont {Wang}\ \emph
  {et~al.}(2022{\natexlab{a}})\citenamefont {Wang}, \citenamefont {Zhou},
  \citenamefont {Zhang}, \citenamefont {Yu},\ and\ \citenamefont
  {Yao}}]{PhysRevB.106.214309}%
  \BibitemOpen
  \bibfield  {author} {\bibinfo {author} {\bibfnamefont {X.}~\bibnamefont
  {Wang}}, \bibinfo {author} {\bibfnamefont {F.}~\bibnamefont {Zhou}}, \bibinfo
  {author} {\bibfnamefont {Z.}~\bibnamefont {Zhang}}, \bibinfo {author}
  {\bibfnamefont {Z.-M.}\ \bibnamefont {Yu}},\ and\ \bibinfo {author}
  {\bibfnamefont {Y.}~\bibnamefont {Yao}},\ }\bibfield  {title} {\bibinfo
  {title} {{Hourglass charge-three Weyl phonons}},\ }\href@noop {} {\bibfield
  {journal} {\bibinfo  {journal} {Phys. Rev. B}\ }\textbf {\bibinfo {volume}
  {106}},\ \bibinfo {pages} {214309} (\bibinfo {year}
  {2022}{\natexlab{a}})}\BibitemShut {NoStop}%
\bibitem [{\citenamefont {Wang}\ \emph
  {et~al.}(2022{\natexlab{b}})\citenamefont {Wang}, \citenamefont {Zhou},
  \citenamefont {Zhang}, \citenamefont {Wu}, \citenamefont {Yu},\ and\
  \citenamefont {Yang}}]{PhysRevB.106.195129}%
  \BibitemOpen
  \bibfield  {author} {\bibinfo {author} {\bibfnamefont {X.}~\bibnamefont
  {Wang}}, \bibinfo {author} {\bibfnamefont {F.}~\bibnamefont {Zhou}}, \bibinfo
  {author} {\bibfnamefont {Z.}~\bibnamefont {Zhang}}, \bibinfo {author}
  {\bibfnamefont {W.}~\bibnamefont {Wu}}, \bibinfo {author} {\bibfnamefont
  {Z.-M.}\ \bibnamefont {Yu}},\ and\ \bibinfo {author} {\bibfnamefont {S.~A.}\
  \bibnamefont {Yang}},\ }\bibfield  {title} {\bibinfo {title} {{Single pair of
  multi-Weyl points in nonmagnetic crystals}},\ }\href@noop {} {\bibfield
  {journal} {\bibinfo  {journal} {Phys. Rev. B}\ }\textbf {\bibinfo {volume}
  {106}},\ \bibinfo {pages} {195129} (\bibinfo {year}
  {2022}{\natexlab{b}})}\BibitemShut {NoStop}%
\bibitem [{\citenamefont {Fan}\ \emph {et~al.}(2023)\citenamefont {Fan},
  \citenamefont {Wan},\ and\ \citenamefont {Tang}}]{PhysRevB.108.104110}%
  \BibitemOpen
  \bibfield  {author} {\bibinfo {author} {\bibfnamefont {D.}~\bibnamefont
  {Fan}}, \bibinfo {author} {\bibfnamefont {X.}~\bibnamefont {Wan}},\ and\
  \bibinfo {author} {\bibfnamefont {F.}~\bibnamefont {Tang}},\ }\bibfield
  {title} {\bibinfo {title} {{Catalog of maximally charged Weyl points hosting
  nearly emanating nodal lines in phonon spectra}},\ }\href@noop {} {\bibfield
  {journal} {\bibinfo  {journal} {Phys. Rev. B}\ }\textbf {\bibinfo {volume}
  {108}},\ \bibinfo {pages} {104110} (\bibinfo {year} {2023})}\BibitemShut
  {NoStop}%
\bibitem [{\citenamefont {Wang}\ \emph {et~al.}(2024)\citenamefont {Wang},
  \citenamefont {Jin}, \citenamefont {Tian}, \citenamefont {Yu}, \citenamefont
  {Zhang}, \citenamefont {Liu},\ and\ \citenamefont
  {Liu}}]{PhysRevB.109.235415}%
  \BibitemOpen
  \bibfield  {author} {\bibinfo {author} {\bibfnamefont {C.}~\bibnamefont
  {Wang}}, \bibinfo {author} {\bibfnamefont {L.}~\bibnamefont {Jin}}, \bibinfo
  {author} {\bibfnamefont {L.}~\bibnamefont {Tian}}, \bibinfo {author}
  {\bibfnamefont {W.-W.}\ \bibnamefont {Yu}}, \bibinfo {author} {\bibfnamefont
  {X.}~\bibnamefont {Zhang}}, \bibinfo {author} {\bibfnamefont
  {G.}~\bibnamefont {Liu}},\ and\ \bibinfo {author} {\bibfnamefont
  {Y.}~\bibnamefont {Liu}},\ }\bibfield  {title} {\bibinfo {title} {{Multiply
  charged topological phonons in
  ${\mathrm{K}}_{2}{\mathrm{Pb}}_{2}{\mathrm{O}}_{3}$}},\ }\href@noop {}
  {\bibfield  {journal} {\bibinfo  {journal} {Phys. Rev. B}\ }\textbf {\bibinfo
  {volume} {109}},\ \bibinfo {pages} {235415} (\bibinfo {year}
  {2024})}\BibitemShut {NoStop}%
\bibitem [{\citenamefont {Liu}\ \emph {et~al.}(2024)\citenamefont {Liu},
  \citenamefont {Ma}, \citenamefont {Sun}, \citenamefont {Zhang}, \citenamefont
  {Ni}, \citenamefont {Meng},\ and\ \citenamefont
  {Zhao}}]{PhysRevB.109.045203}%
  \BibitemOpen
  \bibfield  {author} {\bibinfo {author} {\bibfnamefont {J.}~\bibnamefont
  {Liu}}, \bibinfo {author} {\bibfnamefont {X.}~\bibnamefont {Ma}}, \bibinfo
  {author} {\bibfnamefont {L.}~\bibnamefont {Sun}}, \bibinfo {author}
  {\bibfnamefont {Z.}~\bibnamefont {Zhang}}, \bibinfo {author} {\bibfnamefont
  {Y.}~\bibnamefont {Ni}}, \bibinfo {author} {\bibfnamefont {S.}~\bibnamefont
  {Meng}},\ and\ \bibinfo {author} {\bibfnamefont {M.}~\bibnamefont {Zhao}},\
  }\bibfield  {title} {\bibinfo {title} {{Ideal type-I Weyl phonons in
  ${\mathrm{BAsO}}_{4}$ with fewest Weyl points}},\ }\href@noop {} {\bibfield
  {journal} {\bibinfo  {journal} {Phys. Rev. B}\ }\textbf {\bibinfo {volume}
  {109}},\ \bibinfo {pages} {045203} (\bibinfo {year} {2024})}\BibitemShut
  {NoStop}%
\bibitem [{\citenamefont {Lange}\ \emph {et~al.}(2024)\citenamefont {Lange},
  \citenamefont {Pottecher}, \citenamefont {Robey}, \citenamefont {Monserrat},\
  and\ \citenamefont {Peng}}]{lange2024negative}%
  \BibitemOpen
  \bibfield  {author} {\bibinfo {author} {\bibfnamefont {G.~F.}\ \bibnamefont
  {Lange}}, \bibinfo {author} {\bibfnamefont {J.~D.}\ \bibnamefont
  {Pottecher}}, \bibinfo {author} {\bibfnamefont {C.}~\bibnamefont {Robey}},
  \bibinfo {author} {\bibfnamefont {B.}~\bibnamefont {Monserrat}},\ and\
  \bibinfo {author} {\bibfnamefont {B.}~\bibnamefont {Peng}},\ }\bibfield
  {title} {\bibinfo {title} {{Negative refraction of Weyl phonons at twin
  quartz interfaces}},\ }\href@noop {} {\bibfield  {journal} {\bibinfo
  {journal} {ACS Mater. Lett.}\ }\textbf {\bibinfo {volume} {6}},\ \bibinfo
  {pages} {847} (\bibinfo {year} {2024})}\BibitemShut {NoStop}%
\bibitem [{\citenamefont {Lv}\ \emph {et~al.}(2021)\citenamefont {Lv},
  \citenamefont {Qian},\ and\ \citenamefont {Ding}}]{RevModPhys.93.025002}%
  \BibitemOpen
  \bibfield  {author} {\bibinfo {author} {\bibfnamefont {B.~Q.}\ \bibnamefont
  {Lv}}, \bibinfo {author} {\bibfnamefont {T.}~\bibnamefont {Qian}},\ and\
  \bibinfo {author} {\bibfnamefont {H.}~\bibnamefont {Ding}},\ }\bibfield
  {title} {\bibinfo {title} {{Experimental perspective on three-dimensional
  topological semimetals}},\ }\href@noop {} {\bibfield  {journal} {\bibinfo
  {journal} {Rev. Mod. Phys.}\ }\textbf {\bibinfo {volume} {93}},\ \bibinfo
  {pages} {025002} (\bibinfo {year} {2021})}\BibitemShut {NoStop}%
\bibitem [{\citenamefont {Lv}\ \emph {et~al.}(2015{\natexlab{a}})\citenamefont
  {Lv}, \citenamefont {Xu}, \citenamefont {Weng}, \citenamefont {Ma},
  \citenamefont {Richard}, \citenamefont {Huang}, \citenamefont {Zhao},
  \citenamefont {Chen}, \citenamefont {Matt}, \citenamefont {Bisti} \emph
  {et~al.}}]{lv2015observation}%
  \BibitemOpen
  \bibfield  {author} {\bibinfo {author} {\bibfnamefont {B.}~\bibnamefont
  {Lv}}, \bibinfo {author} {\bibfnamefont {N.}~\bibnamefont {Xu}}, \bibinfo
  {author} {\bibfnamefont {H.}~\bibnamefont {Weng}}, \bibinfo {author}
  {\bibfnamefont {J.}~\bibnamefont {Ma}}, \bibinfo {author} {\bibfnamefont
  {P.}~\bibnamefont {Richard}}, \bibinfo {author} {\bibfnamefont
  {X.}~\bibnamefont {Huang}}, \bibinfo {author} {\bibfnamefont
  {L.}~\bibnamefont {Zhao}}, \bibinfo {author} {\bibfnamefont {G.}~\bibnamefont
  {Chen}}, \bibinfo {author} {\bibfnamefont {C.}~\bibnamefont {Matt}}, \bibinfo
  {author} {\bibfnamefont {F.}~\bibnamefont {Bisti}}, \emph {et~al.},\
  }\bibfield  {title} {\bibinfo {title} {{Observation of Weyl nodes in TaAs}},\
  }\href@noop {} {\bibfield  {journal} {\bibinfo  {journal} {Nat. Phys.}\
  }\textbf {\bibinfo {volume} {11}},\ \bibinfo {pages} {724} (\bibinfo {year}
  {2015}{\natexlab{a}})}\BibitemShut {NoStop}%
\bibitem [{\citenamefont {Lv}\ \emph {et~al.}(2015{\natexlab{b}})\citenamefont
  {Lv}, \citenamefont {Weng}, \citenamefont {Fu}, \citenamefont {Wang},
  \citenamefont {Miao}, \citenamefont {Ma}, \citenamefont {Richard},
  \citenamefont {Huang}, \citenamefont {Zhao}, \citenamefont {Chen},
  \citenamefont {Fang}, \citenamefont {Dai}, \citenamefont {Qian},\ and\
  \citenamefont {Ding}}]{PhysRevX.5.031013}%
  \BibitemOpen
  \bibfield  {author} {\bibinfo {author} {\bibfnamefont {B.~Q.}\ \bibnamefont
  {Lv}}, \bibinfo {author} {\bibfnamefont {H.~M.}\ \bibnamefont {Weng}},
  \bibinfo {author} {\bibfnamefont {B.~B.}\ \bibnamefont {Fu}}, \bibinfo
  {author} {\bibfnamefont {X.~P.}\ \bibnamefont {Wang}}, \bibinfo {author}
  {\bibfnamefont {H.}~\bibnamefont {Miao}}, \bibinfo {author} {\bibfnamefont
  {J.}~\bibnamefont {Ma}}, \bibinfo {author} {\bibfnamefont {P.}~\bibnamefont
  {Richard}}, \bibinfo {author} {\bibfnamefont {X.~C.}\ \bibnamefont {Huang}},
  \bibinfo {author} {\bibfnamefont {L.~X.}\ \bibnamefont {Zhao}}, \bibinfo
  {author} {\bibfnamefont {G.~F.}\ \bibnamefont {Chen}}, \bibinfo {author}
  {\bibfnamefont {Z.}~\bibnamefont {Fang}}, \bibinfo {author} {\bibfnamefont
  {X.}~\bibnamefont {Dai}}, \bibinfo {author} {\bibfnamefont {T.}~\bibnamefont
  {Qian}},\ and\ \bibinfo {author} {\bibfnamefont {H.}~\bibnamefont {Ding}},\
  }\bibfield  {title} {\bibinfo {title} {{Experimental Discovery of Weyl
  Semimetal TaAs}},\ }\href@noop {} {\bibfield  {journal} {\bibinfo  {journal}
  {Phys. Rev. X}\ }\textbf {\bibinfo {volume} {5}},\ \bibinfo {pages} {031013}
  (\bibinfo {year} {2015}{\natexlab{b}})}\BibitemShut {NoStop}%
\bibitem [{\citenamefont {Xu}\ \emph {et~al.}(2015)\citenamefont {Xu},
  \citenamefont {Belopolski}, \citenamefont {Alidoust}, \citenamefont
  {Neupane}, \citenamefont {Bian}, \citenamefont {Zhang}, \citenamefont
  {Sankar}, \citenamefont {Chang}, \citenamefont {Yuan}, \citenamefont {Lee}
  \emph {et~al.}}]{xu2015discovery}%
  \BibitemOpen
  \bibfield  {author} {\bibinfo {author} {\bibfnamefont {S.-Y.}\ \bibnamefont
  {Xu}}, \bibinfo {author} {\bibfnamefont {I.}~\bibnamefont {Belopolski}},
  \bibinfo {author} {\bibfnamefont {N.}~\bibnamefont {Alidoust}}, \bibinfo
  {author} {\bibfnamefont {M.}~\bibnamefont {Neupane}}, \bibinfo {author}
  {\bibfnamefont {G.}~\bibnamefont {Bian}}, \bibinfo {author} {\bibfnamefont
  {C.}~\bibnamefont {Zhang}}, \bibinfo {author} {\bibfnamefont
  {R.}~\bibnamefont {Sankar}}, \bibinfo {author} {\bibfnamefont
  {G.}~\bibnamefont {Chang}}, \bibinfo {author} {\bibfnamefont
  {Z.}~\bibnamefont {Yuan}}, \bibinfo {author} {\bibfnamefont {C.-C.}\
  \bibnamefont {Lee}}, \emph {et~al.},\ }\bibfield  {title} {\bibinfo {title}
  {{Discovery of a Weyl fermion semimetal and topological Fermi arcs}},\
  }\href@noop {} {\bibfield  {journal} {\bibinfo  {journal} {Science}\ }\textbf
  {\bibinfo {volume} {349}},\ \bibinfo {pages} {613} (\bibinfo {year}
  {2015})}\BibitemShut {NoStop}%
\bibitem [{\citenamefont {Yang}\ \emph {et~al.}(2015)\citenamefont {Yang},
  \citenamefont {Liu}, \citenamefont {Sun}, \citenamefont {Peng}, \citenamefont
  {Yang}, \citenamefont {Zhang}, \citenamefont {Zhou}, \citenamefont {Zhang},
  \citenamefont {Guo}, \citenamefont {Rahn} \emph {et~al.}}]{yang2015weyl}%
  \BibitemOpen
  \bibfield  {author} {\bibinfo {author} {\bibfnamefont {L.}~\bibnamefont
  {Yang}}, \bibinfo {author} {\bibfnamefont {Z.}~\bibnamefont {Liu}}, \bibinfo
  {author} {\bibfnamefont {Y.}~\bibnamefont {Sun}}, \bibinfo {author}
  {\bibfnamefont {H.}~\bibnamefont {Peng}}, \bibinfo {author} {\bibfnamefont
  {H.}~\bibnamefont {Yang}}, \bibinfo {author} {\bibfnamefont {T.}~\bibnamefont
  {Zhang}}, \bibinfo {author} {\bibfnamefont {B.}~\bibnamefont {Zhou}},
  \bibinfo {author} {\bibfnamefont {Y.}~\bibnamefont {Zhang}}, \bibinfo
  {author} {\bibfnamefont {Y.}~\bibnamefont {Guo}}, \bibinfo {author}
  {\bibfnamefont {M.}~\bibnamefont {Rahn}}, \emph {et~al.},\ }\bibfield
  {title} {\bibinfo {title} {{Weyl semimetal phase in the non-centrosymmetric
  compound TaAs}},\ }\href@noop {} {\bibfield  {journal} {\bibinfo  {journal}
  {Nat. Phys.}\ }\textbf {\bibinfo {volume} {11}},\ \bibinfo {pages} {728}
  (\bibinfo {year} {2015})}\BibitemShut {NoStop}%
\bibitem [{\citenamefont {Deng}\ \emph {et~al.}(2016)\citenamefont {Deng},
  \citenamefont {Wan}, \citenamefont {Deng}, \citenamefont {Zhang},
  \citenamefont {Ding}, \citenamefont {Wang}, \citenamefont {Yan},
  \citenamefont {Huang}, \citenamefont {Zhang}, \citenamefont {Xu} \emph
  {et~al.}}]{deng2016experimental}%
  \BibitemOpen
  \bibfield  {author} {\bibinfo {author} {\bibfnamefont {K.}~\bibnamefont
  {Deng}}, \bibinfo {author} {\bibfnamefont {G.}~\bibnamefont {Wan}}, \bibinfo
  {author} {\bibfnamefont {P.}~\bibnamefont {Deng}}, \bibinfo {author}
  {\bibfnamefont {K.}~\bibnamefont {Zhang}}, \bibinfo {author} {\bibfnamefont
  {S.}~\bibnamefont {Ding}}, \bibinfo {author} {\bibfnamefont {E.}~\bibnamefont
  {Wang}}, \bibinfo {author} {\bibfnamefont {M.}~\bibnamefont {Yan}}, \bibinfo
  {author} {\bibfnamefont {H.}~\bibnamefont {Huang}}, \bibinfo {author}
  {\bibfnamefont {H.}~\bibnamefont {Zhang}}, \bibinfo {author} {\bibfnamefont
  {Z.}~\bibnamefont {Xu}}, \emph {et~al.},\ }\bibfield  {title} {\bibinfo
  {title} {{Experimental observation of topological Fermi arcs in type-II Weyl
  semimetal MoTe$_{2}$}},\ }\href@noop {} {\bibfield  {journal} {\bibinfo
  {journal} {Nat. Phys.}\ }\textbf {\bibinfo {volume} {12}},\ \bibinfo {pages}
  {1105} (\bibinfo {year} {2016})}\BibitemShut {NoStop}%
\bibitem [{\citenamefont {Wu}\ \emph {et~al.}(2016)\citenamefont {Wu},
  \citenamefont {Mou}, \citenamefont {Jo}, \citenamefont {Sun}, \citenamefont
  {Huang}, \citenamefont {Bud'ko}, \citenamefont {Canfield},\ and\
  \citenamefont {Kaminski}}]{PhysRevB.94.121113}%
  \BibitemOpen
  \bibfield  {author} {\bibinfo {author} {\bibfnamefont {Y.}~\bibnamefont
  {Wu}}, \bibinfo {author} {\bibfnamefont {D.}~\bibnamefont {Mou}}, \bibinfo
  {author} {\bibfnamefont {N.~H.}\ \bibnamefont {Jo}}, \bibinfo {author}
  {\bibfnamefont {K.}~\bibnamefont {Sun}}, \bibinfo {author} {\bibfnamefont
  {L.}~\bibnamefont {Huang}}, \bibinfo {author} {\bibfnamefont {S.~L.}\
  \bibnamefont {Bud'ko}}, \bibinfo {author} {\bibfnamefont {P.~C.}\
  \bibnamefont {Canfield}},\ and\ \bibinfo {author} {\bibfnamefont
  {A.}~\bibnamefont {Kaminski}},\ }\bibfield  {title} {\bibinfo {title}
  {{Observation of Fermi arcs in the type-II Weyl semimetal candidate
  ${\mathrm{WTe}}_{2}$}},\ }\href@noop {} {\bibfield  {journal} {\bibinfo
  {journal} {Phys. Rev. B}\ }\textbf {\bibinfo {volume} {94}},\ \bibinfo
  {pages} {121113(R)} (\bibinfo {year} {2016})}\BibitemShut {NoStop}%
\bibitem [{\citenamefont {Miao}\ \emph {et~al.}(2018)\citenamefont {Miao},
  \citenamefont {Zhang}, \citenamefont {Wang}, \citenamefont {Meyers},
  \citenamefont {Said}, \citenamefont {Wang}, \citenamefont {Shi},
  \citenamefont {Weng}, \citenamefont {Fang},\ and\ \citenamefont
  {Dean}}]{PhysRevLett.121.035302}%
  \BibitemOpen
  \bibfield  {author} {\bibinfo {author} {\bibfnamefont {H.}~\bibnamefont
  {Miao}}, \bibinfo {author} {\bibfnamefont {T.~T.}\ \bibnamefont {Zhang}},
  \bibinfo {author} {\bibfnamefont {L.}~\bibnamefont {Wang}}, \bibinfo {author}
  {\bibfnamefont {D.}~\bibnamefont {Meyers}}, \bibinfo {author} {\bibfnamefont
  {A.~H.}\ \bibnamefont {Said}}, \bibinfo {author} {\bibfnamefont {Y.~L.}\
  \bibnamefont {Wang}}, \bibinfo {author} {\bibfnamefont {Y.~G.}\ \bibnamefont
  {Shi}}, \bibinfo {author} {\bibfnamefont {H.~M.}\ \bibnamefont {Weng}},
  \bibinfo {author} {\bibfnamefont {Z.}~\bibnamefont {Fang}},\ and\ \bibinfo
  {author} {\bibfnamefont {M.~P.~M.}\ \bibnamefont {Dean}},\ }\bibfield
  {title} {\bibinfo {title} {{Observation of double Weyl Phonons in
  Parity-Breaking FeSi}},\ }\href@noop {} {\bibfield  {journal} {\bibinfo
  {journal} {Phys. Rev. Lett.}\ }\textbf {\bibinfo {volume} {121}},\ \bibinfo
  {pages} {035302} (\bibinfo {year} {2018})}\BibitemShut {NoStop}%
\bibitem [{\citenamefont {Li}\ \emph {et~al.}(2023)\citenamefont {Li},
  \citenamefont {Li}, \citenamefont {Tang}, \citenamefont {Tao}, \citenamefont
  {Xue}, \citenamefont {Liu}, \citenamefont {Peng}, \citenamefont {Chen},
  \citenamefont {Guo},\ and\ \citenamefont {Zhu}}]{PhysRevLett.131.116602}%
  \BibitemOpen
  \bibfield  {author} {\bibinfo {author} {\bibfnamefont {J.}~\bibnamefont
  {Li}}, \bibinfo {author} {\bibfnamefont {J.}~\bibnamefont {Li}}, \bibinfo
  {author} {\bibfnamefont {J.}~\bibnamefont {Tang}}, \bibinfo {author}
  {\bibfnamefont {Z.}~\bibnamefont {Tao}}, \bibinfo {author} {\bibfnamefont
  {S.}~\bibnamefont {Xue}}, \bibinfo {author} {\bibfnamefont {J.}~\bibnamefont
  {Liu}}, \bibinfo {author} {\bibfnamefont {H.}~\bibnamefont {Peng}}, \bibinfo
  {author} {\bibfnamefont {X.-Q.}\ \bibnamefont {Chen}}, \bibinfo {author}
  {\bibfnamefont {J.}~\bibnamefont {Guo}},\ and\ \bibinfo {author}
  {\bibfnamefont {X.}~\bibnamefont {Zhu}},\ }\bibfield  {title} {\bibinfo
  {title} {{Direct Observation of Topological Phonons in Graphene}},\
  }\href@noop {} {\bibfield  {journal} {\bibinfo  {journal} {Phys. Rev. Lett.}\
  }\textbf {\bibinfo {volume} {131}},\ \bibinfo {pages} {116602} (\bibinfo
  {year} {2023})}\BibitemShut {NoStop}%
\bibitem [{\citenamefont {Zhang}\ \emph {et~al.}(2019)\citenamefont {Zhang},
  \citenamefont {Miao}, \citenamefont {Wang}, \citenamefont {Lin},
  \citenamefont {Cao}, \citenamefont {Fabbris}, \citenamefont {Said},
  \citenamefont {Liu}, \citenamefont {Lei}, \citenamefont {Fang}, \citenamefont
  {Weng},\ and\ \citenamefont {Dean}}]{PhysRevLett.123.245302}%
  \BibitemOpen
  \bibfield  {author} {\bibinfo {author} {\bibfnamefont {T.~T.}\ \bibnamefont
  {Zhang}}, \bibinfo {author} {\bibfnamefont {H.}~\bibnamefont {Miao}},
  \bibinfo {author} {\bibfnamefont {Q.}~\bibnamefont {Wang}}, \bibinfo {author}
  {\bibfnamefont {J.~Q.}\ \bibnamefont {Lin}}, \bibinfo {author} {\bibfnamefont
  {Y.}~\bibnamefont {Cao}}, \bibinfo {author} {\bibfnamefont {G.}~\bibnamefont
  {Fabbris}}, \bibinfo {author} {\bibfnamefont {A.~H.}\ \bibnamefont {Said}},
  \bibinfo {author} {\bibfnamefont {X.}~\bibnamefont {Liu}}, \bibinfo {author}
  {\bibfnamefont {H.~C.}\ \bibnamefont {Lei}}, \bibinfo {author} {\bibfnamefont
  {Z.}~\bibnamefont {Fang}}, \bibinfo {author} {\bibfnamefont {H.~M.}\
  \bibnamefont {Weng}},\ and\ \bibinfo {author} {\bibfnamefont {M.~P.~M.}\
  \bibnamefont {Dean}},\ }\bibfield  {title} {\bibinfo {title} {{Phononic
  Helical Nodal Lines with $\mathcal{PT}$ Protection in
  ${\mathrm{MoB}}_{2}$}},\ }\href@noop {} {\bibfield  {journal} {\bibinfo
  {journal} {Phys. Rev. Lett.}\ }\textbf {\bibinfo {volume} {123}},\ \bibinfo
  {pages} {245302} (\bibinfo {year} {2019})}\BibitemShut {NoStop}%
\bibitem [{\citenamefont {Jin}\ \emph {et~al.}(2022)\citenamefont {Jin},
  \citenamefont {Hu}, \citenamefont {Liu}, \citenamefont {Li}, \citenamefont
  {Zhang}, \citenamefont {Iida}, \citenamefont {Kamazawa}, \citenamefont
  {Kolesnikov}, \citenamefont {Stone}, \citenamefont {Zhang}, \citenamefont
  {Chen}, \citenamefont {Wang}, \citenamefont {Zaliznyak}, \citenamefont
  {Tranquada}, \citenamefont {Fang},\ and\ \citenamefont
  {Li}}]{PhysRevB.106.224304}%
  \BibitemOpen
  \bibfield  {author} {\bibinfo {author} {\bibfnamefont {Z.}~\bibnamefont
  {Jin}}, \bibinfo {author} {\bibfnamefont {B.}~\bibnamefont {Hu}}, \bibinfo
  {author} {\bibfnamefont {Y.}~\bibnamefont {Liu}}, \bibinfo {author}
  {\bibfnamefont {Y.}~\bibnamefont {Li}}, \bibinfo {author} {\bibfnamefont
  {T.}~\bibnamefont {Zhang}}, \bibinfo {author} {\bibfnamefont
  {K.}~\bibnamefont {Iida}}, \bibinfo {author} {\bibfnamefont {K.}~\bibnamefont
  {Kamazawa}}, \bibinfo {author} {\bibfnamefont {A.~I.}\ \bibnamefont
  {Kolesnikov}}, \bibinfo {author} {\bibfnamefont {M.~B.}\ \bibnamefont
  {Stone}}, \bibinfo {author} {\bibfnamefont {X.}~\bibnamefont {Zhang}},
  \bibinfo {author} {\bibfnamefont {H.}~\bibnamefont {Chen}}, \bibinfo {author}
  {\bibfnamefont {Y.}~\bibnamefont {Wang}}, \bibinfo {author} {\bibfnamefont
  {I.~A.}\ \bibnamefont {Zaliznyak}}, \bibinfo {author} {\bibfnamefont {J.~M.}\
  \bibnamefont {Tranquada}}, \bibinfo {author} {\bibfnamefont {C.}~\bibnamefont
  {Fang}},\ and\ \bibinfo {author} {\bibfnamefont {Y.}~\bibnamefont {Li}},\
  }\bibfield  {title} {\bibinfo {title} {{Chern numbers of topological phonon
  band crossing determined with inelastic neutron scattering}},\ }\href@noop {}
  {\bibfield  {journal} {\bibinfo  {journal} {Phys. Rev. B}\ }\textbf {\bibinfo
  {volume} {106}},\ \bibinfo {pages} {224304} (\bibinfo {year}
  {2022})}\BibitemShut {NoStop}%
\bibitem [{\citenamefont {Zyuzin}\ and\ \citenamefont
  {Burkov}(2012)}]{PhysRevB.86.115133}%
  \BibitemOpen
  \bibfield  {author} {\bibinfo {author} {\bibfnamefont {A.~A.}\ \bibnamefont
  {Zyuzin}}\ and\ \bibinfo {author} {\bibfnamefont {A.~A.}\ \bibnamefont
  {Burkov}},\ }\bibfield  {title} {\bibinfo {title} {{Topological response in
  Weyl semimetals and the chiral anomaly}},\ }\href@noop {} {\bibfield
  {journal} {\bibinfo  {journal} {Phys. Rev. B}\ }\textbf {\bibinfo {volume}
  {86}},\ \bibinfo {pages} {115133} (\bibinfo {year} {2012})}\BibitemShut
  {NoStop}%
\bibitem [{\citenamefont {Son}\ and\ \citenamefont
  {Spivak}(2013)}]{PhysRevB.88.104412}%
  \BibitemOpen
  \bibfield  {author} {\bibinfo {author} {\bibfnamefont {D.~T.}\ \bibnamefont
  {Son}}\ and\ \bibinfo {author} {\bibfnamefont {B.~Z.}\ \bibnamefont
  {Spivak}},\ }\bibfield  {title} {\bibinfo {title} {{Chiral anomaly and
  classical negative magnetoresistance of Weyl metals}},\ }\href@noop {}
  {\bibfield  {journal} {\bibinfo  {journal} {Phys. Rev. B}\ }\textbf {\bibinfo
  {volume} {88}},\ \bibinfo {pages} {104412} (\bibinfo {year}
  {2013})}\BibitemShut {NoStop}%
\bibitem [{\citenamefont {Vazifeh}\ and\ \citenamefont
  {Franz}(2013)}]{PhysRevLett.111.027201}%
  \BibitemOpen
  \bibfield  {author} {\bibinfo {author} {\bibfnamefont {M.~M.}\ \bibnamefont
  {Vazifeh}}\ and\ \bibinfo {author} {\bibfnamefont {M.}~\bibnamefont
  {Franz}},\ }\bibfield  {title} {\bibinfo {title} {{Electromagnetic Response
  of Weyl Semimetals}},\ }\href@noop {} {\bibfield  {journal} {\bibinfo
  {journal} {Phys. Rev. Lett.}\ }\textbf {\bibinfo {volume} {111}},\ \bibinfo
  {pages} {027201} (\bibinfo {year} {2013})}\BibitemShut {NoStop}%
\bibitem [{\citenamefont {Parameswaran}\ \emph {et~al.}(2014)\citenamefont
  {Parameswaran}, \citenamefont {Grover}, \citenamefont {Abanin}, \citenamefont
  {Pesin},\ and\ \citenamefont {Vishwanath}}]{PhysRevX.4.031035}%
  \BibitemOpen
  \bibfield  {author} {\bibinfo {author} {\bibfnamefont {S.~A.}\ \bibnamefont
  {Parameswaran}}, \bibinfo {author} {\bibfnamefont {T.}~\bibnamefont
  {Grover}}, \bibinfo {author} {\bibfnamefont {D.~A.}\ \bibnamefont {Abanin}},
  \bibinfo {author} {\bibfnamefont {D.~A.}\ \bibnamefont {Pesin}},\ and\
  \bibinfo {author} {\bibfnamefont {A.}~\bibnamefont {Vishwanath}},\ }\bibfield
   {title} {\bibinfo {title} {{Probing the Chiral Anomaly with Nonlocal
  Transport in Three-Dimensional Topological Semimetals}},\ }\href@noop {}
  {\bibfield  {journal} {\bibinfo  {journal} {Phys. Rev. X}\ }\textbf {\bibinfo
  {volume} {4}},\ \bibinfo {pages} {031035} (\bibinfo {year}
  {2014})}\BibitemShut {NoStop}%
\bibitem [{\citenamefont {Liu}\ \emph {et~al.}(2013)\citenamefont {Liu},
  \citenamefont {Ye},\ and\ \citenamefont {Qi}}]{PhysRevB.87.235306}%
  \BibitemOpen
  \bibfield  {author} {\bibinfo {author} {\bibfnamefont {C.-X.}\ \bibnamefont
  {Liu}}, \bibinfo {author} {\bibfnamefont {P.}~\bibnamefont {Ye}},\ and\
  \bibinfo {author} {\bibfnamefont {X.-L.}\ \bibnamefont {Qi}},\ }\bibfield
  {title} {\bibinfo {title} {{Chiral gauge field and axial anomaly in a Weyl
  semimetal}},\ }\href@noop {} {\bibfield  {journal} {\bibinfo  {journal}
  {Phys. Rev. B}\ }\textbf {\bibinfo {volume} {87}},\ \bibinfo {pages} {235306}
  (\bibinfo {year} {2013})}\BibitemShut {NoStop}%
\bibitem [{\citenamefont {Huang}\ \emph
  {et~al.}(2015{\natexlab{b}})\citenamefont {Huang}, \citenamefont {Zhao},
  \citenamefont {Long}, \citenamefont {Wang}, \citenamefont {Chen},
  \citenamefont {Yang}, \citenamefont {Liang}, \citenamefont {Xue},
  \citenamefont {Weng}, \citenamefont {Fang}, \citenamefont {Dai},\ and\
  \citenamefont {Chen}}]{PhysRevX.5.031023}%
  \BibitemOpen
  \bibfield  {author} {\bibinfo {author} {\bibfnamefont {X.}~\bibnamefont
  {Huang}}, \bibinfo {author} {\bibfnamefont {L.}~\bibnamefont {Zhao}},
  \bibinfo {author} {\bibfnamefont {Y.}~\bibnamefont {Long}}, \bibinfo {author}
  {\bibfnamefont {P.}~\bibnamefont {Wang}}, \bibinfo {author} {\bibfnamefont
  {D.}~\bibnamefont {Chen}}, \bibinfo {author} {\bibfnamefont {Z.}~\bibnamefont
  {Yang}}, \bibinfo {author} {\bibfnamefont {H.}~\bibnamefont {Liang}},
  \bibinfo {author} {\bibfnamefont {M.}~\bibnamefont {Xue}}, \bibinfo {author}
  {\bibfnamefont {H.}~\bibnamefont {Weng}}, \bibinfo {author} {\bibfnamefont
  {Z.}~\bibnamefont {Fang}}, \bibinfo {author} {\bibfnamefont {X.}~\bibnamefont
  {Dai}},\ and\ \bibinfo {author} {\bibfnamefont {G.}~\bibnamefont {Chen}},\
  }\bibfield  {title} {\bibinfo {title} {{Observation of the
  Chiral-Anomaly-Induced Negative Magnetoresistance in 3D Weyl Semimetal
  TaAs}},\ }\href@noop {} {\bibfield  {journal} {\bibinfo  {journal} {Phys.
  Rev. X}\ }\textbf {\bibinfo {volume} {5}},\ \bibinfo {pages} {031023}
  (\bibinfo {year} {2015}{\natexlab{b}})}\BibitemShut {NoStop}%
\bibitem [{\citenamefont {Kim}\ \emph {et~al.}(2013)\citenamefont {Kim},
  \citenamefont {Kim}, \citenamefont {Wang}, \citenamefont {Sasaki},
  \citenamefont {Satoh}, \citenamefont {Ohnishi}, \citenamefont {Kitaura},
  \citenamefont {Yang},\ and\ \citenamefont {Li}}]{PhysRevLett.111.246603}%
  \BibitemOpen
  \bibfield  {author} {\bibinfo {author} {\bibfnamefont {H.-J.}\ \bibnamefont
  {Kim}}, \bibinfo {author} {\bibfnamefont {K.-S.}\ \bibnamefont {Kim}},
  \bibinfo {author} {\bibfnamefont {J.-F.}\ \bibnamefont {Wang}}, \bibinfo
  {author} {\bibfnamefont {M.}~\bibnamefont {Sasaki}}, \bibinfo {author}
  {\bibfnamefont {N.}~\bibnamefont {Satoh}}, \bibinfo {author} {\bibfnamefont
  {A.}~\bibnamefont {Ohnishi}}, \bibinfo {author} {\bibfnamefont
  {M.}~\bibnamefont {Kitaura}}, \bibinfo {author} {\bibfnamefont
  {M.}~\bibnamefont {Yang}},\ and\ \bibinfo {author} {\bibfnamefont
  {L.}~\bibnamefont {Li}},\ }\bibfield  {title} {\bibinfo {title} {{Dirac
  versus Weyl Fermions in Topological Insulators: Adler-Bell-Jackiw Anomaly in
  Transport Phenomena}},\ }\href@noop {} {\bibfield  {journal} {\bibinfo
  {journal} {Phys. Rev. Lett.}\ }\textbf {\bibinfo {volume} {111}},\ \bibinfo
  {pages} {246603} (\bibinfo {year} {2013})}\BibitemShut {NoStop}%
\bibitem [{\citenamefont {Zhao}\ and\ \citenamefont
  {Yang}(2021)}]{PhysRevLett.126.046401}%
  \BibitemOpen
  \bibfield  {author} {\bibinfo {author} {\bibfnamefont {Y.~X.}\ \bibnamefont
  {Zhao}}\ and\ \bibinfo {author} {\bibfnamefont {S.~A.}\ \bibnamefont
  {Yang}},\ }\bibfield  {title} {\bibinfo {title} {{Index Theorem on Chiral
  Landau Bands for Topological Fermions}},\ }\href@noop {} {\bibfield
  {journal} {\bibinfo  {journal} {Phys. Rev. Lett.}\ }\textbf {\bibinfo
  {volume} {126}},\ \bibinfo {pages} {046401} (\bibinfo {year}
  {2021})}\BibitemShut {NoStop}%
\bibitem [{\citenamefont {Xiong}\ \emph {et~al.}(2015)\citenamefont {Xiong},
  \citenamefont {Kushwaha}, \citenamefont {Liang}, \citenamefont {Krizan},
  \citenamefont {Hirschberger}, \citenamefont {Wang}, \citenamefont {Cava},\
  and\ \citenamefont {Ong}}]{xiong2015evidence}%
  \BibitemOpen
  \bibfield  {author} {\bibinfo {author} {\bibfnamefont {J.}~\bibnamefont
  {Xiong}}, \bibinfo {author} {\bibfnamefont {S.~K.}\ \bibnamefont {Kushwaha}},
  \bibinfo {author} {\bibfnamefont {T.}~\bibnamefont {Liang}}, \bibinfo
  {author} {\bibfnamefont {J.~W.}\ \bibnamefont {Krizan}}, \bibinfo {author}
  {\bibfnamefont {M.}~\bibnamefont {Hirschberger}}, \bibinfo {author}
  {\bibfnamefont {W.}~\bibnamefont {Wang}}, \bibinfo {author} {\bibfnamefont
  {R.~J.}\ \bibnamefont {Cava}},\ and\ \bibinfo {author} {\bibfnamefont
  {N.~P.}\ \bibnamefont {Ong}},\ }\bibfield  {title} {\bibinfo {title}
  {{Evidence for the chiral anomaly in the Dirac semimetal Na$_{3}$Bi}},\
  }\href@noop {} {\bibfield  {journal} {\bibinfo  {journal} {Science}\ }\textbf
  {\bibinfo {volume} {350}},\ \bibinfo {pages} {413} (\bibinfo {year}
  {2015})}\BibitemShut {NoStop}%
\bibitem [{\citenamefont {K\"onye}\ \emph {et~al.}(2021)\citenamefont
  {K\"onye}, \citenamefont {Bouhon}, \citenamefont {Fulga}, \citenamefont
  {Slager}, \citenamefont {van~den Brink},\ and\ \citenamefont
  {Facio}}]{PhysRevResearch.3.L042017}%
  \BibitemOpen
  \bibfield  {author} {\bibinfo {author} {\bibfnamefont {V.}~\bibnamefont
  {K\"onye}}, \bibinfo {author} {\bibfnamefont {A.}~\bibnamefont {Bouhon}},
  \bibinfo {author} {\bibfnamefont {I.~C.}\ \bibnamefont {Fulga}}, \bibinfo
  {author} {\bibfnamefont {R.-J.}\ \bibnamefont {Slager}}, \bibinfo {author}
  {\bibfnamefont {J.}~\bibnamefont {van~den Brink}},\ and\ \bibinfo {author}
  {\bibfnamefont {J.~I.}\ \bibnamefont {Facio}},\ }\bibfield  {title} {\bibinfo
  {title} {{Chirality flip of Weyl nodes and its manifestation in strained
  ${\mathrm{MoTe}}_{2}$}},\ }\href@noop {} {\bibfield  {journal} {\bibinfo
  {journal} {Phys. Rev. Res.}\ }\textbf {\bibinfo {volume} {3}},\ \bibinfo
  {pages} {L042017} (\bibinfo {year} {2021})}\BibitemShut {NoStop}%
\bibitem [{\citenamefont {Thareja}\ \emph {et~al.}()\citenamefont {Thareja},
  \citenamefont {Pantano}, \citenamefont {Vekhter},\ and\ \citenamefont
  {Gayles}}]{thareja}%
  \BibitemOpen
  \bibfield  {author} {\bibinfo {author} {\bibfnamefont {E.}~\bibnamefont
  {Thareja}}, \bibinfo {author} {\bibfnamefont {G.}~\bibnamefont {Pantano}},
  \bibinfo {author} {\bibfnamefont {I.}~\bibnamefont {Vekhter}},\ and\ \bibinfo
  {author} {\bibfnamefont {J.}~\bibnamefont {Gayles}},\ }\bibfield  {title}
  {\bibinfo {title} {{Tuning Quantum States at Chirality-Reversed Planar
  Interface in Weyl Semimetals using an Interstitial Layer}},\ }\href@noop {}
  {\bibinfo  {journal} {arXiv:2501.05594}\ }\BibitemShut {NoStop}%
\bibitem [{\citenamefont {Yoshikawa}\ \emph {et~al.}(2022)\citenamefont
  {Yoshikawa}, \citenamefont {Ogawa}, \citenamefont {Hirai}, \citenamefont
  {Fujiwara}, \citenamefont {Ikeda}, \citenamefont {Tsukazaki},\ and\
  \citenamefont {Shimano}}]{yoshikawa2022non}%
  \BibitemOpen
\bibfield  {journal} {  }\bibfield  {author} {\bibinfo {author} {\bibfnamefont
  {N.}~\bibnamefont {Yoshikawa}}, \bibinfo {author} {\bibfnamefont
  {K.}~\bibnamefont {Ogawa}}, \bibinfo {author} {\bibfnamefont
  {Y.}~\bibnamefont {Hirai}}, \bibinfo {author} {\bibfnamefont
  {K.}~\bibnamefont {Fujiwara}}, \bibinfo {author} {\bibfnamefont
  {J.}~\bibnamefont {Ikeda}}, \bibinfo {author} {\bibfnamefont
  {A.}~\bibnamefont {Tsukazaki}},\ and\ \bibinfo {author} {\bibfnamefont
  {R.}~\bibnamefont {Shimano}},\ }\bibfield  {title} {\bibinfo {title}
  {{Non-volatile chirality switching by all-optical magnetization reversal in
  ferromagnetic Weyl semimetal Co$_{3}$Sn$_{2}$S$_{2}$}},\ }\href@noop {}
  {\bibfield  {journal} {\bibinfo  {journal} {Commun. Phys.}\ }\textbf
  {\bibinfo {volume} {5}},\ \bibinfo {pages} {328} (\bibinfo {year}
  {2022})}\BibitemShut {NoStop}%
\bibitem [{\citenamefont {Liu}\ \emph {et~al.}(2016)\citenamefont {Liu},
  \citenamefont {Kim}, \citenamefont {Tan},\ and\ \citenamefont
  {Rappe}}]{liu2016strain}%
  \BibitemOpen
  \bibfield  {author} {\bibinfo {author} {\bibfnamefont {S.}~\bibnamefont
  {Liu}}, \bibinfo {author} {\bibfnamefont {Y.}~\bibnamefont {Kim}}, \bibinfo
  {author} {\bibfnamefont {L.~Z.}\ \bibnamefont {Tan}},\ and\ \bibinfo {author}
  {\bibfnamefont {A.~M.}\ \bibnamefont {Rappe}},\ }\bibfield  {title} {\bibinfo
  {title} {{Strain-induced ferroelectric topological insulator}},\ }\href@noop
  {} {\bibfield  {journal} {\bibinfo  {journal} {Nano Lett.}\ }\textbf
  {\bibinfo {volume} {16}},\ \bibinfo {pages} {1663} (\bibinfo {year}
  {2016})}\BibitemShut {NoStop}%
\bibitem [{\citenamefont {Narayan}(2015)}]{PhysRevB.92.220101}%
  \BibitemOpen
  \bibfield  {author} {\bibinfo {author} {\bibfnamefont {A.}~\bibnamefont
  {Narayan}},\ }\bibfield  {title} {\bibinfo {title} {{Class of Rashba
  ferroelectrics in hexagonal semiconductors}},\ }\href@noop {} {\bibfield
  {journal} {\bibinfo  {journal} {Phys. Rev. B}\ }\textbf {\bibinfo {volume}
  {92}},\ \bibinfo {pages} {220101(R)} (\bibinfo {year} {2015})}\BibitemShut
  {NoStop}%
\bibitem [{\citenamefont {Di~Sante}\ \emph {et~al.}(2016)\citenamefont
  {Di~Sante}, \citenamefont {Barone}, \citenamefont {Stroppa}, \citenamefont
  {Garrity}, \citenamefont {Vanderbilt},\ and\ \citenamefont
  {Picozzi}}]{PhysRevLett.117.076401}%
  \BibitemOpen
  \bibfield  {author} {\bibinfo {author} {\bibfnamefont {D.}~\bibnamefont
  {Di~Sante}}, \bibinfo {author} {\bibfnamefont {P.}~\bibnamefont {Barone}},
  \bibinfo {author} {\bibfnamefont {A.}~\bibnamefont {Stroppa}}, \bibinfo
  {author} {\bibfnamefont {K.~F.}\ \bibnamefont {Garrity}}, \bibinfo {author}
  {\bibfnamefont {D.}~\bibnamefont {Vanderbilt}},\ and\ \bibinfo {author}
  {\bibfnamefont {S.}~\bibnamefont {Picozzi}},\ }\bibfield  {title} {\bibinfo
  {title} {{Intertwined Rashba, Dirac, and Weyl fermions in hexagonal
  hyperferroelectrics}},\ }\href@noop {} {\bibfield  {journal} {\bibinfo
  {journal} {Phys. Rev. Lett.}\ }\textbf {\bibinfo {volume} {117}},\ \bibinfo
  {pages} {076401} (\bibinfo {year} {2016})}\BibitemShut {NoStop}%
\bibitem [{\citenamefont {Monserrat}\ \emph {et~al.}(2017)\citenamefont
  {Monserrat}, \citenamefont {Bennett}, \citenamefont {Rabe},\ and\
  \citenamefont {Vanderbilt}}]{PhysRevLett.119.036802}%
  \BibitemOpen
  \bibfield  {author} {\bibinfo {author} {\bibfnamefont {B.}~\bibnamefont
  {Monserrat}}, \bibinfo {author} {\bibfnamefont {J.~W.}\ \bibnamefont
  {Bennett}}, \bibinfo {author} {\bibfnamefont {K.~M.}\ \bibnamefont {Rabe}},\
  and\ \bibinfo {author} {\bibfnamefont {D.}~\bibnamefont {Vanderbilt}},\
  }\bibfield  {title} {\bibinfo {title} {{Antiferroelectric Topological
  Insulators in Orthorhombic $A\mathrm{MgBi}$ Compounds ($A=\mathrm{Li}$, Na,
  K)}},\ }\href@noop {} {\bibfield  {journal} {\bibinfo  {journal} {Phys. Rev.
  Lett.}\ }\textbf {\bibinfo {volume} {119}},\ \bibinfo {pages} {036802}
  (\bibinfo {year} {2017})}\BibitemShut {NoStop}%
\bibitem [{\citenamefont {Mao}\ \emph {et~al.}(2023)\citenamefont {Mao},
  \citenamefont {Li}, \citenamefont {Zou}, \citenamefont {Dai}, \citenamefont
  {Huang},\ and\ \citenamefont {Niu}}]{PhysRevB.107.045125}%
  \BibitemOpen
  \bibfield  {author} {\bibinfo {author} {\bibfnamefont {N.}~\bibnamefont
  {Mao}}, \bibinfo {author} {\bibfnamefont {R.}~\bibnamefont {Li}}, \bibinfo
  {author} {\bibfnamefont {X.}~\bibnamefont {Zou}}, \bibinfo {author}
  {\bibfnamefont {Y.}~\bibnamefont {Dai}}, \bibinfo {author} {\bibfnamefont
  {B.}~\bibnamefont {Huang}},\ and\ \bibinfo {author} {\bibfnamefont
  {C.}~\bibnamefont {Niu}},\ }\bibfield  {title} {\bibinfo {title}
  {{Ferroelectric higher-order topological insulator in two dimensions}},\
  }\href@noop {} {\bibfield  {journal} {\bibinfo  {journal} {Phys. Rev. B}\
  }\textbf {\bibinfo {volume} {107}},\ \bibinfo {pages} {045125} (\bibinfo
  {year} {2023})}\BibitemShut {NoStop}%
\bibitem [{\citenamefont {Yang}\ \emph {et~al.}(2025)\citenamefont {Yang},
  \citenamefont {Zhang}, \citenamefont {Li}, \citenamefont {Zhang},
  \citenamefont {Yu}, \citenamefont {Huang},\ and\ \citenamefont {Yao}}]{jsm}%
  \BibitemOpen
  \bibfield  {author} {\bibinfo {author} {\bibfnamefont {N.-J.}\ \bibnamefont
  {Yang}}, \bibinfo {author} {\bibfnamefont {J.-M.}\ \bibnamefont {Zhang}},
  \bibinfo {author} {\bibfnamefont {X.-P.}\ \bibnamefont {Li}}, \bibinfo
  {author} {\bibfnamefont {Z.}~\bibnamefont {Zhang}}, \bibinfo {author}
  {\bibfnamefont {Z.-M.}\ \bibnamefont {Yu}}, \bibinfo {author} {\bibfnamefont
  {Z.}~\bibnamefont {Huang}},\ and\ \bibinfo {author} {\bibfnamefont
  {Y.}~\bibnamefont {Yao}},\ }\bibfield  {title} {\bibinfo {title} {{Sliding
  Ferroelectrics Induced Hybrid-Order Topological Phase Transitions}},\
  }\href@noop {} {\bibfield  {journal} {\bibinfo  {journal} {Phys. Rev. Lett.}\
  }\textbf {\bibinfo {volume} {134}},\ \bibinfo {pages} {256602} (\bibinfo
  {year} {2025})}\BibitemShut {NoStop}%
\bibitem [{\citenamefont {Zhao}\ \emph {et~al.}(2018)\citenamefont {Zhao},
  \citenamefont {Dong}, \citenamefont {Gao}, \citenamefont {Xu}, \citenamefont
  {Xu}, \citenamefont {Yan}, \citenamefont {Zhao}, \citenamefont {Liu},
  \citenamefont {Yan}, \citenamefont {Zhang} \emph
  {et~al.}}]{zhao2018reversible}%
  \BibitemOpen
  \bibfield  {author} {\bibinfo {author} {\bibfnamefont {X.-W.}\ \bibnamefont
  {Zhao}}, \bibinfo {author} {\bibfnamefont {S.-N.}\ \bibnamefont {Dong}},
  \bibinfo {author} {\bibfnamefont {G.-Y.}\ \bibnamefont {Gao}}, \bibinfo
  {author} {\bibfnamefont {Z.-X.}\ \bibnamefont {Xu}}, \bibinfo {author}
  {\bibfnamefont {M.}~\bibnamefont {Xu}}, \bibinfo {author} {\bibfnamefont
  {J.-M.}\ \bibnamefont {Yan}}, \bibinfo {author} {\bibfnamefont {W.-Y.}\
  \bibnamefont {Zhao}}, \bibinfo {author} {\bibfnamefont {Y.-K.}\ \bibnamefont
  {Liu}}, \bibinfo {author} {\bibfnamefont {S.-Y.}\ \bibnamefont {Yan}},
  \bibinfo {author} {\bibfnamefont {J.-X.}\ \bibnamefont {Zhang}}, \emph
  {et~al.},\ }\bibfield  {title} {\bibinfo {title} {{Reversible and nonvolatile
  manipulation of the electronic transport properties of topological insulators
  by ferroelectric polarization switching}},\ }\href@noop {} {\bibfield
  {journal} {\bibinfo  {journal} {npj Quantum Mater.}\ }\textbf {\bibinfo
  {volume} {3}},\ \bibinfo {pages} {52} (\bibinfo {year} {2018})}\BibitemShut
  {NoStop}%
\bibitem [{\citenamefont {Sharma}\ \emph {et~al.}(2019)\citenamefont {Sharma},
  \citenamefont {Xiang}, \citenamefont {Shao}, \citenamefont {Zhang},
  \citenamefont {Tsymbal}, \citenamefont {Hamilton},\ and\ \citenamefont
  {Seidel}}]{sharma2019room}%
  \BibitemOpen
  \bibfield  {author} {\bibinfo {author} {\bibfnamefont {P.}~\bibnamefont
  {Sharma}}, \bibinfo {author} {\bibfnamefont {F.-X.}\ \bibnamefont {Xiang}},
  \bibinfo {author} {\bibfnamefont {D.-F.}\ \bibnamefont {Shao}}, \bibinfo
  {author} {\bibfnamefont {D.}~\bibnamefont {Zhang}}, \bibinfo {author}
  {\bibfnamefont {E.~Y.}\ \bibnamefont {Tsymbal}}, \bibinfo {author}
  {\bibfnamefont {A.~R.}\ \bibnamefont {Hamilton}},\ and\ \bibinfo {author}
  {\bibfnamefont {J.}~\bibnamefont {Seidel}},\ }\bibfield  {title} {\bibinfo
  {title} {{A room-temperature ferroelectric semimetal}},\ }\href@noop {}
  {\bibfield  {journal} {\bibinfo  {journal} {Sci. Adv.}\ }\textbf {\bibinfo
  {volume} {5}},\ \bibinfo {pages} {eaax5080} (\bibinfo {year}
  {2019})}\BibitemShut {NoStop}%
\bibitem [{\citenamefont {Li}\ \emph {et~al.}(2016{\natexlab{a}})\citenamefont
  {Li}, \citenamefont {Xu}, \citenamefont {He}, \citenamefont {Ullah},
  \citenamefont {Li}, \citenamefont {Liu}, \citenamefont {Li}, \citenamefont
  {Franchini}, \citenamefont {Weng},\ and\ \citenamefont {Chen}}]{li2016weyl}%
  \BibitemOpen
  \bibfield  {author} {\bibinfo {author} {\bibfnamefont {R.}~\bibnamefont
  {Li}}, \bibinfo {author} {\bibfnamefont {Y.}~\bibnamefont {Xu}}, \bibinfo
  {author} {\bibfnamefont {J.}~\bibnamefont {He}}, \bibinfo {author}
  {\bibfnamefont {S.}~\bibnamefont {Ullah}}, \bibinfo {author} {\bibfnamefont
  {J.}~\bibnamefont {Li}}, \bibinfo {author} {\bibfnamefont {J.-M.}\
  \bibnamefont {Liu}}, \bibinfo {author} {\bibfnamefont {D.}~\bibnamefont
  {Li}}, \bibinfo {author} {\bibfnamefont {C.}~\bibnamefont {Franchini}},
  \bibinfo {author} {\bibfnamefont {H.}~\bibnamefont {Weng}},\ and\ \bibinfo
  {author} {\bibfnamefont {X.-Q.}\ \bibnamefont {Chen}},\ }\bibfield  {title}
  {\bibinfo {title} {{Weyl ferroelectric semimetal}},\ }\href@noop {}
  {\bibfield  {journal} {\bibinfo  {journal} {arXiv:1610.07142}\ } (\bibinfo
  {year} {2016}{\natexlab{a}})}\BibitemShut {NoStop}%
\bibitem [{\citenamefont {Bradley}\ and\ \citenamefont
  {Cracknell}(2009)}]{Bradley2009Mathematical-Oxford}%
  \BibitemOpen
  \bibfield  {author} {\bibinfo {author} {\bibfnamefont {C.}~\bibnamefont
  {Bradley}}\ and\ \bibinfo {author} {\bibfnamefont {A.}~\bibnamefont
  {Cracknell}},\ }\href@noop {} {\emph {\bibinfo {title} {{The Mathematical
  Theory of Symmetry in Solids: Representation Theory for Point Groups and
  Space Groups}}}}\ (\bibinfo  {publisher} {Oxford University Press, Oxford,
  U.K.},\ \bibinfo {year} {2009})\BibitemShut {NoStop}%
\bibitem [{sm()}]{sm}%
  \BibitemOpen
  \href@noop {} {\bibinfo  {journal} {See Supplemental Material at [URL to be
  added by publisher] for a more detailed table of candidate space groups,
  computational methods, the whole phonon spectrum of K$_{2}$ZnBr$_{4}$ and
  other materials in the $A$$_{2}$$BX$$_{4}$ family (where $A$ = Rb, K; $B$ =
  Zn, Co; $X$ = Cl, Br, I), the Wilson loop spectrum, and the symmetry analysis
  and $k\cdot p$ model of Weyl points. Also included are chirality-switchable
  Weyl phonon materials with high Chern numbers ($C$ = 2 and 3), namely
  Y(OH)$_{3}$ and LiIO$_{3}$, as well as the symmetry requirements for the
  nonlinear phonon Hall effect in K$_{2}$ZnBr$_{4}$, and the effects of
  temperature and anharmonicity on phonons. These discussions include
  Refs.~\cite{PhysRevB.107.L241107,s1,s2,s3,s4,s5,s6,s7,s8,s9}}\ }\BibitemShut
  {NoStop}%
\bibitem [{\citenamefont {Shimizu}\ \emph {et~al.}(1990)\citenamefont
  {Shimizu}, \citenamefont {Yamaguchi}, \citenamefont {Suzuki}, \citenamefont
  {Takashige},\ and\ \citenamefont {Sawada}}]{shimizu1990new}%
  \BibitemOpen
\bibfield  {journal} {  }\bibfield  {author} {\bibinfo {author} {\bibfnamefont
  {F.}~\bibnamefont {Shimizu}}, \bibinfo {author} {\bibfnamefont
  {T.}~\bibnamefont {Yamaguchi}}, \bibinfo {author} {\bibfnamefont
  {H.}~\bibnamefont {Suzuki}}, \bibinfo {author} {\bibfnamefont
  {M.}~\bibnamefont {Takashige}},\ and\ \bibinfo {author} {\bibfnamefont
  {S.}~\bibnamefont {Sawada}},\ }\bibfield  {title} {\bibinfo {title} {{New
  Ferroelectric K$_{2}$ZnBr$_{4}$}},\ }\href@noop {} {\bibfield  {journal}
  {\bibinfo  {journal} {J. Phys. Soc. Jpn.}\ }\textbf {\bibinfo {volume}
  {59}},\ \bibinfo {pages} {1936} (\bibinfo {year} {1990})}\BibitemShut
  {NoStop}%
\bibitem [{\citenamefont {Fang}\ \emph {et~al.}(2016)\citenamefont {Fang},
  \citenamefont {Lu}, \citenamefont {Liu},\ and\ \citenamefont
  {Fu}}]{fang2016topological}%
  \BibitemOpen
  \bibfield  {author} {\bibinfo {author} {\bibfnamefont {C.}~\bibnamefont
  {Fang}}, \bibinfo {author} {\bibfnamefont {L.}~\bibnamefont {Lu}}, \bibinfo
  {author} {\bibfnamefont {J.}~\bibnamefont {Liu}},\ and\ \bibinfo {author}
  {\bibfnamefont {L.}~\bibnamefont {Fu}},\ }\bibfield  {title} {\bibinfo
  {title} {{Topological semimetals with helicoid surface states}},\ }\href@noop
  {} {\bibfield  {journal} {\bibinfo  {journal} {Nat. Phys.}\ }\textbf
  {\bibinfo {volume} {12}},\ \bibinfo {pages} {936} (\bibinfo {year}
  {2016})}\BibitemShut {NoStop}%
\bibitem [{\citenamefont {Luo}\ \emph {et~al.}(2023)\citenamefont {Luo},
  \citenamefont {Ji}, \citenamefont {Chen}, \citenamefont {Xu}, \citenamefont
  {Zhang}, \citenamefont {Xiang},\ and\ \citenamefont
  {Bellaiche}}]{PhysRevB.107.L241107}%
  \BibitemOpen
  \bibfield  {author} {\bibinfo {author} {\bibfnamefont {W.}~\bibnamefont
  {Luo}}, \bibinfo {author} {\bibfnamefont {J.}~\bibnamefont {Ji}}, \bibinfo
  {author} {\bibfnamefont {P.}~\bibnamefont {Chen}}, \bibinfo {author}
  {\bibfnamefont {Y.}~\bibnamefont {Xu}}, \bibinfo {author} {\bibfnamefont
  {L.}~\bibnamefont {Zhang}}, \bibinfo {author} {\bibfnamefont
  {H.}~\bibnamefont {Xiang}},\ and\ \bibinfo {author} {\bibfnamefont
  {L.}~\bibnamefont {Bellaiche}},\ }\bibfield  {title} {\bibinfo {title}
  {{Nonlinear phonon Hall effects in ferroelectrics: Existence and nonvolatile
  electrical control}},\ }\href@noop {} {\bibfield  {journal} {\bibinfo
  {journal} {Phys. Rev. B}\ }\textbf {\bibinfo {volume} {107}},\ \bibinfo
  {pages} {L241107} (\bibinfo {year} {2023})}\BibitemShut {NoStop}%
\bibitem [{\citenamefont {Chen}\ \emph {et~al.}(2024)\citenamefont {Chen},
  \citenamefont {Wu}, \citenamefont {Sun}, \citenamefont {Yang},\ and\
  \citenamefont {Zhang}}]{chen2024electrically}%
  \BibitemOpen
  \bibfield  {author} {\bibinfo {author} {\bibfnamefont {H.}~\bibnamefont
  {Chen}}, \bibinfo {author} {\bibfnamefont {W.}~\bibnamefont {Wu}}, \bibinfo
  {author} {\bibfnamefont {K.}~\bibnamefont {Sun}}, \bibinfo {author}
  {\bibfnamefont {S.~A.}\ \bibnamefont {Yang}},\ and\ \bibinfo {author}
  {\bibfnamefont {L.}~\bibnamefont {Zhang}},\ }\bibfield  {title} {\bibinfo
  {title} {{Electrically controllable chiral phonons in ferroelectric
  materials}},\ }\href@noop {} {\bibfield  {journal} {\bibinfo  {journal}
  {Appl. Phys. Lett.}\ }\textbf {\bibinfo {volume} {124}} (\bibinfo {year}
  {2024})}\BibitemShut {NoStop}%
\bibitem [{\citenamefont {Hirayama}\ \emph {et~al.}(2015)\citenamefont
  {Hirayama}, \citenamefont {Okugawa}, \citenamefont {Ishibashi}, \citenamefont
  {Murakami},\ and\ \citenamefont {Miyake}}]{PhysRevLett.114.206401}%
  \BibitemOpen
  \bibfield  {author} {\bibinfo {author} {\bibfnamefont {M.}~\bibnamefont
  {Hirayama}}, \bibinfo {author} {\bibfnamefont {R.}~\bibnamefont {Okugawa}},
  \bibinfo {author} {\bibfnamefont {S.}~\bibnamefont {Ishibashi}}, \bibinfo
  {author} {\bibfnamefont {S.}~\bibnamefont {Murakami}},\ and\ \bibinfo
  {author} {\bibfnamefont {T.}~\bibnamefont {Miyake}},\ }\bibfield  {title}
  {\bibinfo {title} {{Weyl node and spin texture in trigonal tellurium and
  selenium}},\ }\href@noop {} {\bibfield  {journal} {\bibinfo  {journal} {Phys.
  Rev. Lett.}\ }\textbf {\bibinfo {volume} {114}},\ \bibinfo {pages} {206401}
  (\bibinfo {year} {2015})}\BibitemShut {NoStop}%
\bibitem [{\citenamefont {Zhang}\ \emph
  {et~al.}(2020{\natexlab{b}})\citenamefont {Zhang}, \citenamefont {Zhao},
  \citenamefont {Li}, \citenamefont {Wang}, \citenamefont {Xie}, \citenamefont
  {Cheng}, \citenamefont {Li}, \citenamefont {Lin}, \citenamefont {Xi},
  \citenamefont {Ke} \emph {et~al.}}]{zhang2020magnetotransport}%
  \BibitemOpen
  \bibfield  {author} {\bibinfo {author} {\bibfnamefont {N.}~\bibnamefont
  {Zhang}}, \bibinfo {author} {\bibfnamefont {G.}~\bibnamefont {Zhao}},
  \bibinfo {author} {\bibfnamefont {L.}~\bibnamefont {Li}}, \bibinfo {author}
  {\bibfnamefont {P.}~\bibnamefont {Wang}}, \bibinfo {author} {\bibfnamefont
  {L.}~\bibnamefont {Xie}}, \bibinfo {author} {\bibfnamefont {B.}~\bibnamefont
  {Cheng}}, \bibinfo {author} {\bibfnamefont {H.}~\bibnamefont {Li}}, \bibinfo
  {author} {\bibfnamefont {Z.}~\bibnamefont {Lin}}, \bibinfo {author}
  {\bibfnamefont {C.}~\bibnamefont {Xi}}, \bibinfo {author} {\bibfnamefont
  {J.}~\bibnamefont {Ke}}, \emph {et~al.},\ }\bibfield  {title} {\bibinfo
  {title} {{Magnetotransport signatures of Weyl physics and discrete scale
  invariance in the elemental semiconductor tellurium}},\ }\href@noop {}
  {\bibfield  {journal} {\bibinfo  {journal} {Proc. Natl. Acad. Sci. USA}\
  }\textbf {\bibinfo {volume} {117}},\ \bibinfo {pages} {11337} (\bibinfo
  {year} {2020}{\natexlab{b}})}\BibitemShut {NoStop}%
\bibitem [{\citenamefont {Li}\ \emph {et~al.}(2016{\natexlab{b}})\citenamefont
  {Li}, \citenamefont {Li}, \citenamefont {Kim}, \citenamefont {Balents},
  \citenamefont {Yu},\ and\ \citenamefont {Chen}}]{li2016weylmagnons}%
  \BibitemOpen
  \bibfield  {author} {\bibinfo {author} {\bibfnamefont {F.-Y.}\ \bibnamefont
  {Li}}, \bibinfo {author} {\bibfnamefont {Y.-D.}\ \bibnamefont {Li}}, \bibinfo
  {author} {\bibfnamefont {Y.~B.}\ \bibnamefont {Kim}}, \bibinfo {author}
  {\bibfnamefont {L.}~\bibnamefont {Balents}}, \bibinfo {author} {\bibfnamefont
  {Y.}~\bibnamefont {Yu}},\ and\ \bibinfo {author} {\bibfnamefont
  {G.}~\bibnamefont {Chen}},\ }\bibfield  {title} {\bibinfo {title} {Weyl
  magnons in breathing pyrochlore antiferromagnets},\ }\href@noop {} {\bibfield
   {journal} {\bibinfo  {journal} {Nat. Commun.}\ }\textbf {\bibinfo {volume}
  {7}},\ \bibinfo {pages} {12691} (\bibinfo {year}
  {2016}{\natexlab{b}})}\BibitemShut {NoStop}%
\bibitem [{\citenamefont {Mook}\ \emph {et~al.}(2016)\citenamefont {Mook},
  \citenamefont {Henk},\ and\ \citenamefont {Mertig}}]{PhysRevLett.117.157204}%
  \BibitemOpen
  \bibfield  {author} {\bibinfo {author} {\bibfnamefont {A.}~\bibnamefont
  {Mook}}, \bibinfo {author} {\bibfnamefont {J.}~\bibnamefont {Henk}},\ and\
  \bibinfo {author} {\bibfnamefont {I.}~\bibnamefont {Mertig}},\ }\bibfield
  {title} {\bibinfo {title} {{Tunable magnon Weyl points in ferromagnetic
  pyrochlores}},\ }\href@noop {} {\bibfield  {journal} {\bibinfo  {journal}
  {Phys. Rev. Lett.}\ }\textbf {\bibinfo {volume} {117}},\ \bibinfo {pages}
  {157204} (\bibinfo {year} {2016})}\BibitemShut {NoStop}%
\bibitem [{\citenamefont {Kresse}\ and\ \citenamefont
  {Furthm\"uller}(1996)}]{s1}%
  \BibitemOpen
  \bibfield  {author} {\bibinfo {author} {\bibfnamefont {G.}~\bibnamefont
  {Kresse}}\ and\ \bibinfo {author} {\bibfnamefont {J.}~\bibnamefont
  {Furthm\"uller}},\ }\bibfield  {title} {\bibinfo {title} {{Efficient
  iterative schemes for $ab$ $initio$ total-energy calculations using a
  plane-wave basis set}},\ }\href@noop {} {\bibfield  {journal} {\bibinfo
  {journal} {Phys. Rev. B}\ }\textbf {\bibinfo {volume} {54}},\ \bibinfo
  {pages} {11169} (\bibinfo {year} {1996})}\BibitemShut {NoStop}%
\bibitem [{\citenamefont {Perdew}\ \emph {et~al.}(1996)\citenamefont {Perdew},
  \citenamefont {Burke},\ and\ \citenamefont {Ernzerhof}}]{s2}%
  \BibitemOpen
  \bibfield  {author} {\bibinfo {author} {\bibfnamefont {J.~P.}\ \bibnamefont
  {Perdew}}, \bibinfo {author} {\bibfnamefont {K.}~\bibnamefont {Burke}},\ and\
  \bibinfo {author} {\bibfnamefont {M.}~\bibnamefont {Ernzerhof}},\ }\bibfield
  {title} {\bibinfo {title} {{Generalized Gradient Approximation Made
  Simple}},\ }\href@noop {} {\bibfield  {journal} {\bibinfo  {journal} {Phys.
  Rev. Lett.}\ }\textbf {\bibinfo {volume} {77}},\ \bibinfo {pages} {3865}
  (\bibinfo {year} {1996})}\BibitemShut {NoStop}%
\bibitem [{\citenamefont {Henkelman}\ \emph {et~al.}(2000)\citenamefont
  {Henkelman}, \citenamefont {Uberuaga},\ and\ \citenamefont
  {J{\'o}nsson}}]{s3}%
  \BibitemOpen
  \bibfield  {author} {\bibinfo {author} {\bibfnamefont {G.}~\bibnamefont
  {Henkelman}}, \bibinfo {author} {\bibfnamefont {B.~P.}\ \bibnamefont
  {Uberuaga}},\ and\ \bibinfo {author} {\bibfnamefont {H.}~\bibnamefont
  {J{\'o}nsson}},\ }\bibfield  {title} {\bibinfo {title} {{A climbing image
  nudged elastic band method for finding saddle points and minimum energy
  paths}},\ }\href@noop {} {\bibfield  {journal} {\bibinfo  {journal} {J. Chem.
  Phys.}\ }\textbf {\bibinfo {volume} {113}},\ \bibinfo {pages} {9901}
  (\bibinfo {year} {2000})}\BibitemShut {NoStop}%
\bibitem [{\citenamefont {Togo}\ and\ \citenamefont {Tanaka}(2015)}]{s4}%
  \BibitemOpen
  \bibfield  {author} {\bibinfo {author} {\bibfnamefont {A.}~\bibnamefont
  {Togo}}\ and\ \bibinfo {author} {\bibfnamefont {I.}~\bibnamefont {Tanaka}},\
  }\bibfield  {title} {\bibinfo {title} {{First principles phonon calculations
  in materials science}},\ }\href@noop {} {\bibfield  {journal} {\bibinfo
  {journal} {Scr. Mater.}\ }\textbf {\bibinfo {volume} {108}},\ \bibinfo
  {pages} {1} (\bibinfo {year} {2015})}\BibitemShut {NoStop}%
\bibitem [{\citenamefont {Wu}\ \emph {et~al.}(2018)\citenamefont {Wu},
  \citenamefont {Zhang}, \citenamefont {Song}, \citenamefont {Troyer},\ and\
  \citenamefont {Soluyanov}}]{s5}%
  \BibitemOpen
  \bibfield  {author} {\bibinfo {author} {\bibfnamefont {Q.}~\bibnamefont
  {Wu}}, \bibinfo {author} {\bibfnamefont {S.}~\bibnamefont {Zhang}}, \bibinfo
  {author} {\bibfnamefont {H.-F.}\ \bibnamefont {Song}}, \bibinfo {author}
  {\bibfnamefont {M.}~\bibnamefont {Troyer}},\ and\ \bibinfo {author}
  {\bibfnamefont {A.~A.}\ \bibnamefont {Soluyanov}},\ }\bibfield  {title}
  {\bibinfo {title} {{WannierTools: An open-source software package for novel
  topological materials}},\ }\href@noop {} {\bibfield  {journal} {\bibinfo
  {journal} {Comput. Phys. Commun.}\ }\textbf {\bibinfo {volume} {224}},\
  \bibinfo {pages} {405} (\bibinfo {year} {2018})}\BibitemShut {NoStop}%
\bibitem [{\citenamefont {Beall}\ \emph {et~al.}(1977)\citenamefont {Beall},
  \citenamefont {Milligan},\ and\ \citenamefont {Wolcott}}]{s6}%
  \BibitemOpen
  \bibfield  {author} {\bibinfo {author} {\bibfnamefont {G.}~\bibnamefont
  {Beall}}, \bibinfo {author} {\bibfnamefont {W.}~\bibnamefont {Milligan}},\
  and\ \bibinfo {author} {\bibfnamefont {H.~A.}\ \bibnamefont {Wolcott}},\
  }\bibfield  {title} {\bibinfo {title} {{Structural trends in the lanthanide
  trihydroxides}},\ }\href@noop {} {\bibfield  {journal} {\bibinfo  {journal}
  {J. Inorg. Nucl. Chem.}\ }\textbf {\bibinfo {volume} {39}},\ \bibinfo {pages}
  {65} (\bibinfo {year} {1977})}\BibitemShut {NoStop}%
\bibitem [{\citenamefont {Coquet}\ \emph {et~al.}(1983)\citenamefont {Coquet},
  \citenamefont {Crettez}, \citenamefont {Pannetier}, \citenamefont
  {Bouillot},\ and\ \citenamefont {Damien}}]{s7}%
  \BibitemOpen
  \bibfield  {author} {\bibinfo {author} {\bibfnamefont {E.}~\bibnamefont
  {Coquet}}, \bibinfo {author} {\bibfnamefont {J.}~\bibnamefont {Crettez}},
  \bibinfo {author} {\bibfnamefont {J.}~\bibnamefont {Pannetier}}, \bibinfo
  {author} {\bibfnamefont {J.}~\bibnamefont {Bouillot}},\ and\ \bibinfo
  {author} {\bibfnamefont {J.}~\bibnamefont {Damien}},\ }\bibfield  {title}
  {\bibinfo {title} {{Effect of temperature on interatomic distances in
  pyroelectric $\alpha$-LiIO3}},\ }\href@noop {} {\bibfield  {journal}
  {\bibinfo  {journal} {Acta Crystallogr., Sect. B}\ }\textbf {\bibinfo
  {volume} {39}},\ \bibinfo {pages} {408} (\bibinfo {year} {1983})}\BibitemShut
  {NoStop}%
\bibitem [{\citenamefont {Carreras}\ \emph {et~al.}(2017)\citenamefont
  {Carreras}, \citenamefont {Togo},\ and\ \citenamefont {Tanaka}}]{s8}%
  \BibitemOpen
  \bibfield  {author} {\bibinfo {author} {\bibfnamefont {A.}~\bibnamefont
  {Carreras}}, \bibinfo {author} {\bibfnamefont {A.}~\bibnamefont {Togo}},\
  and\ \bibinfo {author} {\bibfnamefont {I.}~\bibnamefont {Tanaka}},\
  }\bibfield  {title} {\bibinfo {title} {Dynaphopy: A code for extracting
  phonon quasiparticles from molecular dynamics simulations},\ }\href@noop {}
  {\bibfield  {journal} {\bibinfo  {journal} {Comput. Phys. Commun.}\ }\textbf
  {\bibinfo {volume} {221}},\ \bibinfo {pages} {221} (\bibinfo {year}
  {2017})}\BibitemShut {NoStop}%
\bibitem [{\citenamefont {Abrahams}(1994)}]{s9}%
  \BibitemOpen
  \bibfield  {author} {\bibinfo {author} {\bibfnamefont {S.~C.}\ \bibnamefont
  {Abrahams}},\ }\bibfield  {title} {\bibinfo {title} {{Experimental (155 K)
  and predicted (151 K) Curie temperature ($T$$_{C}$) of K$_{2}$ZnBr$_{4}$:
  structural confirmation of ferroelectric state below $T$$_{C}$}},\
  }\href@noop {} {\bibfield  {journal} {\bibinfo  {journal} {Acta Crystallogr.,
  Sect. B}\ }\textbf {\bibinfo {volume} {50}},\ \bibinfo {pages} {257}
  (\bibinfo {year} {1994})}\BibitemShut {NoStop}%
\end{thebibliography}%


\begin{thebibliography}{10}%
\makeatletter
\providecommand \@ifxundefined [1]{%
 \@ifx{#1\undefined}
}%
\providecommand \@ifnum [1]{%
 \ifnum #1\expandafter \@firstoftwo
 \else \expandafter \@secondoftwo
 \fi
}%
\providecommand \@ifx [1]{%
 \ifx #1\expandafter \@firstoftwo
 \else \expandafter \@secondoftwo
 \fi
}%
\providecommand \natexlab [1]{#1}%
\providecommand \enquote  [1]{``#1''}%
\providecommand \bibnamefont  [1]{#1}%
\providecommand \bibfnamefont [1]{#1}%
\providecommand \citenamefont [1]{#1}%
\providecommand \href@noop [0]{\@secondoftwo}%
\providecommand \href [0]{\begingroup \@sanitize@url \@href}%
\providecommand \@href[1]{\@@startlink{#1}\@@href}%
\providecommand \@@href[1]{\endgroup#1\@@endlink}%
\providecommand \@sanitize@url [0]{\catcode `\\12\catcode `\$12\catcode
  `\&12\catcode `\#12\catcode `\^12\catcode `\_12\catcode `\%12\relax}%
\providecommand \@@startlink[1]{}%
\providecommand \@@endlink[0]{}%
\providecommand \url  [0]{\begingroup\@sanitize@url \@url }%
\providecommand \@url [1]{\endgroup\@href {#1}{\urlprefix }}%
\providecommand \urlprefix  [0]{URL }%
\providecommand \Eprint [0]{\href }%
\providecommand \doibase [0]{https://doi.org/}%
\providecommand \selectlanguage [0]{\@gobble}%
\providecommand \bibinfo  [0]{\@secondoftwo}%
\providecommand \bibfield  [0]{\@secondoftwo}%
\providecommand \translation [1]{[#1]}%
\providecommand \BibitemOpen [0]{}%
\providecommand \bibitemStop [0]{}%
\providecommand \bibitemNoStop [0]{.\EOS\space}%
\providecommand \EOS [0]{\spacefactor3000\relax}%
\providecommand \BibitemShut  [1]{\csname bibitem#1\endcsname}%
\let\auto@bib@innerbib\@empty
%</preamble>
\bibitem [{\citenamefont {Kresse}\ and\ \citenamefont
  {Furthm\"uller}(1996)}]{1}%
  \BibitemOpen
  \bibfield  {author} {\bibinfo {author} {\bibfnamefont {G.}~\bibnamefont
  {Kresse}}\ and\ \bibinfo {author} {\bibfnamefont {J.}~\bibnamefont
  {Furthm\"uller}},\ }\bibfield  {title} {\bibinfo {title} {{Efficient
  iterative schemes for ab initio total-energy calculations using a plane-wave
  basis set}},\ }\href@noop {} {\bibfield  {journal} {\bibinfo  {journal}
  {Phys. Rev. B}\ }\textbf {\bibinfo {volume} {54}},\ \bibinfo {pages} {11169}
  (\bibinfo {year} {1996})}\BibitemShut {NoStop}%
\bibitem [{\citenamefont {Perdew}\ \emph {et~al.}(1996)\citenamefont {Perdew},
  \citenamefont {Burke},\ and\ \citenamefont {Ernzerhof}}]{2}%
  \BibitemOpen
  \bibfield  {author} {\bibinfo {author} {\bibfnamefont {J.~P.}\ \bibnamefont
  {Perdew}}, \bibinfo {author} {\bibfnamefont {K.}~\bibnamefont {Burke}},\ and\
  \bibinfo {author} {\bibfnamefont {M.}~\bibnamefont {Ernzerhof}},\ }\bibfield
  {title} {\bibinfo {title} {{Generalized gradient approximation made
  simple}},\ }\href@noop {} {\bibfield  {journal} {\bibinfo  {journal} {Phys.
  Rev. Lett.}\ }\textbf {\bibinfo {volume} {77}},\ \bibinfo {pages} {3865}
  (\bibinfo {year} {1996})}\BibitemShut {NoStop}%
\bibitem [{\citenamefont {Henkelman}\ \emph {et~al.}(2000)\citenamefont
  {Henkelman}, \citenamefont {Uberuaga},\ and\ \citenamefont
  {J{\'o}nsson}}]{3}%
  \BibitemOpen
  \bibfield  {author} {\bibinfo {author} {\bibfnamefont {G.}~\bibnamefont
  {Henkelman}}, \bibinfo {author} {\bibfnamefont {B.~P.}\ \bibnamefont
  {Uberuaga}},\ and\ \bibinfo {author} {\bibfnamefont {H.}~\bibnamefont
  {J{\'o}nsson}},\ }\bibfield  {title} {\bibinfo {title} {{A climbing image
  nudged elastic band method for finding saddle points and minimum energy
  paths}},\ }\href@noop {} {\bibfield  {journal} {\bibinfo  {journal} {J. Chem.
  Phys.}\ }\textbf {\bibinfo {volume} {113}},\ \bibinfo {pages} {9901}
  (\bibinfo {year} {2000})}\BibitemShut {NoStop}%
\bibitem [{\citenamefont {Togo}\ and\ \citenamefont {Tanaka}(2015)}]{4}%
  \BibitemOpen
  \bibfield  {author} {\bibinfo {author} {\bibfnamefont {A.}~\bibnamefont
  {Togo}}\ and\ \bibinfo {author} {\bibfnamefont {I.}~\bibnamefont {Tanaka}},\
  }\bibfield  {title} {\bibinfo {title} {{First principles phonon calculations
  in materials science}},\ }\href@noop {} {\bibfield  {journal} {\bibinfo
  {journal} {Scr. Mater.}\ }\textbf {\bibinfo {volume} {108}},\ \bibinfo
  {pages} {1} (\bibinfo {year} {2015})}\BibitemShut {NoStop}%
\bibitem [{\citenamefont {Wu}\ \emph {et~al.}(2018)\citenamefont {Wu},
  \citenamefont {Zhang}, \citenamefont {Song}, \citenamefont {Troyer},\ and\
  \citenamefont {Soluyanov}}]{5}%
  \BibitemOpen
  \bibfield  {author} {\bibinfo {author} {\bibfnamefont {Q.}~\bibnamefont
  {Wu}}, \bibinfo {author} {\bibfnamefont {S.}~\bibnamefont {Zhang}}, \bibinfo
  {author} {\bibfnamefont {H.-F.}\ \bibnamefont {Song}}, \bibinfo {author}
  {\bibfnamefont {M.}~\bibnamefont {Troyer}},\ and\ \bibinfo {author}
  {\bibfnamefont {A.~A.}\ \bibnamefont {Soluyanov}},\ }\bibfield  {title}
  {\bibinfo {title} {{WannierTools: An open-source software package for novel
  topological materials}},\ }\href@noop {} {\bibfield  {journal} {\bibinfo
  {journal} {Comput. Phys. Commun.}\ }\textbf {\bibinfo {volume} {224}},\
  \bibinfo {pages} {405} (\bibinfo {year} {2018})}\BibitemShut {NoStop}%
\bibitem [{\citenamefont {Beall}\ \emph {et~al.}(1977)\citenamefont {Beall},
  \citenamefont {Milligan},\ and\ \citenamefont
  {Wolcott}}]{beall1977structural}%
  \BibitemOpen
  \bibfield  {author} {\bibinfo {author} {\bibfnamefont {G.}~\bibnamefont
  {Beall}}, \bibinfo {author} {\bibfnamefont {W.}~\bibnamefont {Milligan}},\
  and\ \bibinfo {author} {\bibfnamefont {H.~A.}\ \bibnamefont {Wolcott}},\
  }\bibfield  {title} {\bibinfo {title} {{Structural trends in the lanthanide
  trihydroxides}},\ }\href@noop {} {\bibfield  {journal} {\bibinfo  {journal}
  {J. Inorg. Nucl. Chem.}\ }\textbf {\bibinfo {volume} {39}},\ \bibinfo {pages}
  {65} (\bibinfo {year} {1977})}\BibitemShut {NoStop}%
\bibitem [{\citenamefont {Coquet}\ \emph {et~al.}(1983)\citenamefont {Coquet},
  \citenamefont {Crettez}, \citenamefont {Pannetier}, \citenamefont
  {Bouillot},\ and\ \citenamefont {Damien}}]{coquet1983effect}%
  \BibitemOpen
  \bibfield  {author} {\bibinfo {author} {\bibfnamefont {E.}~\bibnamefont
  {Coquet}}, \bibinfo {author} {\bibfnamefont {J.}~\bibnamefont {Crettez}},
  \bibinfo {author} {\bibfnamefont {J.}~\bibnamefont {Pannetier}}, \bibinfo
  {author} {\bibfnamefont {J.}~\bibnamefont {Bouillot}},\ and\ \bibinfo
  {author} {\bibfnamefont {J.}~\bibnamefont {Damien}},\ }\bibfield  {title}
  {\bibinfo {title} {{Effect of temperature on interatomic distances in
  pyroelectric $\alpha$-LiIO3}},\ }\href@noop {} {\bibfield  {journal}
  {\bibinfo  {journal} {Acta Crystallogr. B}\ }\textbf {\bibinfo {volume}
  {39}},\ \bibinfo {pages} {408} (\bibinfo {year} {1983})}\BibitemShut
  {NoStop}%
\bibitem [{\citenamefont {Luo}\ \emph {et~al.}(2023)\citenamefont {Luo},
  \citenamefont {Ji}, \citenamefont {Chen}, \citenamefont {Xu}, \citenamefont
  {Zhang}, \citenamefont {Xiang},\ and\ \citenamefont
  {Bellaiche}}]{PhysRevB.107.L241107}%
  \BibitemOpen
  \bibfield  {author} {\bibinfo {author} {\bibfnamefont {W.}~\bibnamefont
  {Luo}}, \bibinfo {author} {\bibfnamefont {J.}~\bibnamefont {Ji}}, \bibinfo
  {author} {\bibfnamefont {P.}~\bibnamefont {Chen}}, \bibinfo {author}
  {\bibfnamefont {Y.}~\bibnamefont {Xu}}, \bibinfo {author} {\bibfnamefont
  {L.}~\bibnamefont {Zhang}}, \bibinfo {author} {\bibfnamefont
  {H.}~\bibnamefont {Xiang}},\ and\ \bibinfo {author} {\bibfnamefont
  {L.}~\bibnamefont {Bellaiche}},\ }\bibfield  {title} {\bibinfo {title}
  {{Nonlinear phonon Hall effects in ferroelectrics: Existence and nonvolatile
  electrical control}},\ }\href@noop {} {\bibfield  {journal} {\bibinfo
  {journal} {Phys. Rev. B}\ }\textbf {\bibinfo {volume} {107}},\ \bibinfo
  {pages} {L241107} (\bibinfo {year} {2023})}\BibitemShut {NoStop}%
\bibitem [{\citenamefont {Carreras}\ \emph {et~al.}(2017)\citenamefont
  {Carreras}, \citenamefont {Togo},\ and\ \citenamefont
  {Tanaka}}]{carreras2017dynaphopy}%
  \BibitemOpen
  \bibfield  {author} {\bibinfo {author} {\bibfnamefont {A.}~\bibnamefont
  {Carreras}}, \bibinfo {author} {\bibfnamefont {A.}~\bibnamefont {Togo}},\
  and\ \bibinfo {author} {\bibfnamefont {I.}~\bibnamefont {Tanaka}},\
  }\bibfield  {title} {\bibinfo {title} {Dynaphopy: A code for extracting
  phonon quasiparticles from molecular dynamics simulations},\ }\href@noop {}
  {\bibfield  {journal} {\bibinfo  {journal} {Comput. Phys. Commun.}\ }\textbf
  {\bibinfo {volume} {221}},\ \bibinfo {pages} {221} (\bibinfo {year}
  {2017})}\BibitemShut {NoStop}%
\bibitem [{\citenamefont {Abrahams}(1994)}]{abrahams1994experimental}%
  \BibitemOpen
  \bibfield  {author} {\bibinfo {author} {\bibfnamefont {S.~C.}\ \bibnamefont
  {Abrahams}},\ }\bibfield  {title} {\bibinfo {title} {{Experimental (155 K)
  and predicted (151 K) Curie temperature ($T$c) of K2ZnBr4: structural
  confirmation of ferroelectric state below $T$c}},\ }\href@noop {} {\bibfield
  {journal} {\bibinfo  {journal} {Acta Crystallogr. B}\ }\textbf {\bibinfo
  {volume} {50}},\ \bibinfo {pages} {257} (\bibinfo {year} {1994})}\BibitemShut
  {NoStop}%
\end{thebibliography}%
\end{document}

\title{Supplementary materials
	 for $``$Ferroelectrically Controlled Chirality Switching of Weyl quasiparticles$"$}
\author{Zeling Li}
%\thanks{These authors contributed equally to this work.}
\address{Research Center for Quantum Physics and Technologies, School of Physical Science and Technology, Inner Mongolia University, Hohhot 010021, China}
\address{Key Laboratory of Semiconductor Photovoltaic Technology and Energy Materials at Universities of Inner Mongolia Autonomous Region, Inner Mongolia University, Hohhot 010021, China}

\author{Yu Liu}
%\thanks{These authors contributed equally to this work.}
\address{Research Center for Quantum Physics and Technologies, School of Physical Science and Technology, Inner Mongolia University, Hohhot 010021, China}
\address{Key Laboratory of Semiconductor Photovoltaic Technology and Energy Materials at Universities of Inner Mongolia Autonomous Region, Inner Mongolia University, Hohhot 010021, China}

\author{Le Du}
\address{Research Center for Quantum Physics and Technologies, School of Physical Science and Technology, Inner Mongolia University, Hohhot 010021, China}
\address{Key Laboratory of Semiconductor Photovoltaic Technology and Energy Materials at Universities of Inner Mongolia Autonomous Region, Inner Mongolia University, Hohhot 010021, China}

\author{Fengyu Li}
\address{Research Center for Quantum Physics and Technologies, School of Physical Science and Technology, Inner Mongolia University, Hohhot 010021, China}
\address{Key Laboratory of Semiconductor Photovoltaic Technology and Energy Materials at Universities of Inner Mongolia Autonomous Region, Inner Mongolia University, Hohhot 010021, China}

\author{Zhifeng Liu}
\address{Research Center for Quantum Physics and Technologies, School of Physical Science and Technology, Inner Mongolia University, Hohhot 010021, China}
\address{Inner Mongolia Key Laboratory of Microscale Physics and Atom Innovation, Inner Mongolia University, Hohhot 010021, China}

\author{Lei Li}
\address{Research Center for Quantum Physics and Technologies, School of Physical Science and Technology, Inner Mongolia University, Hohhot 010021, China}

\author{Lei Wang}
\email{lwang@imu.edu.cn}
\address{Research Center for Quantum Physics and Technologies, School of Physical Science and Technology, Inner Mongolia University, Hohhot 010021, China}
\address{Inner Mongolia Key Laboratory of Microscale Physics and Atom Innovation, Inner Mongolia University, Hohhot 010021, China}

\author{Botao Fu}
\email{fubotao2008@gmail.com}
\address{College of Physics and Electronic Engineering, Center for Computational Sciences, Sichuan Normal University, Chengdu 610068, China}

\author{Xiao-Ping Li}
\email{xpli@imu.edu.cn}
\address{Research Center for Quantum Physics and Technologies, School of Physical Science and Technology, Inner Mongolia University, Hohhot 010021, China}
\address{Key Laboratory of Semiconductor Photovoltaic Technology and Energy Materials at Universities of Inner Mongolia Autonomous Region, Inner Mongolia University, Hohhot 010021, China}

\maketitle

\section{Candidate space groups and $k\cdot p$ models}
In this section, we give a more detailed and comprehensive explanation of the candidate space groups (SGs) presented in Table I of the main text. The supplementary information includes the point groups (PG), Brillouin zones (BZ), little groups (LG) and their irreducible representations (IRRs) of high-symmetry points or paths related to Weyl points. These results are summarized in Table~\ref{tabs1}. Based on these information, we construct $k\cdot p$ models to characterize these Weyl points. Here, we take two $k\cdot p$ Hamiltonians in Table~\ref{tabs1} as examples to describe the working procedure. 

For Weyl points on high-symmetry paths, we consider SG 4 as an illustration,  whose candidate material is K$_{2}$ZnBr$_{4}$ as discussed in the main text. For the $Y$-$C$ path, there are two distinct one-dimensional IRRs, $R_{1}$ and $R_{2}$, which cross each other to form $R_{1}\oplus R_{2}$, thereby protecting the existence of Weyl points. Alternatively, we can analyze the band crossing from the perspective of symmetry operations. The $Y$-$C$ path is invariant under a twofold screw-rotation $\widetilde{C_{2y}}$, and thus its bands can be labeled by the eigenvalues of $\widetilde{C_{2y}}$. When band inversion occurs, the crossing of bands with distinct eigenvalues gives rise to a Weyl point, which is protected by the $\widetilde{C_{2y}}$ symmetry. Moreover, the symmetry operators of the little group at a generic point on $Y$-$C$ can be generated by $\widetilde{C_{2y}}=\left\{ C_{2y}|0\frac{1}{2}0\right\}$, and it satisfies $\widetilde{C_{2y}}^{2}=e^{-i2\pi u}$ $\left[u\in\left(0,\frac{1}{2}\right)\right]$.  The Bloch states at the generic point can be chosen as the eigenstates of $\widetilde{C_{2y}}$, denoted as $\left|c_{2y}\right\rangle$ with $c_{2y}=\pm e{}^{-i\pi u}$ the eigenvalue of $\widetilde{C_{2y}}$.  

Under the basis of $\left\{ \left|+e^{-i\pi u}\right\rangle ,\left|-e^{-i\pi u}\right\rangle \right\}$, the matrix representations of generators can be expressed as
\begin{eqnarray}
\widetilde{C_{2y}}	=	e^{-i\pi u}\sigma_{3}. 
\end{eqnarray}
With the standard approach, the effective Hamiltonian of the Weyl point around the $Y$-$C$ path, to the leading order, takes the following form:
\begin{eqnarray}\label{h1}
H_{1}	=	\sigma_{0}(c_{1}+c_{2}k_{y})+c_{3}\sigma_{3}k_{y}+\sum_{j=1}^{2}\sigma_{j}(c_{j,1}k_{x}+c_{j,2}k_{z}),
\end{eqnarray}
with $\sigma_{i}(i=1,2,3)$ the Pauli matrix and $\sigma_{0}$ the identity
matrix. Where $c_{i=1,2,3}$, $c_{j,1}$ and $c_{j,2}$ ($j$ = 1, 2) are real parameters. The Hamiltonian $H_{1}$ exhibits linear band splitting along all directions, which describe a C-1 WP (Note that symmetry can constrain the magnitude of the chirality but not its sign at Weyl point).

For Weyl points at high-symmetry points, we take the $\Gamma$ point in SG 75 as an example. The little group of the $\Gamma$ point hosts a two-dimensional IRR $R_{2}R_{4}$, which enforces the emergence of the Weyl point. The generators of the co-little group of the $\Gamma$ point are $C_{4z}$ and time-reversal symmetry $\mathcal{T}$.  The generators satisfy the following algebra
\begin{eqnarray}
C^{4}_{4z}	=	1,\,\,\,\left[C_{4z},\mathcal{T}\right]=0.
\end{eqnarray}
The Bloch states at $\Gamma$ point can be chosen as the eigenstates of $C_{4z}$, denoted as $\left|c_{4z}\right\rangle$ with $c_{4z}=$ $1$, $i$, $-1$, and $-i$. One can find that
\begin{eqnarray}
C_{4z}\mathcal{T}\left|\pm i\right\rangle 	=	\mathcal{T}C_{4z}\left|\pm i\right\rangle =\mp i\mathcal{T}\left|\pm i\right\rangle ,
\end{eqnarray}
which indicates that the two states $\left|+i\right\rangle$ and $\left|-i\right\rangle$ would be degenerate at the same energy.
  
Under the duet basis of $\left\{ \left|+i\right\rangle ,\left|-i\right\rangle \right\}$, the symmetry generators are expressed as follows:
\begin{eqnarray}
C_{4z}	=	i\sigma_{3},    \mathcal{T}	=	\sigma_{1}\mathcal{K}
\end{eqnarray}
with $\mathcal{K}$ as the complex conjugate operator. The effective Hamiltonian up to the leading order is
\begin{eqnarray}\label{h2}
H_{6}	=	\left(c_{1}+c_{2}k^{2}+c_{3}k_{z}^{2}\right)\sigma_{0}+\sum_{i=1}^{2}\left[c_{i,1}k_{x}k_{y}+c_{i,2}\left(k_{x}^{2}-k_{y}^{2}\right)\right]\sigma_{i}+c_{4}\sigma_{3}k_{z},
\end{eqnarray}
with $k^{2}=k_{x}^{2}+k_{y}^{2}+k_{z}^{2}$. Here, $c_{i=1,2,3,4}$, $c_{i,1}$ and $c_{i,2}$ ($i$ = 1, 2) are real parameters. One can observe that Hamiltonian $H_{6}$ exhibits linear band dispersion along the $k_{z}$ direction and quadratic dispersion for $k_{x}$-$k_{y}$ plane, which indicates the presence of a C-2 WP at the $\Gamma$ point. This procedure is applied to all candidate SGs, and the result is presented in the sixth column of Table~\ref{tabs1}. The explicit expressions of these $k\cdot p$ Hamiltonians are presented in Table~\ref{tabs2}. Based on these Hamiltonians, the species of Weyl points can be readily identified.

\begin{table}[t]
\caption{Candidate space groups (SGs) for symmetry-protected Weyl points in polar groups of spinless systems. PG denotes the point group of SGs. LG denotes the little group at high symmetry points/lines. BZ denotes the Brillouin zone of candidate SGs. IRR denotes the irreducible representation of the LG related to Weyl points, with the low-energy effective models given in the “$k\cdot p$ Hamiltonian” column and their expressions listed in Table \ref{tabs2}. Species stands the type of Weyl points, the charge-one (charge-two, charge-three) Weyl points are denoted as C-1 WP (C-2 WP, C-3 WP).}
		\renewcommand{\arraystretch}{1.5}
		\resizebox{\textwidth}{!}{%
			\begin{tabular}{lllllll}
				\hline
				\hline
				PG & SG (position) & Generators of LG & BZ & IRR & $k \mathbin{\cdot} p$ Hamiltonian & Species \\ \hline
				\multirow[t]{2}{*}{$\mathrm{C}_{2}$} & 3-4 ($\Gamma$-$Z$, $B$-$D$, $Y$-$C$, $A$-$E$) & $\mathrm{C}_{2z}$ & $\mathrm{\Gamma}_m$ & $R_1$$\oplus$$R_2$ & $H_1$ & C-1 WP \\ 
				& 5 ($\Gamma$-$Z$, $A$-$M$) & $\mathrm{C}_{2z}$ & $\mathrm{\Gamma}^{b}_{m}$ & $R_1$$\oplus$$R_2$ & $H_1$  & C-1 WP \\ \hline
				\multirow[t]{2}{*}{$\mathrm{C}_{2v}$} & 35-37 ($S$-$R$) & $\mathrm{C}_{2z}$ & $\mathrm{\Gamma}^{b}_{o}$ & $R_1$$\oplus$$R_2$ & $H_1$ & C-1 WP \\ 
				& 44-46 ($T$-$W$) & $\mathrm{C}_{2z}$ & $\mathrm{\Gamma}^{o}_{v}$ & $R_1$$\oplus$$R_2$ & $H_1$ & C-1 WP \\ \hline
				\multirow[t]{10}{*}{$\mathrm{C}_{4}$} & 75-78 ($\Gamma$-$Z$, $M$-$A$) & $\mathrm{C}^{+}_{4z}$ & $\mathrm{\Gamma}_{q}$ & $R_1$$\oplus$$R_2$, $R_2$$\oplus$$R_3$, $R_3$$\oplus$$R_4$  & $H_2$ & C-1 WP \\ & & & & $R_1$$\oplus$$R_4$ & $H_3$ & C-1 WP \\
				& 75-78 ($X$-$R$) & $\mathrm{C}_{2z}$ & $\mathrm{\Gamma}_{q}$ & $R_1$$\oplus$$R_2$ & $H_1$ & C-1 WP \\ 
				& 79-80 ($\Gamma$-$Z$, $Z$-$V$) & $\mathrm{C}^{+}_{4z}$ & $\mathrm{\Gamma}^{v}_{q}$ & $R_1$$\oplus$$R_2$, $R_2$$\oplus$$R_3$, $R_3$$\oplus$$R_4$ & $H_2$ & C-1 WP \\ 
				& & & & $R_1$$\oplus$$R_4$ & $H_3$ & C-1 WP \\
				& 79-80 ($X$-$P$) & $\mathrm{C}_{2z}$ &$\mathrm{\Gamma}^{v}_{q}$ & $R_1$$\oplus$$R_2$  & $H_1$ & C-1 WP \\ 
				& 80 ($P$) & $\mathrm{C}_{2z}$, $\mathrm{C}^{+}_{4z}$$\mathcal{T}$ & $\mathrm{\Gamma}^{v}_{q}$ & $R_1$$R_2$ & $H_4$ & C-1 WP \\ 
				& 75-78 ($\Gamma$-$Z$, $M$-$A$) & $\mathrm{C}^{+}_{4z}$ & $\mathrm{\Gamma}_{q}$ & $R_1$$\oplus$$R_3$, $R_2$$\oplus$$R_4$ & $H_5$ & C-2 WP \\ 
				& 79-80 ($\Gamma$-$Z$, $Z$-$V$) & $\mathrm{C}^{+}_{4z}$ & $\mathrm{\Gamma}^{v}_{q}$ & $R_1$$\oplus$$R_3$, $R_2$$\oplus$$R_4$  & $H_5$ & C-2 WP \\ 
				& 75, 77 ($\Gamma$, $M$, $Z$, $A$); 76, 78 ($\Gamma$, $M$) & $\mathrm{C}^{+}_{4z}$, $\mathcal{T}$ & $\mathrm{\Gamma}_{q}$ & $R_2$$R_4$ & $H_6$ & C-2 WP \\ 
				& 79 ($\Gamma$, $Z$), 80 ($\Gamma$, $Z$) & $\mathrm{C}^{+}_{4z}$, $\mathcal{T}$ & $\mathrm{\Gamma}^{v}_{q}$ & $R_2$$R_4$ & $H_6$ & C-2 WP \\ 
				& 79 ($P$) & $\mathrm{C}_{2z}$, $\mathrm{C}^{+}_{4z}$$\mathcal{T}$ & $\mathrm{\Gamma}^{v}_{q}$ & $R_2$$R_2$ & $H_7$ & C-2 WP \\ \hline
				\multirow[t]{4}{*}{$\mathrm{C}_{3}$} & 143-145 ($\Gamma$-$A$, $K$-$H$) & $\mathrm{C}^{+}_{3}$ & $\mathrm{\Gamma}_{h}$ & $R_1$$\oplus$$R_2$, $R_2$$\oplus$$R_3$ & $H_8$ & C-1 WP \\ 
				& & & & $R_1$$\oplus$$R_3$ & $H_9$ & C-1 WP \\
				& 146 ($\Gamma$-$Z$, $Z$-$P$) & $\mathrm{C}^{+}_{3}$ & $\mathrm{\Gamma}_{rh}$ & $R_1$$\oplus$$R_2$, $R_2$$\oplus$$R_3$ & $H_8$ & C-1 WP \\ 
				& & & & $R_1$$\oplus$$R_3$ & $H_9$ & C-1 WP \\ 
				& 143-145 ($\Gamma$, $A$) & $\mathrm{C}^{+}_{3}$, $\mathcal{T}$ & $\mathrm{\Gamma}_{h}$ & $R_2$$R_3$ & $H_{10}$ & C-2 WP \\ 
				& 146 ($\Gamma$, $Z$) & $\mathrm{C}^{+}_{3}$, $\mathcal{T}$ & $\mathrm{\Gamma}_{rh}$ & $R_2$$\oplus$$R_3$ & $H_{10}$ & C-2 WP \\ \hline
				\multirow[t]{2}{*}{$\mathrm{C}_{3v}$} & \multirow[t]{2}{*}{156, 158 ($K$-$H$)} & \multirow[t]{2}{*}{$\mathrm{C}^{+}_{3}$} & \multirow[t]{2}{*}{$\mathrm{\Gamma}_{h}$} & $R_1$$\oplus$$R_2$, $R_2$$\oplus$$R_3$ & $H_8$ & C-1 WP \\ 
				&  &  &  & $R_1$$\oplus$$R_3$ & $H_9$ & C-1 WP \\ \hline
				\multirow[t]{11}{*}{$\mathrm{C}_{6}$} & \multirow[t]{2}{*}{168-173 ($\Gamma$-$A$)} & \multirow[t]{2}{*}{$\mathrm{C}^{+}_{6}$} & \multirow[t]{11}{*}{$\mathrm{\Gamma}_{h}$} & $R_1$$\oplus$$R_2$, $R_2$$\oplus$$R_3$, $R_3$$\oplus$$R_4$, $R_4$$\oplus$$R_5$, $R_5$$\oplus$$R_6$ & $H_{11}$ & C-1 WP \\ 
				&  &  &  & $R_1$$\oplus$$R_6$ & $H_{12}$ & C-1 WP \\ 
				& 168-173 ($M$-$L$) & $\mathrm{C}_{2}$ & \multirow[t]{11}{*}{$\mathrm{\Gamma}_{h}$} & $R_1$$\oplus$$R_2$ & $H_{13}$ & C-1 WP \\ 
				& \multirow[t]{2}{*}{168-173 ($K$-$H$)} & \multirow[t]{2}{*}{$\mathrm{C}^{+}_{3}$} & \multirow[t]{11}{*}{$\mathrm{\Gamma}_{h}$} & $R_1$$\oplus$$R_2$, $R_2$$\oplus$$R_3$ & $H_{11}$ & C-1 WP \\ 
				&  &  &  & $R_1$$\oplus$$R_3$ & $H_{12}$ & C-1 WP \\  
				& 168, 171, 172 ($K$, $H$); 169, 170, 173 ($K$) & $\mathrm{C}^{+}_{3}$, $\mathrm{C}^{+}_{6}$$\mathcal{T}$ & \multirow[t]{11}{*}{$\mathrm{\Gamma}_{h}$} & $R_2$$R_3$ & $H_{14}$ & C-1 WP \\
				& \multirow[t]{2}{*}{168-173 ($\Gamma$-$A$)} & $\mathrm{C}^{+}_{6}$ & \multirow[t]{11}{*}{$\mathrm{\Gamma}_{h}$} & $R_1$$\oplus$$R_3$, $R_2$$\oplus$$R_4$, $R_3$$\oplus$$R_5$, $R_4$$\oplus$$R_6$ & $H_{15}$ & C-2 WP \\  
				&  &  &  & $R_1$$\oplus$$R_5$, $R_2$$\oplus$$R_6$ & $H_{15}$ & C-2 WP \\  
				& \multirow[t]{2}{*}{168, 171, 172 ($\Gamma$, $A$); 169, 170, 173 ($\Gamma$)} & $\mathrm{C}^{+}_{6}$, $\mathcal{T}$ & \multirow[t]{11}{*}{$\mathrm{\Gamma}_{h}$} & $R_2$$R_6$ & $H_{16}$ & C-2 WP \\ 
				&  &  &  & $R_3$$R_5$ & $H_{17}$ & C-2 WP \\  
				& 168-173 ($\Gamma$-$A$) & $\mathrm{C}^{+}_{6}$ & \multirow[t]{11}{*}{$\mathrm{\Gamma}_{h}$} & $R_1$$\oplus$$R_4$, $R_2$$\oplus$$R_5$, $R_3$$\oplus$$R_6$ & $H_{18}$ & C-3 WP \\ \hline\hline
			\end{tabular}%
		}
	\label{tabs1}
	\end{table}

\begin{table}[ht]
\caption{The expression of the Hamiltonian in the “$k\cdot p$ Hamiltonian” column of Table~\ref{tabs1}
}
		\renewcommand{\arraystretch}{1.5}
		\centering
		\normalsize % Reduce font size
		\resizebox{\textwidth}{!}{%
			\begin{tabular}{ll}
				\hline
				\hline % Double line at top
				$k \mathbin{\cdot} p$ Hamiltonian &  \\
				\midrule
				$H_1$ & $\sigma_0(c_1+c_2k_z)+c_3\sigma_3k_z+\sum_{i=1}^2\sigma_i(c_{i,1}k_x+c_{i,2}k_y)$ \\
				$H_2$ & \(\sigma_0 (c_1 + c_2 k_z)+c_3\sigma_3 k_z+(\alpha_1\sigma_+k_- + h.c.)\)  \\
				$H_3$ & \(\sigma_0 (c_1 + c_2 k_z)+c_3\sigma_3 k_z+(\alpha_1\sigma_+k_+ + h.c.)\)  \\
				$H_4$ & \( c_1\sigma_0 + c_3(\sigma_2 k_x+\sigma_1 k_y)+c_4(\sigma_1 k_x - \sigma_2 k_y)+c_2\sigma_3 k_z\)  \\
				$H_5$ & \( \sigma_0 (c_1 + c_2 k_z + c_3k^2 + c_4k_z^2)+ \sigma_3 (c_5 k_z + c_6k^2 + c_7k_z^2)+ [\sigma_+ (\alpha_1k_+^2 + \alpha_2k_-^2)+ h.c.]\)  \\
				$H_6$ & \( (c_1 + c_2k^2 + c_3k_z^2)\sigma_0 + \sum_{i = 1}^{2}[c_{i,1}k_xk_y + c_{i,2}(k_x^2 - k_y^2)]\sigma_i + c_4\sigma_3k_z\)  \\
				$H_7$ & \( (c_1 + c_2k^2 + c_3k_z^2)\sigma_0 + \sum_{i = 1}^{3}[c_{i,1}k_xk_y + c_{i,2}(k_x^2 - k_y^2)+c_{i,3}k_z]\sigma_i\)  \\
				$H_8$ & \( \sigma_0 (c_1 + c_2 k_z)+c_3\sigma_3 k_z + c_5(\sigma_1 k_y - \sigma_2 k_x)+c_6(\sigma_1 k_x + \sigma_2 k_y)\)  \\
				$H_9$ & \( \sigma_0 (c_1 + c_2 k_z)+c_3\sigma_3 k_z + c_5(\sigma_2 k_x + \sigma_1 k_y)+c_6(\sigma_1 k_x - \sigma_2 k_y)\)  \\
				$H_{10}$ & \( [c_1 + c_2(k_x^2 + k_y^2)+c_3k_z^2]\sigma_0 + c_4\sigma_3k_z + [\sigma_+(\alpha_1k_zk_- + \alpha_2k_+^2)+h.c.]\)  \\
				$H_{11}$ & \( \sigma_0 (c_1 + c_2 k_z)+c_3\sigma_3 k_z + (\alpha_1k_-\sigma_+ + h.c.)\)  \\
				$H_{12}$ & \( \sigma_0 (c_1 + c_2 k_z)+c_3\sigma_3 k_z + (\alpha_1k_+\sigma_+ + h.c.)\)  \\
				$H_{13}$ & \( \sigma_0 (c_1 + c_2 k_z)+c_3\sigma_3 k_z+\sigma_1 (c_1 k_x + c_2 k_y)+\sigma_2 (c_3 k_x + c_4 k_y)\)  \\
				$H_{14}$ & \( c_1\sigma_0 + c_2(\sigma_1 k_x + \sigma_2 k_y)+c_3(\sigma_1 k_y - \sigma_2 k_x)+c_4\sigma_3 k_z\) \\
				$H_{15}$ &  \( \sigma_0c_1 + \sum_{i = 0,3}\sigma_0 [c_{i,1}k^2 + c_{i,2}k_z^2 + c_{i,3}k_z]+ (\alpha_1\sigma_+k_-^2 + h.c.)\) \\
				$H_{16}$ & \( [c_1 + c_2(k_x^2 + k_y^2)+c_3k_z^2]\sigma_0 + c_4\sigma_3k_z + (\alpha_1\sigma_+k_+^2 + h.c.)\)  \\
				$H_{17}$ &  \( [c_1 + c_2(k_x^2 + k_y^2)+c_3k_z^2]\sigma_0 + c_4\sigma_3k_z + (\alpha_1\sigma_+k_-^2 + h.c.)\) \\
				$H_{18}$ & \( \sigma_0c_1 + \sum_{i = 0,3}\sigma_0 \big[c_{i,1}k^2 + c_{i,2}k_z^2 + c_{i,3}k_z(1 + c_{i,4}k^2 + c_{i,5}k_z^2)\big]+ \big[(\alpha_1k_+^3 + \alpha_2k_-^3)\sigma_+ + h.c.\big]\) \\
				\hline
				\hline
			\end{tabular}%
		}
		\label{tabs2}
	\end{table}

\begin{figure}[b]
\centering
\includegraphics[width=1.0\columnwidth]{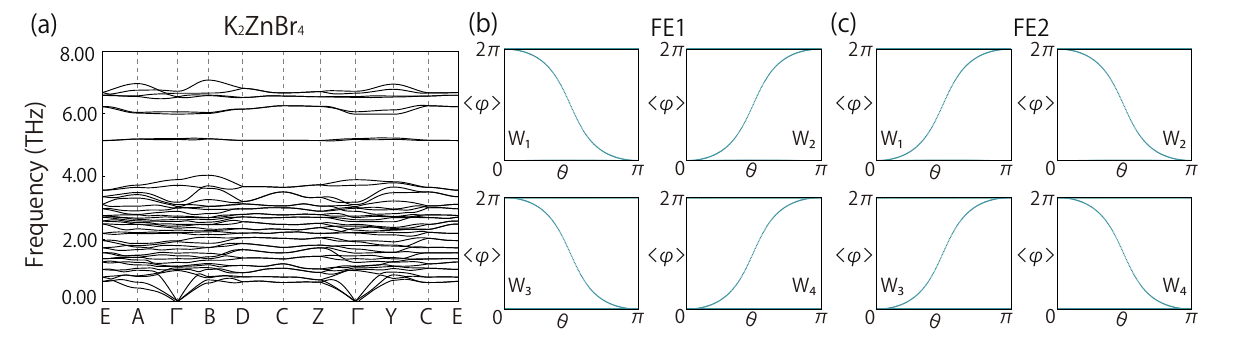} \caption{(a) The calculated phonon dispersion for K$_{2}$ZnBr$_{4}$ along the high-symmetry paths. (b) The calculated Wilson loop of the four Weyl point phonons ($W_{i=,1,2,3,4}$) in the FE1 phase and  (c) the FE2 phase. ~\label{figs1}}
\end{figure}

\section{Methods}
The first-principles calculations based on the density functional theory (DFT) were performed using the Vienna ab initio simulation package (VASP)~\cite{1}. The exchange-correlation potential was treated within the generalized gradient approximation (GGA) with Perdew-Burke-Ernzerhof (PBE)~\cite{2}. The cutoff energy was set to 500 eV for $A_{2}BX_{4}$ ($A$ = Rb, K; $B$ = Zn, Co and $X$ = Cl, Br, I), and 520 eV for Y(OH)$_{3}$, and 650 eV for LiIO$_{3}$.   A $\Gamma$-centered \emph{k}-mesh of $3\times3\times2$ [$5\times4\times4$ / $4\times3\times3$ / $5\times4\times4$ / $4\times3\times3$ / $4\times3\times3$ / $4\times3\times3$ / $5\times4\times4$ / $5\times4\times4$ / $5\times4\times4$ / $4\times4\times4$ / $4\times4\times6$] for K$_{2}$ZnBr$_{4}$ [K$_{2}$ZnI$_{4}$ / Rb$_{2}$ZnCl$_{4}$ / Rb$_{2}$ZnBr$_{4}$ / Rb$_{2}$CoCl$_{4}$ / Rb$_{2}$CoBr$_{4}$ / K$_{2}$ZnCl$_{4}$ / K$_{2}$CoCl$_{4}$ / K$_{2}$CoBr$_{4}$, K$_{2}$CoI$_{4}$ / Y(OH)$_{3}$ / LiIO$_{3}$] was used for BZ sampling. The energy and force convergence criteria are set to be 10$^{-7}$ eV and 10$^{-3}$ eV/$\mathring{\mathrm{A}}$, respectively. The energy barriers of the various kinetic processes in ferroelectric (FE) materials are calculated using the climbing image nudged elastic band (CINEB) method~\cite{3}. The force constants were obtained through VASP combined with the PHONOPY package~\cite{4}. The surface states are calculated using the iterative Green's function method based on the phononic tight-binding Hamiltonian constructed with the WannierTools software package~\cite{5}. 
\begin{figure}[H]
\centering
\includegraphics[width=0.9\columnwidth]{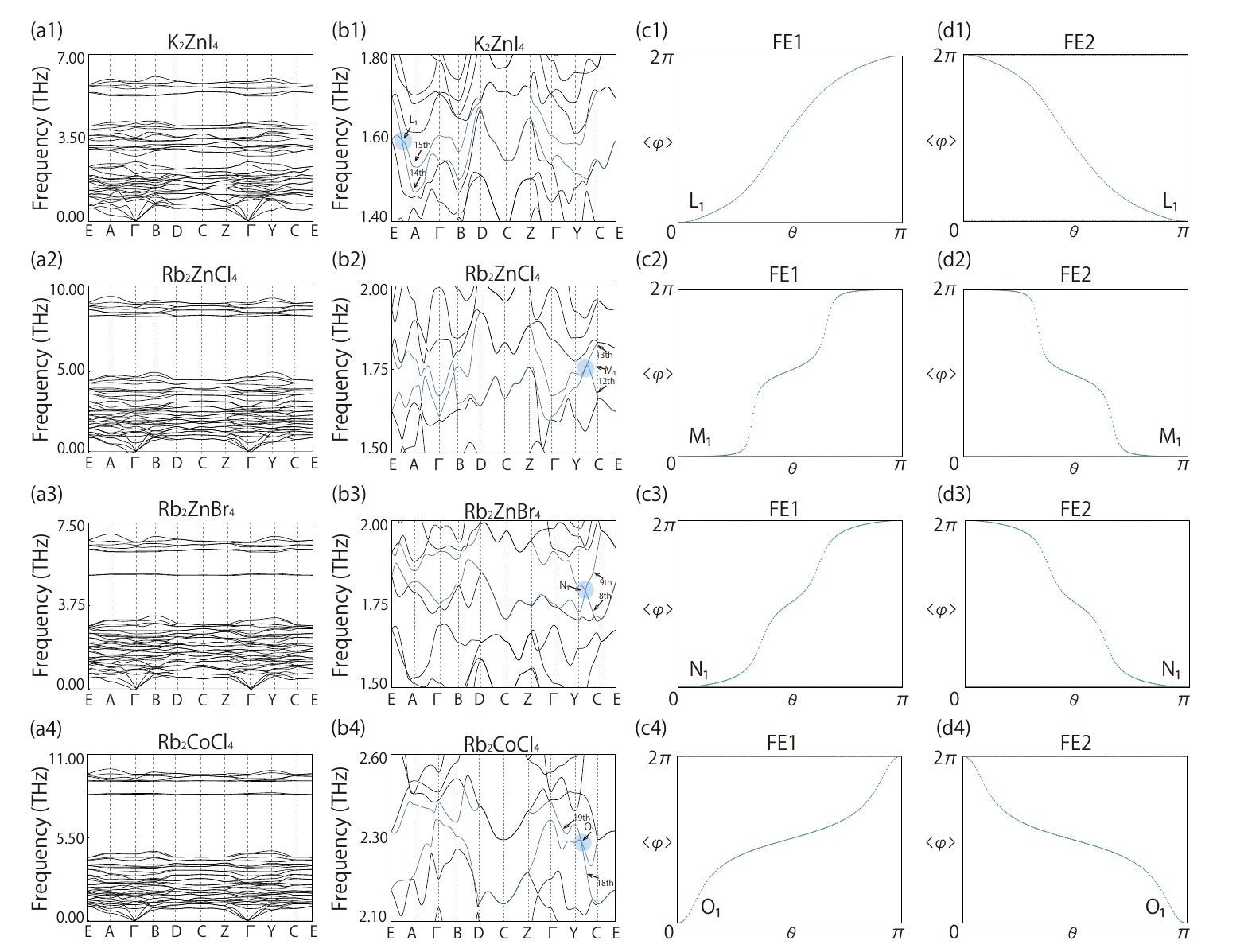} \caption{(a1)-(a4) The calculated phonon dispersion for K$_{2}$ZnI$_{4}$, Rb$_{2}$ZnCl$_{4}$, Rb$_{2}$ZnBr$_{4}$, Rb$_{2}$CoCl$_{4}$ along the high-symmetry paths. (b1)-(b4) show the enlarged phonon dispersions. The phonon branches forming the Weyl phonons are highlighted by blue curves. (c1)-(c4) show the Wilson loops of Weyl points $L_{1}$, $M_{1}$, $N_{1}$, $O_{1}$ in the FE1 phase, respectively. (d1)-(d4) denote the corresponding Wilson loops in the FE2 phase.
 ~\label{figs2}}
\end{figure}
\section{Ferroelectric controllable Weyl phonon materials}
Here, we present more candidates of ferroelectric switchable Weyl phonon materials with chirality $C=1,2$, and 3.
\subsection{C-1 WP}
In the main text, we have shown the phonon spectrum of K$_{2}$ZnBr$_{4}$ within a specific frequency range. Here, we present the whole phonon dispersion, as shown in Fig.~\ref{figs1}(a). The chirality of the Weyl point (W$_{i=1,2,3,4}$) mentioned in the text is calculated using the Wilson loop method. Specifically, a sphere parametrized by azimuth angle $\varphi$ and elevation angle $\theta$ is chosen to enclose the Weyl point. By calculating the evolution of the Wannier center $\left\langle \varphi\right\rangle $ with respect to $\theta$, the chirality can be obtained. The results of two ferroelectric (FE) phases are shown in Fig.~\ref{figs1}(b) and~\ref{figs1}(c).

\begin{figure}[h]
\centering
\includegraphics[width=0.9\columnwidth]{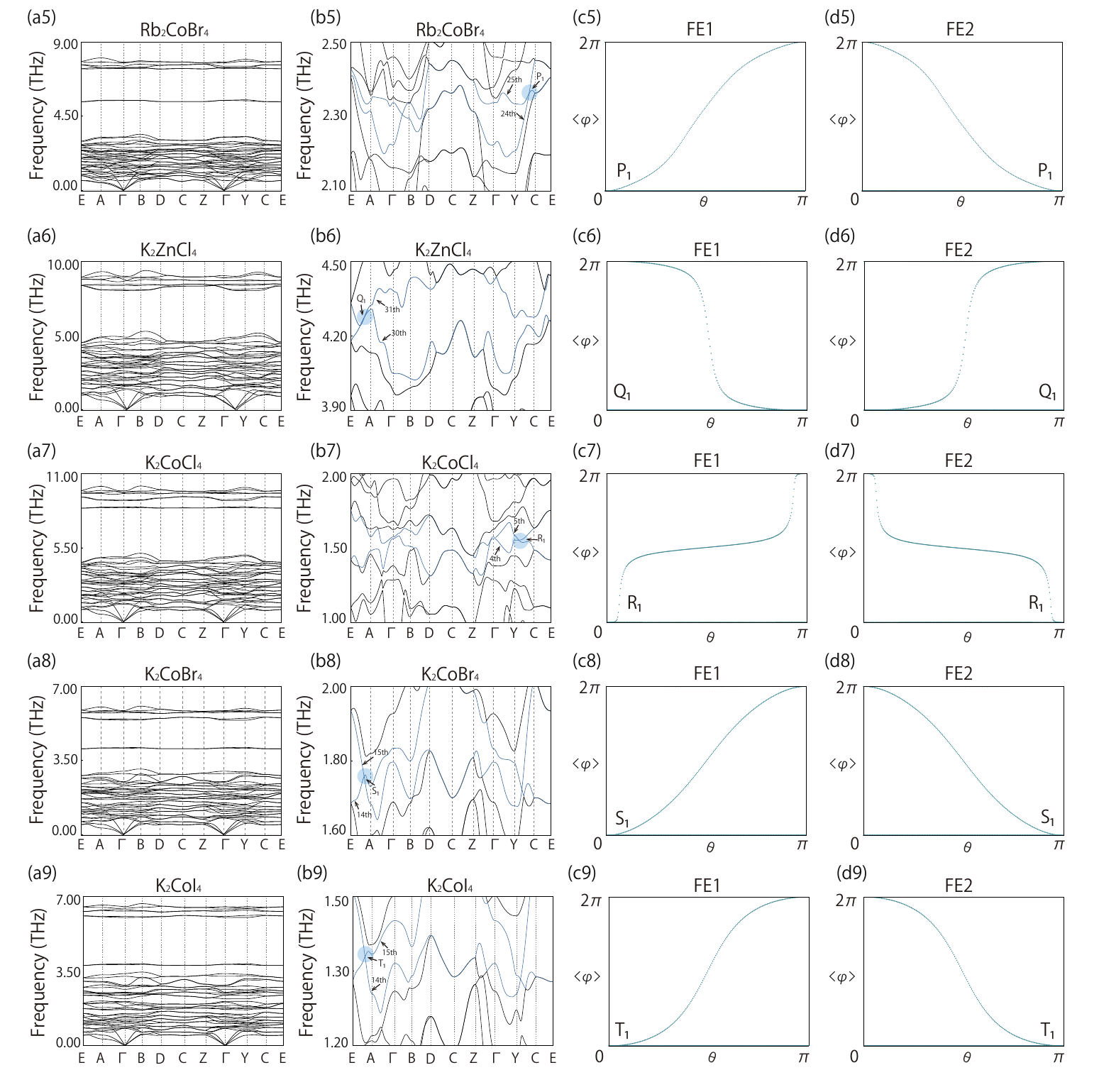} \caption{(a5)-(a9) The calculated phonon dispersion for Rb$_{2}$CoBr$_{4}$, K$_{2}$ZnCl$_{4}$, K$_{2}$CoCl$_{4}$, K$_{2}$CoBr$_{4}$, K$_{2}$CoI$_{4}$ along the high-symmetry paths. (b5)-(b9) show the enlarged phonon dispersions. The phonon branches forming the Weyl phonons are highlighted by blue curves. (c5)-(c9) show the Wilson loops of Weyl points $P_{1}$, $Q_{1}$, $R_{1}$, $S_{1}$, $T_{1}$ in the FE1 phase, respectively. (d5)-(d9) denote the corresponding Wilson loops in the FE2 phase. ~\label{figs3}}
\end{figure}

Moreover, the phonon spectra of other ferroelectric materials in the  $A_{2}BX_{4}$ ($A$ = Rb, K; $B$ = Zn, Co and $X$ = Cl, Br, I) family, including K$_{2}$ZnI$_{4}$, Rb$_{2}$ZnCl$_{4}$, Rb$_{2}$ZnBr$_{4}$, Rb$_{2}$CoCl$_{4}$, Rb$_{2}$CoBr$_{4}$, K$_{2}$ZnCl$_{4}$, K$_{2}$CoCl$_{4}$, K$_{2}$CoBr$_{4}$ and K$_{2}$CoI$_{4}$, are also presented in Fig.~\ref{figs2} and Fig.~\ref{figs3}.\\
\\
\\
%\newpage
\subsection{C-2 and C-3 WPs}
\begin{figure}[h]
\centering
\includegraphics[width=0.85\columnwidth]{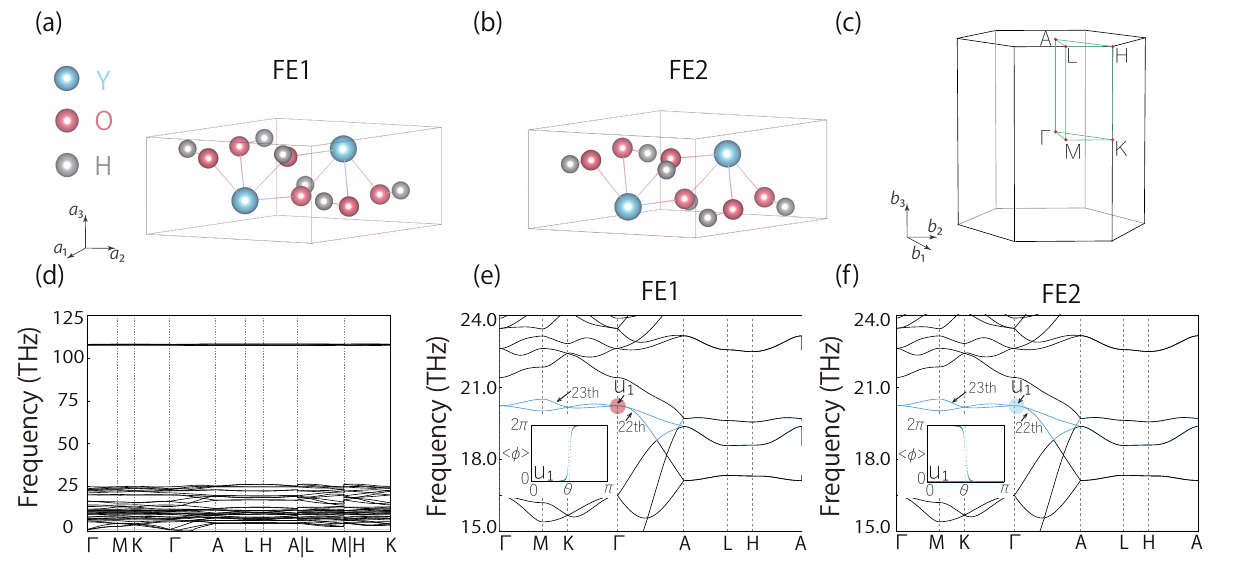} \caption{The crystal structure of Y(OH)$_{3}$ for (a) the FE1 phase and (b) the FE2 phase. (c) Bulk BZ of Y(OH)$_{3}$. (d) Calculated phonon spectrum of Y(OH)$_{3}$ along high-symmetry paths. (e) and (f) respectively show the enlarged phonon dispersions of FE1 and FE2 phases. The insets in (e) and (f) show the obtained Wilson loops of Weyl point U$_{1}$ at the $\Gamma$ point. ~\label{figs4}}
\end{figure}

In addition, we have also identified candidates for ferroelectric switchable Weyl phonon materials with high Chern numbers, hosting C-2 WP and C-3 WP. For C-2 WP, we demonstrate that Y(OH)$_{3}$ can serve as a concrete material example, which has been experimentally synthesized~\cite{beall1977structural}. The crystal Y(OH)$_{3}$ belongs to SG 173, one of the candidates in Table~\ref{tabs1}, and its crystal structure is illustrated in Figs.~\ref{figs4}(a)-~\ref{figs4}(b)
. The obtained phonon dispersion of Y(OH)$_{3}$ along the high-symmetry paths is shown in Figs.~\ref{figs4}(d)-~\ref{figs4}(f). Obviously, we find that a Weyl point exists at the $\Gamma$ points.  With the help of Wilson loop method, we determine that the chirality of this Weyl point is $C=2$ [see Fig.~\ref{figs4}(e)].  After the ferroelectric polarization reversal, the chirality of the Weyl point becomes $C=-2$ [see Fig.~\ref{figs4}(f)].

\begin{figure}[h]
\centering
\includegraphics[width=0.85\columnwidth]{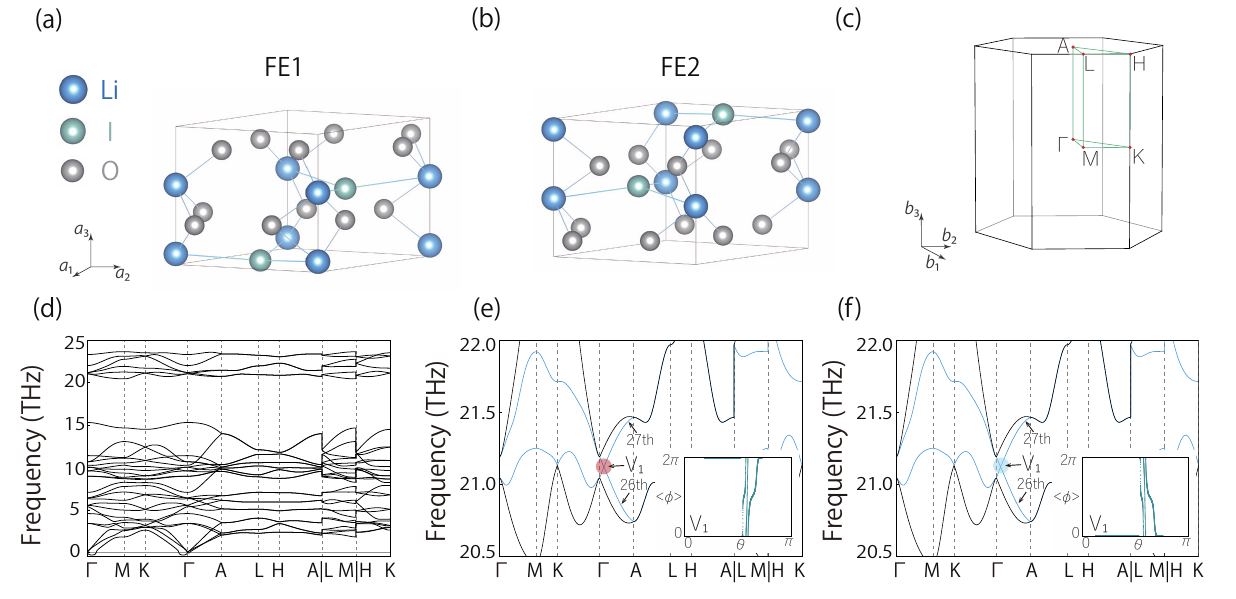} \caption{The crystal structure of LiIO$_{3}$ for (a) the FE1 phase and (b) the FE2 phase. (c) Bulk BZ of LiIO$_{3}$. (d) Calculated phonon spectrum of LiIO$_{3}$ along high-symmetry paths. (e) and (f) respectively show the enlarged phonon dispersions of FE1 and FE2 phases. The insets in (e) and (f) show the obtained Wilson loops of Weyl point V$_{1}$ at the $\Gamma$-$A$ path. ~\label{figs5}}
\end{figure}

In the case of C-3 WP, we demonstrate that LiIO$_{3}$ with SG 173 can serve as a concrete material example, which has been successfully synthesized experimentally~\cite{coquet1983effect}. The crystal structure of LiIO$_{3}$ is presented in Figs.~\ref{figs5}(a)-~\ref{figs5}(b), and the calculated phonon spectrum is shown in Figs.~\ref{figs5}(d)-~\ref{figs5}(f). One can observe a band crossing with hourglass-type on $\Gamma$-$Z$ path. The result of Wilson loop, which surrounding band crossing, indicates it is a C-3 WP with $C=3$ [see Fig.~\ref{figs5}(e)]. Similarly, the sign of the chirality of the C-3 WP can also be manipulated by reversing the ferroelectric polarization [see Fig.~\ref{figs5}(f)] (It is worth noting that, owing to the Nielsen-Ninomiya no-go theorem, the net topological charge of the system would be zero. Hence, besides the discussed $C=2$ and $C=3$ Weyl points, there must exist other WPs in the system to balance the chirality. These are not further discussed in detail here).

\section{Nonlinear phonon Hall effect in K$_{2}$ZnBr$_{4}$}

The nonlinear phonon Hall effect (NPHE) has been proposed to exist in non-centrosymmetric phonon systems~\cite{PhysRevB.107.L241107}, and the second order of the temperature gradient induced thermal current can be written as
\begin{eqnarray}\label{h1}
j_{a}	=	\chi_{adb}\nabla_{d}T\nabla_{b}T,
\end{eqnarray}
with 
\begin{eqnarray}\label{h1}
\chi_{adb}	=	\frac{\tau}{V\hbar^{2}}\sum_{\boldsymbol{q},n}\frac{E_{n,\boldsymbol{q}}^{3}}{T^{2}}\frac{\partial\rho_{0}}{\partial q_{d}}\Omega_{n,ab}(\boldsymbol{q}).
\end{eqnarray}
Here, $T$ is the temperature and $V$ is the material's volume. $k_{B}$ is the Boltzmann constant, and $\hbar$ is the reduced Planck constant. $\boldsymbol{q}$ is the phonon Bloch wave vector, and $n$ is the band index. The phonon relaxation time $\tau$ depends on $\boldsymbol{q}$ and $n$. $\rho_{0}$ and $E_{n,\boldsymbol{q}}$ are the phonon distribution function and energy, respectively. $\Omega$ represents the phonon Berry curvature. $a$, $d$ and $b$ (equivalent to $x$, $y$, and $z$) serve as Cartesian coordinates. The ferroelectric phase of K$_{2}$ZnBr$_{4}$ has the symmetry of space group P2$_{1}$ (No.4) and C$_{2}$ point group. This symmetry implies that the response coefficient $\chi_{adb}$ has the form of 
\begin{eqnarray*}
\chi_{adb} & = & \left[\begin{array}{ccccccccc}
0 & 0 & 0 & 0 & 0 & \chi_{xzy} & 0 & \chi_{xyz} & 0\\
\chi_{yxx} & 0 & \chi_{yzx} & 0 & 0 & 0 & \chi_{yxz} & 0 & \chi_{yzz}\\
0 & \chi_{zyx} & 0 & \chi_{zxy} & 0 & 0 & 0 & 0 & 0
\end{array}\right].
\end{eqnarray*}
In the main text, we choose the  $\chi_{yxx}$ component as an example to construct an electrically controlled nonlinear thermal current device.

\begin{figure}[h]
\centering
\includegraphics[width=0.8\columnwidth]{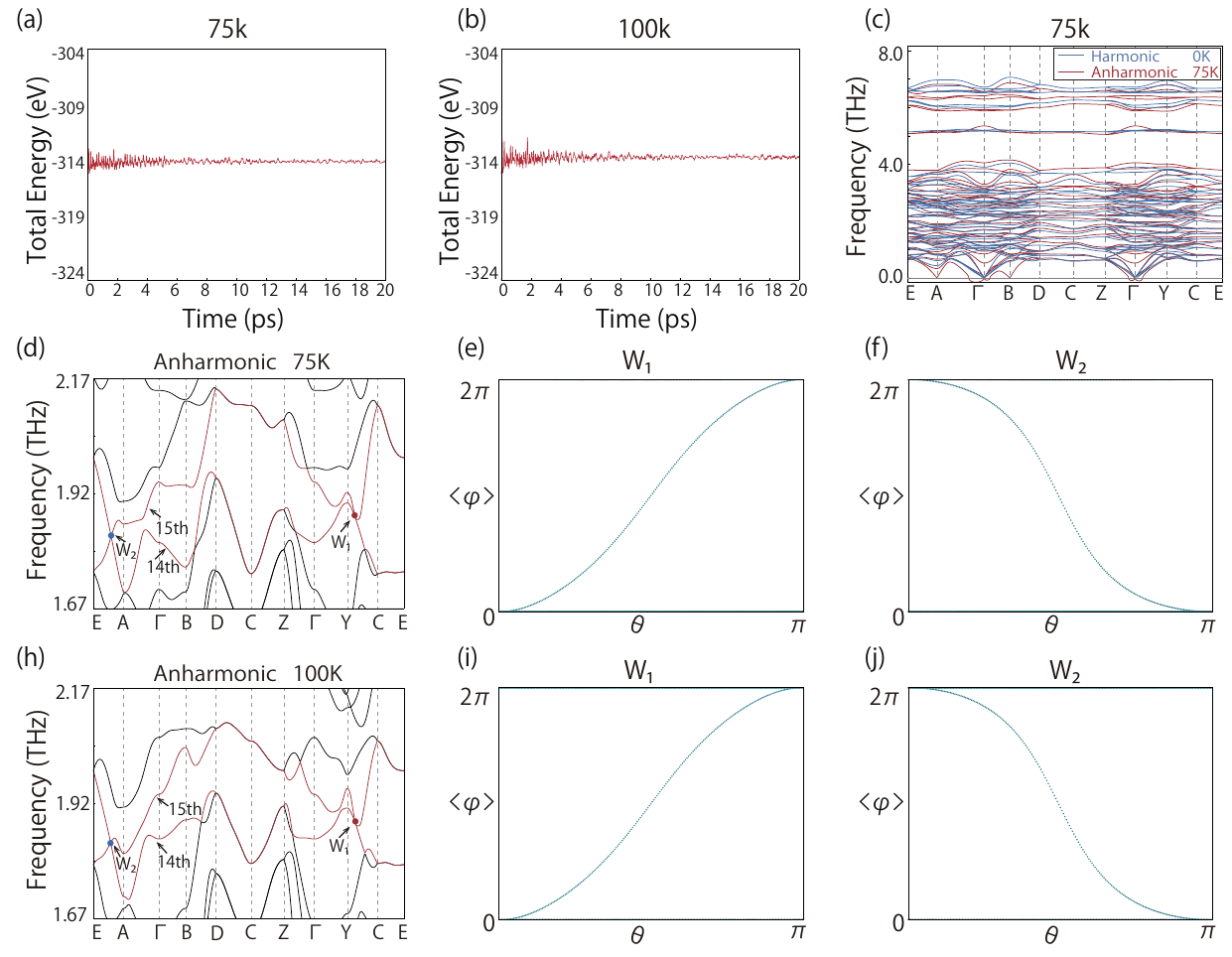} \caption{(a) and (b) show the total energy of the K$_{2}$ZnBr$_{4}$ structure as a function of simulation time. (c) Harmonic phonon dispersion (0 K) and anharmonic phonon dispersion at 75 K for K$_{2}$ZnBr$_{4}$ in FE2 phase. (d) and (h) show enlarged views of anharmonic phonon dispersion with Weyl points at 75 K and 100 K, respectively. The phonon branches forming the Weyl points are highlighted by red curves. (e)-(f) and (i)-(j) present the Wilson loops corresponding to the Weyl points shown in (d) and (h), respectively. ~\label{figs6}}
\end{figure}

\section{Temperature and anharmonicity effects on phonons}
We investigate the stability of Weyl points in K$_{2}$ZnBr$_{4}$ under temperature fluctuations and lattice anharmonicity. This topic is frequently discussed in phononic studies of materials and will be beneficial for further experimental observations. The renormalized phonon spectra, including anharmonicity and temperature effects, can be obtained by processing molecular dynamics trajectories using the DYNAPHOPY code~\cite{carreras2017dynaphopy}. Molecular dynamics simulations were performed with \textit{ab initio} molecular dynamics (AIMD) capabilities implemented in VASP using $2\times2\times2$ supercells. 

Since the experimental transition temperature of K$_{2}$ZnBr$_{4}$ from the FE to paraelectric phase is $T_{c}=$155 K~\cite{abrahams1994experimental}, we selected AIMD simulations at 75 K and 100 K below $T_{c}$ as illustrative cases. Each molecular dynamics simulation lasted 20 ps with a time step of 2 fs, and the results are shown in Figs.~\ref{figs6}(a)-~\ref{figs6}(b). One observes that the total energy has a convergent result with no violent fluctuations. Furthermore, the anharmonic phonon spectrum at finite temperature can be obtained by extracting renormalized force constants from AIMD simulations.  As shown in Fig.~\ref{figs6}(c), we use K$_{2}$ZnBr$_{4}$ at 75 K as an example to compare the anharmonic phonon spectrum with the harmonic counterpart (at 0 K). It is evident that the anharmonic phonon spectrum exhibits a noticeable spectral reconstruction. 

We are primarily concerned with the robustness of the Weyl points against temperature and lattice anharmonicity, so we show the renormalized phonon spectrum of K$_{2}$ZnBr$_{4}$ in the FE2 phase within the 1.6-2.1 THz range at 75 K and 100 K, as shown in Figs.~\ref{figs6}(d) and~\ref{figs6}(h). We can observe that both Weyl points along $E-A$ and $Y-C$ remain stable. Interestingly, under the influence of temperature fluctuations and lattice anharmonicity, the Weyl point $W_{1}$ along the $Y-C$ path shifts toward the $Y$ point, thereby changing it from a type-I to a type-II Weyl point. Furthermore, we also calculate the topological charge of these Weyl points in the renormalized phonon spectrum [see Figs.~\ref{figs6}(e), ~\ref{figs6}(f), ~\ref{figs6}(i), and ~\ref{figs6}(j)], and the result is consistent with the harmonic case at 0 K.

%
%\newpage

%\bibliographystyle{aipnum4-1}
%\newpage
\bibliography{supref}